\documentclass[11pt]{article}

\usepackage[utf8]{inputenc}
\usepackage{latexsym}
\usepackage{amssymb,amsmath}
\usepackage{amsthm}
\usepackage{thmtools,thm-restate}
\usepackage{amsfonts}
\usepackage{parskip}
\usepackage{physics} 
\usepackage{authblk}
\usepackage{xcolor}
\usepackage[margin=1in]{geometry}
\usepackage{graphicx}
\usepackage[numbers,sort&compress]{natbib}  
\usepackage{breqn}
\usepackage{blkarray}
\usepackage{makecell}
\usepackage{float}
\usepackage{appendix}
\usepackage{footnote}
\usepackage{caption}
\usepackage{subcaption}

\usepackage[colorinlistoftodos]{todonotes}

\usepackage{array}
\usepackage{setspace}
\usepackage[colorlinks=true, allcolors=blue]{hyperref}

\numberwithin{equation}{section}

\newcommand{\bbm}{\begin{bmatrix}}
\newcommand{\bpm}{\begin{pmatrix}}
\newcommand{\ebm}{\end{bmatrix}}
\newcommand{\epm}{\end{pmatrix}}

 \newcommand{\del}[2]{\frac{\partial #1}{\partial #2}}
 \newcommand{\dsdel}[2]{\displaystyle\frac{\partial #1}{\partial #2}}

 \newcommand{\doubledelsame}[2]{\displaystyle\frac{\partial^2 #1}{\partial #2^2}}

\newcommand{\revision}[1]{#1}

\usepackage{projmacros}

\author[1,*]{Daniel B. Cooney}
\author[2,*]{Dylan H. Morris}
\author[3]{Simon A. Levin}
\author[3]{Daniel I. Rubenstein}
\author[4,5,6]{Pawel Romanczuk}

\affil[1]{\small{Department of Mathematics, University of Pennsylvania, Philadelphia, USA}}
\affil[2]{\small{Department of Ecology \& Evolutionary Biology, University of California, Los Angeles, USA}}
\affil[3]{\small{Department of Ecology \& Evolution Biology, Princeton University, Princeton, USA}}
\affil[4]{\small{Institute for Theoretical Biology, Department of Biology, Humboldt-Universität zu Berlin, 10115 Berlin, Germany}}
\affil[5]{\small{Bernstein Center for Computational Neuroscience Berlin, 10115 Berlin, Germany}}
\affil[6]{\small{Science of Intelligence, Research Cluster of Excellence, Marchstr. 23, 10587 Berlin, Germany, https://www.scienceofintelligence.de}}

\affil[*]{These authors contributed equally and should be considered joint first authors; correspondence to dbcooney@sas.upenn.edu, dylan@dylanhmorris.com}

\title{Social dilemmas of sociality due to beneficial and costly contagion}
\date{\today}

\begin{document}

\theoremstyle{plain}
\newtheorem{theorem}{Theorem}
\newtheorem{proposition}{Proposition}[section]
\theoremstyle{remark}
\newtheorem{remark}{Remark}[section]

\maketitle

\begin{abstract}
Levels of sociality in nature vary widely. Some species are solitary; others live in family groups; some form complex multi-family societies. Increased levels of social interaction can allow for the spread of useful innovations and beneficial information, but can also facilitate the spread of harmful contagions, such as infectious diseases. It is natural to assume that these contagion processes shape the evolution of complex social systems, but an explicit account of the dynamics of sociality under selection pressure imposed by contagion remains elusive.

We consider a model for the evolution of sociality strategies in the presence of both a beneficial and costly contagion. We study the dynamics of this model at three timescales: using a susceptible-infectious-susceptible (SIS) model to describe contagion spread for given sociality strategies, a replicator equation to study the changing fractions of two different levels of sociality, and an adaptive dynamics approach to study the long-time evolution of the population level of sociality.

For a wide range of assumptions about the benefits and costs of infection, we identify a social dilemma: the evolutionarily-stable sociality strategy (ESS) is distinct from the collective optimum---the level of sociality that would be best for all individuals. In particular, the ESS level of social interaction is greater (respectively less) than the social optimum when the good contagion spreads more (respectively less) readily than the bad contagion.

Our results shed light on how contagion shapes the evolution of social interaction, but reveals that evolution may not necessarily lead populations to social structures that are good for any or all.
\end{abstract}

\hypersetup{linkbordercolor=black, linkcolor = black}
\begin{spacing}{0.01}
\renewcommand{\baselinestretch}{0.1}\normalsize
\tableofcontents
\addtocontents{toc}{\protect\setcounter{tocdepth}{2}}
\end{spacing}
\singlespacing

\section{Introduction}

\subsection{The evolution of sociality}
Social interaction has evolved many times \cite{rubenstein2017evolution}. Evolutionary explanations for social behavior cite its many benefits, such as the sharing of collective information. But social living has costs. For one, it can facilitate the spread of infectious diseases \cite{alexander1974evolution}. A basic premise of disease biology and modeling is that higher levels of contact among individuals facilitates disease spread \cite{kermack1927contribution}. But beneficial contagion is also possible, such as the adoption of useful innovations \cite{bass1969new,rogers2010diffusion} or the spread of social information \cite{romano2020stemming}. 
Evolutionary tradeoffs between contagious benefits and contagious costs of social behavior have been hypothesized \cite{romano2020stemming} and modeled theoretically for specific animal systems, such as bats \cite{kashima2013fission}, but a general, formal theory of how selection imposed by contagions shapes the rate social interaction itself is needed.  

In this paper, we formulate a simple and flexible framework for modeling the evolution of sociality strategies given the benefits and costs of social transmission. We use an adaptive dynamics approach  \cite{brannstrom2013hitchhiker,diekmann2002beginners,geritz1998evolutionarily} to model the evolution of sociality in the presence of contagious benefits and costs. In this framework, epidemiology (contagion processes), competition (success or failure of different sociality strategies), and trait evolution (change in the overall population sociality level) occur on separate timescales: contagion is fastest, then competition, then trait evolution.

We find that the evolutionary consequences of beneficial and costly contagion for social behavior depend not only on the benefits and harms of the two contagions but also on their relative transmissibility. When the two contagions are not precisely equal in their transmissibilty, a social dilemma occurs. Individual-level selection on sociality drives the population to an equilibrium at which all individuals are worse off than they would be if they all could agree on some socially-optimal shared social interaction level: either with less access to the good contagion (e.g. less informed) or with more exposure to the bad (e.g. sicker). In some cases, beneficial sociality vanishes entirely: in an attempt to avoid disease, individuals reduce their interaction to the point that the beneficial contagion cannot spread at all. We explore ways in which this evolutionary problem---analogous to a prisoners' dilemma---can be mitigated.

Our results reveal that evolutionary dynamics can produce social behavior, but that one should not assume that such behavior is optimal for the population---or even ideal for the individual. Rather, evolved levels of social interaction must be stable against evolutionary invasion---even if that means an outcome that is sub-optimal for all members of the social group.

\subsection{Existing related work}
A substantial theoretical and empirical literature has addressed the evolution of sociality \cite{rubenstein2017evolution}, but the role of contagion processes \revision{has received minimal theoretical attention}, despite the fact that contagion is a near-ubiquitous property of social networks. It has often been proposed that animals may modulate their social interactions to avoid infectious diseases \cite{romano2020stemming}, but few models exist that study how this disease avoidance might trade off against the benefits of social contact.

The infectious disease literature includes adaptive behavior models, in which individuals change their behavior during epidemics to avoid \cite{fenichel2011adaptive,morin2013sir,morin2014disease} or seek out \cite{berdahl2019dynamics} the contagion. Similarly, Reluga \cite{reluga2009sis} provides a game-theoretic treatment of disease avoidance in a two-population disease model. \revision{Other authors have modeled} social contagion processes that modulate simultaneously-spreading infection processes  \cite{Zhan2018CouplingDO,perra2011towards,peng2021multilayer}\revision{: examples include models of} the simultaneous spread of an interaction strategy alongside an infectious disease  \cite{tanaka2002coevolution,bauch2005imitation,papst2022modeling,cascante2022disease}, and of the spread of a social contagion of awareness or fear (and thus disease-avoiding behaviors) alongside an infectious epidemic \cite{greenhalgh2015awareness,perra2011towards}.

Much of this literature focuses on strategic or contagious behavior changes that occur on the same timescale as the disease outbreak. We are interested in the long-term evolution of social behavior itself. \revision{When the costs and benefits of contagion drive evolution, what are the consequences for levels of social behavior?}

Recent work has explored the interplay between infectious disease and the behavioral evolution of social interactions \cite{nunn2015sociality}, including models showing that division-of-labor can emerge in the presence of disease risk \cite{udiani2020disease}, that serial monogamy can arise under selection imposed by sexually-transmitted pathogens \cite{mcleod2014sexually}, and that fission-fusion dynamics in bat colonies can result from tradeoffs between information about roosting sites and risks from pathogen infection \cite{kashima2013fission}. A common theme across these systems is the tradeoff between the benefits of social learning and the costs of infection \cite{evans2020infected}, and past studies have also explored the coevolution of host sociality and pathogen virulence \cite{bonds2005higher,ashby2020social}. Most relevant to the current paper is a model proposed by Ashby and Farine \revision{\cite{ashby2020social}}, who studied the evolution of social interaction rates in the presence of an informational contagion that conferred decreased mortality to an infectious disease spreading simultaneously in the population. \revision{The main contributions of our paper are (1)} our formulation of a flexible framework for exploring the tradeoff between the costs of a bad contagion and the benefits of a good contagion, \revision{and (2)} showing how the resulting evolution of social interaction rates may produce sub-optimal outcomes for the population.

\subsection{Structure of the paper}
In this paper, we study the epidemiological and evolutionary dynamics of sociality strategies over three progressively longer timescales. \revision{We first consider a fast epidemiological timescale, and study how the good and bad contagion converge to either endemic or disease-free equilibria depending on the levels of social interaction in the population, including when two sociality strategies exist in fixed proportions in the population.} Next, we consider how the utility of individuals obtained from steady-state levels of the good and bad contagion can influence the evolution of sociality strategies, using a replicator equation framework to see how the fraction of resident and mutant sociality strategies will evolve on an intermediate evolutionary timescale. Finally, we consider a slower evolutionary timescale in which the quantitative resident level of sociality can evolve, and employ the framework of adaptive dynamics to explore the long-run evolutionarily-stable sociality strategies. For both the replicator equation and adaptive dynamics frameworks, we consider the tension between individual and social utility optimization, asking how the levels of sociality chosen by self-interested individuals compare to the balance of good and bad contagion that would provided by a central planner.

The remainder of the paper is structured as follows. In Section \ref{sec:baselinemodel}, we introduce our model of the spread of good and bad contagions via social interaction, formulate a utility function balancing the benefits of the desirable contagion and the costs of the undesirable contagion, and calculate the socially-optimal level of sociality.

In Section \ref{sec:dimorphicevolutionary}, we analyze the evolutionary competition between pairs of sociality strategies, showing with a replicator equation that the socially-optimal sociality strategy can be invaded by strategies that leave the population worse off in the long-run.

In Section \ref{sec:adaptivedynamics}, we characterize the long-term evolution of the level of sociality using adaptive dynamics, finding that the evolutionarily-stable sociality strategy disagrees with the social optimum whenever the good or bad contagion have different intrinsic transmissibilities. In Section \ref{sec:discussion}, we discuss the implications of these social dilemmas and our other results in light of existing work in evolutionary game theory, social evolution, and behavioral epidemiology. 

In the Appendix, we provide detailed analysis of contagion and evolutionary dynamics for populations featuring two sociality strategies (Appendix A); extend our characterization of evolutionary stability and attempts to mitigate social dilemmas with assortative interaction rules (Appendix B), and present derivations for example linear and Cobb-Douglas utility functions from
expected payoffs achieved by individuals at equilibrium states of the contagion dynamics (Appendix C).

\section{Good and bad contagion dynamics} \label{sec:baselinemodel}

When they interact with one another, individuals are potentially exposed to contagion. We consider two contagion processes: a beneficial (``good'') contagion process $g$ and a harmful (``bad'') contagion process $b$. For example, the good contagion could be the spread of beneficial social information and the bad contagion could be the spread of a harmful infectious disease. We assume that each contagion is governed by Susceptible-Infectious-Susceptible (SIS) dynamics: individuals who are not currently infectious are susceptible ($S$), and can become infectious ($I$) through interactions with currently infectious individuals. 

We call an individual's infection status for each contagion their ``state''. Four states are possible: susceptible to both good and bad, infectious with the good but not the bad, infectious with the bad but not the good, and infectious with both. We assume that the good and bad contagions spread entirely independently, so an individual's probability of becoming infected with the bad contagion in a given social interaction does not depend on whether that individual is currently susceptible or infectious with the good contagion, and vice versa. This contrasts with situations such as contagious fear spreading alongside a pathogen, in which one's state of fear may affect one's probability of contracting the pathogen \revision{(as considered by Perra and coauthors \cite{perra2011towards} and by Epstein and coauthors \cite{epstein2008coupled,epstein2021triple}, for example)}.

Since the contagions spread independently, we can characterize population based upon the fractions susceptible to ($S^{(g)}$) infectious with ($I^{(g)}$) the good contagion, and the corresponding fractions $S^{(b)}$ and $I^{(b)}$ for the bad contagion. \revision{We assume that individuals have social interactions with rate $\sigma$, and that this rate of social interaction is fixed throughout the course of the contagion dynamics. This allows us to consider contagion dynamics that occur on a faster timescale than the evolutionary dynamics governing the strategies for social interaction. For the good and bad contagions, we assume that an interaction between a susceptible and infectious individual results in transmission of the contagion with probability $p_g$ and $p_b$, respectively.} For each contagion $x \in \{g,b\}$, susceptible individuals $S^{(x)}$ \revision{therefore} become infected at a rate $\revision{\sigma p_x} I^{(x)}$, \revision{and we further assume that} infected individuals \revision{spontaneously} recover and return to the susceptible state at rate $\gamma_x$. The contagion dynamics thus obey the following ordinary differential equations: 
\begin{subequations} \label{eq:baselineSISmonomorphic}
\begin{align}
    \dv{S^{(g)}}{t} &= - \revision{\sigma p_g} S^{(g)} I^{(g)}   
    + \gamma_g I^{(g)} \\[1em]
    \dv{I^{(g)}}{t} &= \revision{\sigma p_g} S^{(g)} I^{(g)} 
    - \gamma_g I^{(g)} \\[1em]
    \dv{S^{(b)}}{t} &= -\revision{\sigma p_b} S^{(b)} I^{(b)} 
    + \gamma_b I^{(b)} \\[1em]
    \dv{I^{(b)}}{t} &=  \revision{\sigma p_b} S^{(b)} I^{(b)} 
   - \gamma_b I^{(b)}
\end{align}
\end{subequations}

The two contagion processes have basic reproduction numbers
\begin{equation} \label{eq:RbandRg}
\Rg{} = \frac{\revision{\sigma p_g}}{\gamma_g} \: \: \mathrm{and} \: \: \Rb{}  = \frac{\revision{\sigma p_b}}{\gamma_b}.
\end{equation}

\revision{We can rearrange Equation \eqref{eq:RbandRg} to obtain the following relationship between $\Rb{}$ and $\Rg{}$:}
\begin{equation} \label{eq:RbintermsofRg}
    \Rb{} = c \Rg{} , \: \: \mathrm{where} \: \:
c = \frac{p_b \gamma_g}{p_g \gamma_b}.
\end{equation}

The parameter $c$ can be thought of as the relative transmissability of the bad contagion compared to the good. Using Equations \eqref{eq:RbandRg} and \eqref{eq:RbintermsofRg}, we can describe the impact of the social contact rate $\sigma$ on the dynamics of the good and bad contagion through the parameters $\Rg{}$ and $c$. For simplicity, we will now primarily characterize the sociality strategies of individuals through the resulting reproduction number $\Rg{}$. \revision{We use this parametrization of the contagion dynamics with the right-hand side expressed in terms of $c$ and $\Rg{}$ in order to have a compact representation of how the spread of the coupled contagions depend social interaction rates (encoded by $\Rg{} = \frac{\sigma p_g}{\gamma_g}$) and the relative infectiousness of the good and bad contagion (as encoded by $c$).} 
 \revision{Notably, derivatives of the social utility and invasion fitness with respect to $\Rg{}$ will be proportional to the derivatives of the same functions with respect to $\sigma$, and so social optima and evolutionarily stable strategies expressed in terms of the reproduction number $\Rg{}$ will directly correspond to equivalent quantities expressed in terms of the social interaction rate $\sigma$.} %

After applying the formula of Equations \eqref{eq:RbandRg} and Equations \eqref{eq:RbintermsofRg} to Equation \eqref{eq:baselineSISmonomorphic}, we can divide through by the recovery rates $\gamma_g$ and $\gamma_b$ and use the fact that $I^{(g)} + S^{(g)} = 1$ and $I^{(b)} + S^{(b)} = 1$ to express the dynamics of the good and bad contagion through the pair of decoupled differential equations: 
\begin{subequations} \label{eq:SISmonomorphicR}
\begin{align}
   \revision{\frac{1}{\gamma_g}}  \dsddt{I^{(g)}} &= \Rg{} (1 - I^{(g)}) I^{(g)} - I^{(g)} \\[1em]
     \revision{\frac{1}{\gamma_b}}  \dsddt{I^{(b)}} &= c\Rg{} (1 - I^{(b)}) I^{(b)} - I^{(b)}.
\end{align}
\end{subequations}

Equation \eqref{eq:SISmonomorphicR} shows that, for a given value of $\Rg{} \ge 0$, the good and the bad contagions have globally-stable equilibria $\ibar{}^{(g)}$ and $\ibar{}^{(b)}$ given by
\begin{subequations} \label{eq:equilibriapiecewise}
\begin{align} \label{eq:goodequilibriapiecewise}
\ibar{}^{(g)}(\Rg{}) &= 
\left\{\begin{array}{cr}
    1 - \ds\frac{1}{\Rg{}} & \Rg{} > 1 \\[0.75em]
    0 & \Rg{} \le 1
\end{array} \right. \\[1.5em] \label{eq:badequilibriapiecewise}
\ibar{}^b(\Rg{}) &= 
\left\{\begin{array}{cr}
    1 - \ds\frac{1}{c\Rg{}} & c \Rg{} > 1 \\[0.75em]
    0 & c\Rg{} \le 1
\end{array} \right.
\end{align}
\label{eqn:homogeneous-sis-equilibria}
\end{subequations}

For examples of how these equilibria vary with \Rg{} and $c$, see Figure \ref{fig:cobb-douglas-example}d--f.

\subsection{Benefits and costs of sociality}
 \label{sec:utilityonepop} 
To understand how sociality evolves given the benefits of the good contagion and the costs of the bad contagion, we quantify those benefits and costs with a utility or fitness function, $U$. We assume that an individual's utility depends on the equilibrium prevalences of the two contagions (Equation \eqref{eq:SISmonomorphicR}), so we have $U(\ihat{g}, \revision{\shat{b}})$. 

One interpretation of the equilibrium fractions infected $\ihat{g}$ and $\shat{b}$ is as each individual's long-run fraction of time spent \revision{infected with the good contagion and susceptible to the bad contagion}, respectively. We also show (Appendix C) that some linear and Cobb-Douglas utility functions can be derived by calculating an individual's expected utility at the endemic equilibrium, where $\ihat{g}$ and \revision{$\shat{b}$} are treated as the \revision{respective probabilities of being infected with the good contagion and susceptible to the bad contagion}.

Implicit in this choice of utility functions is the assumption that the spreading processes for the good and bad contagions are fast relative to the rates of reproduction or social learning in the population (which can lead to changes in sociality $\sigma$). That is, we assume that contagion dynamics happen on a faster timescale than do births and deaths, and ignore demographic effects in our coupled SIS contagion models. To weigh the benefits of the good contagion against the costs of the bad contagion, we would like a utility function that increases the more time one spends infected with the good contagion (\ihat{g})  and also increases the more time one remains uninfected (i.e. susceptible) with the bad contagion (\shat{b}).

\subsection{Cobb-Douglas utility functions}
To capture this key property of always preferring greater \ihat{g} and \shat{b}, we explore as a first example the Cobb-Douglas family of utility functions:

\begin{equation} \label{eq:CDexample}
    U(\ibar{}^{(g)}, \sbar{}^{(b)})  = (\ibar{}^{(g)})^\alpha (\sbar{}^{(b)})^{1 - \alpha},
\end{equation}
where the parameter $\alpha \in [0,1]$ measures the importance placed upon seeking out the benefits of the good contagion relative to avoiding the harms of the bad (Figure \ref{fig:cobb-douglas-example}a--c). 

This $\alpha$ allows us to explore how our evolutionary dynamics depend upon the relative importance of the good and bad contagions for utility or fitness, and how this relative importance interacts with their relative transmissibility $c$ in shaping desirable and evolutionarily-stable rates of social interaction. The Cobb-Douglas utility functions allow us to compute explicit socially-optimal and evolutionarily-stable levels of sociality, but we show in Section \ref{sec:adaptivedynamics} that many of our qualitative results generalize to broader classes of utility functions with similar properties.

Noting from Equation \eqref{eq:equilibriapiecewise} that both $\ibar{}^{(g)}$ and $\sbar{}^{(b)}$ are functions of $\Rg{}$, we can write $U$ as a function of \revision{$\Rg{}$}. 
\revision{Plugging in} the equilibria from Equation \eqref{eq:equilibriapiecewise}, we write the Cobb-Douglas utility $U(\Rg{}$) as:

\begin{equation} \label{eq:CDmonopiecewise}
U(\Rg{}) = \left\{
     \begin{array}{cr}
      0 & : \Rg{} \leq 1  \\[1em]
       \left( 1 - \dfrac{1}{\Rg{}} \right)^{\alpha} & : 1 \leq  \Rg{} \leq  \dfrac{1}{c} \\[2em]
      \left( 1 - \dfrac{1}{\Rg{}} \right)^{\alpha} \left(\dfrac{1}{c \Rg{}} \right)^{1-\alpha} & : \Rg{} \geq  \dfrac{1}{c}   
     \end{array}
   \right.    
\end{equation}

These three cases come from the fact that the good contagion will not spread if $\Rg{} \leq 1$, the bad contagion will not spread if $\Rb{} = c \Rg{} \leq 1$, and otherwise both spread. \revision{Notably, the middle case in which $1 \leq \Rg{} \leq \frac{1}{c}$ cannot occur when $c > 1$. When $c > 1$, the bad contagion spreads more readily than the good contagion, so the bad contagion will always be present at positive endemic equilibrium when the good contagion survives in the long-time limit (provided both contagions are present in the initial population).}

\revision{We also note} that $U(\Rg{}) \geq 0$ for $\Rg{} \geq 1$, and that utility is minimized when $\Rg{} \leq 1$ and correspondingly $U(\Rg{}) = 0$.

\subsection{Socially-optimal levels of sociality}
\begin{figure}[!ht]
    \centering
    \includegraphics[width = \textwidth]{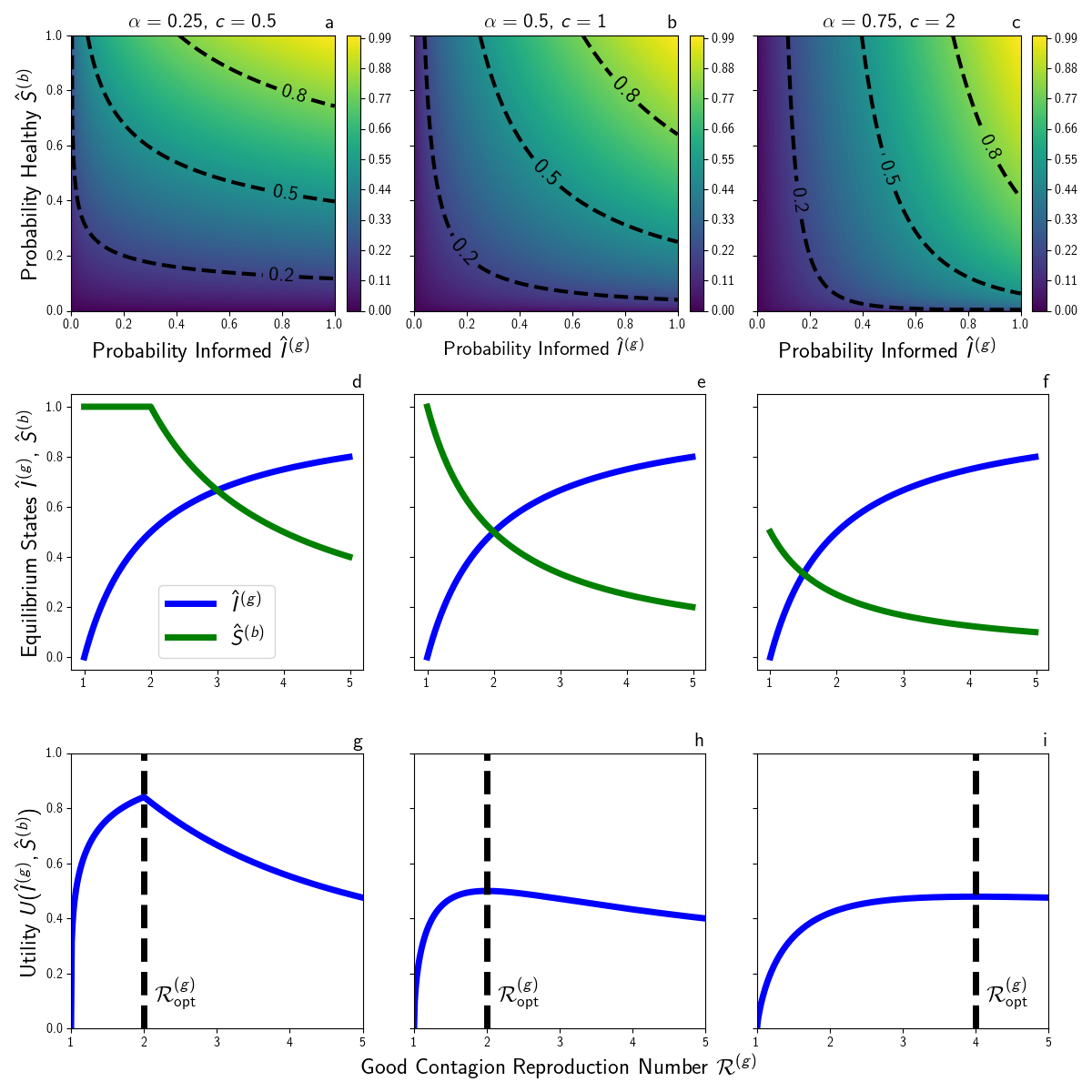}
    \caption{Example heatmaps of Cobb-Douglas utility for various weights $\alpha$ of emphasis on acquiring the good contagion versus avoiding the bad contagion (\textbf{a}--\textbf{c}),
    endemic equilibria $\hat{I}^{(g)}$ and $\hat{S}^{(b)}$ as a function of sociality strategy \Rg{} for different values of relative transmissibility $c$ (\textbf{d}--\textbf{f}), and resultant overall utilities as a function of \Rg{} given $\alpha$ and $c$ (\textbf{g}--\textbf{i}). Note that in certain cases (e.g. \textbf{g}) utility is maximized by setting $\Rg{} = \frac{1}{c}$, the maximal degree of sociality at which the bad contagion fails to spread. \revision{Vertical dashed lines in Panels \textbf{g}--\textbf{i} correspond to the socially-optimal sociality strategy $\Ropt$.}}
    \label{fig:cobb-douglas-example}
\end{figure}

For a population with monomorphic sociality (i.e. one \Rg{} shared by all individuals), there exists a socially-optimal $\Rg{}$ that maximizes each individual's fitness (Figure \ref{fig:cobb-douglas-example}g--i).

\revision{Using the piecewise characterization of the Cobb-Douglas utility function from Equation \eqref{eq:CDmonopiecewise}, We can calculate that the reproduction $\Ropt$ maximizing the social utility $U(\Rg{})$ is given by}
\begin{equation} \label{eq:RoptCD}
    \Ropt =  \max\left( \frac{1}{c}, \; \frac{1}{1 - \alpha} \right).%
\end{equation}
\revision{For completeness, we present the derivation of this socially optimal level of sociality $\Ropt$ in Appendix B.1.}

\revision{From Equation \eqref{eq:RoptCD}}, we see that $\Ropt > 1$ provided that $\alpha > 0$, so the social optimum under the Cobb-Douglas utility features involves socializing enough to achieve good contagion transmissions whenever individual utility places any weight on the good contagion. \revision{Furthermore, because the bad contagion cannot spread when $\Rb{} = c \Rg{} \leq 1$, the bad contagion will be absent from the population when $\Rg{} \leq \frac{1}{c}$. As a result, we see from Equation \eqref{eq:RoptCD} that the socially optimal interaction rate features elimination of the bad contagion when $\alpha \leq 1 - c$, which corresponds to the case in which the bad contagion spreads less readily than the good contagion and social utility places a sufficiently small relative weight upon acquiring the good contagion. By contrast, when $\alpha > 1 - c$, the socially optimal level of sociality $\Ropt$ will allow for the long-time survival of both the good and bad contagion.}

We further illustrate our results in Figure \ref{fig:cobb-douglas-example}, showing the Cobb-Douglas utility, the endemic equilibria, and the socially-optimal level of sociality for various $c$ and $\alpha$.  

\section{Evolutionary Dynamics with Two Levels of Sociality}
\label{sec:dimorphicevolutionary}

Having established the socially-optimal rate of social interaction for a monomorphic population, we now consider which sociality levels (``strategies'') can succeed under evolutionary competition, if individual utility or reproductive fitness depends on infection with the good and bad contagions.

We can consider a scenario in which individuals reproduce proportional to their utility, so that the proportion of individuals with sociality strategies leading to higher utility will increase over time. Alternatively, we could consider individuals who engage in social learning, imitating the sociality strategies of peers who are obtaining higher utilities. In either case, the fraction $f$ of individuals following one strategy may change over time.

To analyze this competition, we first study pairwise competition between two sociality strategies $m$ and $r$. First, we introduce our models of good and bad contagion dynamics in the presence of two sociality strategies (Section \ref{sec:dimorphiccontagion}), and derive equilibrium contagion levels. Next, we use these equilibria and our utility functions to study how the fractions of the population following each strategy will evolve over time under a replicator equation modeling evolutionary competition (Section \ref{sec:twosocialitycompetition}). 

\subsubsection{Contagion Dynamics for Two Levels of Sociality} \label{sec:dimorphiccontagion}

We assume that, on the epidemic time scale, there is a fixed fraction $f$ of individuals following a strategy of interest (``mutant strategy'') $m$ with social interaction rate $\sigma_m$. The remaining fraction $1-f$ of individuals follow a different strategy (``resident strategy'') $r$ with interaction rate $\sigma_r$. \revision{For both the mutant and resident strategies, we assume that the social interaction rates $\sigma_m$ and $\sigma_r$ are fixed throughout the course of the contagion dynamics.} As in the monomorphic case, \revision{we can describe} good contagion reproduction numbers $\Rg{m} = \tfrac{\sigma_m p_g}{\gamma_g}$ and $\Rg{r} = \tfrac{\sigma_r p_g}{\gamma_g}$ for the resident and mutant strategies, and resultant bad contagion reproduction numbers $\Rb{m} = c \Rg{m}$ and $\Rb{r} = c \Rg{r}$. As in Section \ref{sec:baselinemodel}, we use these relations to parametrize the resident and mutant sociality strategies via their respective reproduction numbers under the good contagion $\Rr^{(g)}$ and $\Rmg$. Under the \revision{assumption that the probability of social interactions with individuals following the resident and mutant strategies is derived from unbiased sampling of the pool of available contacts}, we show in Appendix A that the level of good contagion in the resident and mutant populations evolves according to

\begin{subequations} \label{eqn:full-epidemic-dynamics-r0}
\begin{align}
    \revision{\frac{1}{\gamma_g}} \dv{I^{(g)}_r}{t} &= \Rrg\Big[\frac{\Rrg(1-f) I_r^{(g)} + \Rmg f I^{(g)}_m}{\Rrg(1-f) + \Rmg f}\Big] \left(1 - I^{(g)}_r\right) - I^{(g)}_{r}\label{eq:rgooddimorphic}\\[1.5em]
     \revision{\frac{1}{\gamma_g}} \revision{\dv{I^{(g)}_m}{t}} &= \Rmg \Big[\frac{\Rrg(1-f) I^{(g)}_r + \revision{\Rmg f} I^{(g)}_m}{\Rrg(1-f) + \Rmg f}\Big] \left(1 - I^{(g)}_m \right) - I^{(g)}_m, \label{eq:mgooddimorphic}
\end{align}
with the corresponding system of ODEs holding for the bad contagion
\begin{align}
   \revision{\frac{1}{\gamma_b}}  \dv{I^{(b)}_r}{t} &= c \Rrg\Big[\frac{ \Rrg(1-f) I_r^{(b)} +  \Rmg f I^{(b)}_m}{\Rrg(1-f) + \Rmg f}\Big] \left(1 - I^{(b)}_r\right) - I^{(b)}_{r} \label{eq:rbaddimorphic}\\[1.5em] 
    \revision{\frac{1}{\gamma_b}}  \revision{\dv{I^{(b)}_m}{t}} &= c\Rmg \Big[\frac{\Rrg(1-f) I^{(b)}_r + \Rmg f I^{(b)}_m}{\Rrg(1-f) + \Rmg f}\Big] \left(1 - I^{(b)}_m \right) - I^{(b)}_m.\label{eq:mbaddimorphic}
\end{align}
\end{subequations}
\revision{In particular, the terms in square brackets in Equations \eqref{eqn:full-epidemic-dynamics-r0} describe the probability that a given social contact takes place with an infectious individual, and we note that this probability depends on the relative chance of interacting with individuals following the resident and mutant strategy due to both the relative abundance of these strategies and the relative social contact rates characterized by the strategies.}

We show (see Appendix A) that the overall good contagion dynamics for the resident and mutant populations combined has a basic reproduction number given by
\begin{equation}\label{eq:netreproductionmain}
    \Rnet^{(g)} = \frac{f\left(\Rmg\right)^2 + (1-f)\left(\Rrg\right)^2}{f \Rmg + (1-f)\Rrg}.
\end{equation}
It follows from a result of Hethcote and Yorke on two population SIS dynamics that the disease-free equilibrium $(\hat{I}^{(g)}_m, \hat{I}^{(g)}_r) = (0, 0)$ of \revision{Equations \eqref{eq:rgooddimorphic} and \eqref{eq:mgooddimorphic}} is globally stable when $\Rnet^{(g)} < 1$, whereas, \revision{when $\Rnet^{(g)} > 1$, there exists a unique endemic equilibrium $(\hat{I}^{(g)}_m, \hat{I}^{(g)}_r) \ne (0,0)$ that is the achieved in the long-time limit for any initial condition starting with any presence of the good contagion in the population (i.e. for any initial state other than the disease-free equilibrium $(0,0)$)}. An analogous result holds for the equilibria of the bad contagion dynamics with $\Rnet^{(b)} = c \Rnet^{(g)}$.

\subsubsection{Competition Between Two Sociality Strategies} \label{sec:twosocialitycompetition}

Since we assume contagion dynamics occur on a much faster timescale than evolutionary competition, we can explore the evolutionary consequences of fitness/utility that depend on equilibrium fractions susceptible to the bad contagion $\hat{S}_{\cdot}^{(b)}(\Rmg,\Rrg,f)$ and infected with the good contagion $\hat{I}_{\cdot}^{(g)}(\Rmg,\Rrg,f)$. 

For a given mutant fraction $f$, we define utility functions $U_m(f)$ and $U_r(f)$ for mutant and resident individuals, respectively:

\begin{equation} \label{eq:dimorphicutility}
\begin{aligned}
    U_m(f) &= U\left(\hat{I}_m^{(g)}(\Rmg,\Rrg,f), \hat{S}_m^{(b)}(\Rmg,\Rrg,f) \right) \\
     U_r(f) &= U\left(\hat{I}_r^{(g)}(\Rmg,\Rrg,f), \hat{S}_r^b(\Rmg,\Rrg,f) \right).
    \end{aligned}
\end{equation}

To encode our assumption that an individual's chance of being reproducing or being imitated is proportional to its utility, we model $f(t)$ according to the following replicator equation:

\begin{equation}
    \dv{f}{t} = f(1-f)\Big[U_m(f) - U_r(f)\Big]
\label{eqn:replicator}
\end{equation}

This replicator equation can be derived from a variety of individual-based models of social imitation \cite{sandholm2010population} as well as from models of reproduction-based natural selection \cite{hofbauer1998evolutionary}. 

Equation \ref{eqn:replicator} has three types of biologically plausible equilibria: $\bar{f}=0$, $\bar{f}=1$ (population takeover by the resident or mutant, respectively), and interior frequencies $\bar{f} \in (0,1)$ where $U_m(\bar{f}) = U_r(\bar{f})$ (coexistence of the two strategies).

\begin{figure}[ht!]
    \centering
     \begin{subfigure}[h]{0.48 \textwidth}
         \centering
         \includegraphics[width = \linewidth]{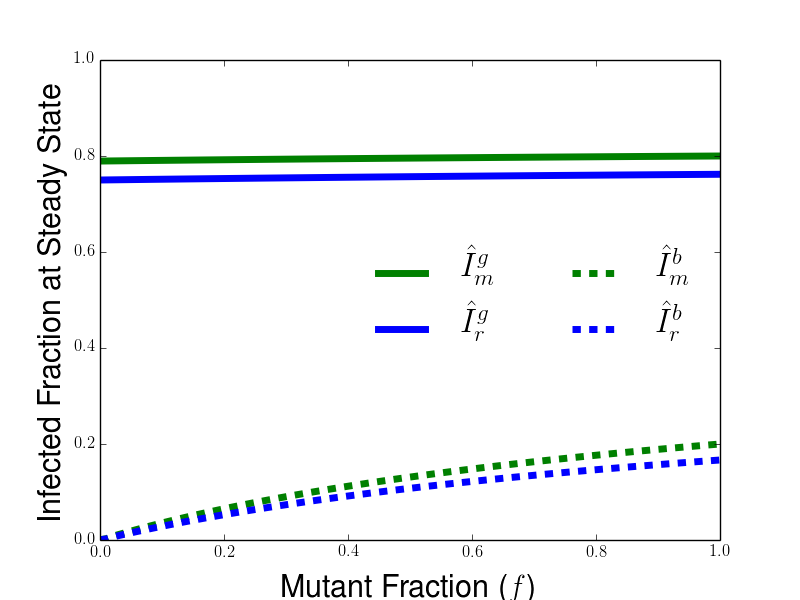}
         \vspace{-5mm}
         \caption{Contagion Equilibria ($c = 0.25$)}
         \label{fig:Endemiccquarter}
     \end{subfigure}
      \begin{subfigure}[h]{0.48 \textwidth}
         \centering
         \includegraphics[width = \linewidth]{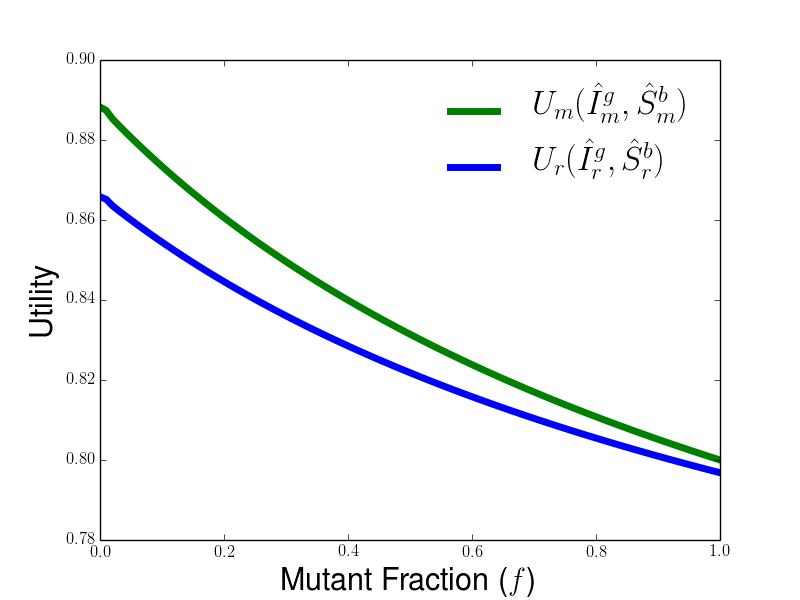}
         \vspace{-5mm}
         \caption{Resident and mutant utility ($c = 0.25$)}
         \label{fig:Utilitycquarter}
     \end{subfigure}
     
     \begin{subfigure}[h]{0.48 \textwidth}
         \centering
         \includegraphics[width = \linewidth]{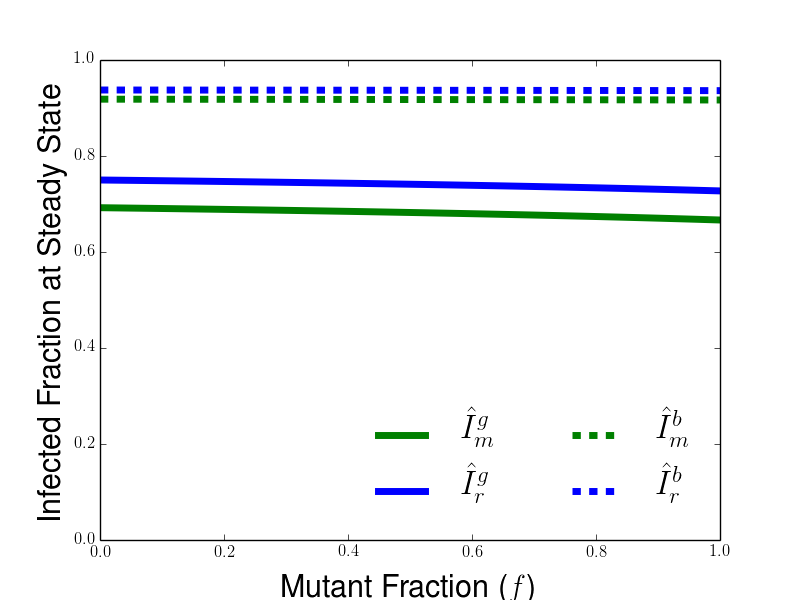}
         \vspace{-5mm}
         \caption{Contagion Equilibria ($c = 4$)}
         \label{fig:Endemicc4}
     \end{subfigure}
      \begin{subfigure}[h]{0.48 \textwidth}
         \centering
         \includegraphics[width = \linewidth]{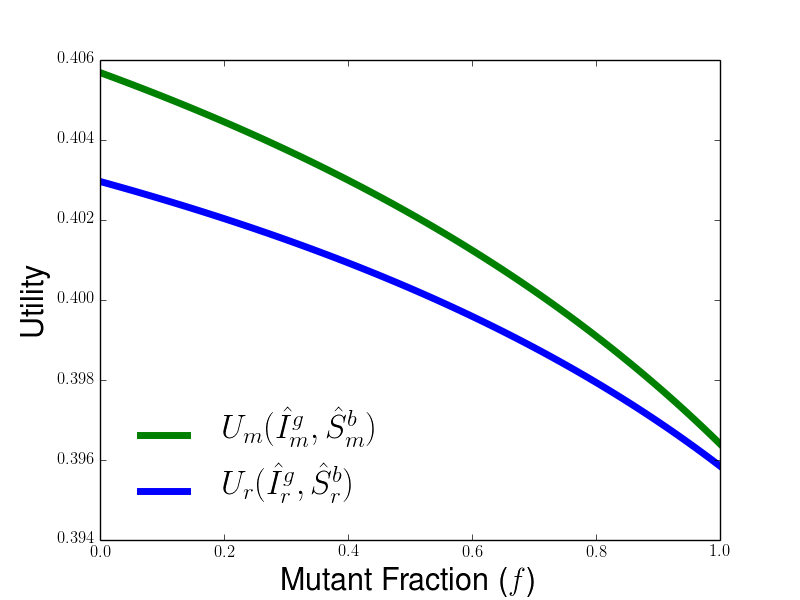}
         \vspace{-5mm}
         \caption{Resident and mutant utility ($c = 4$)}
         \label{fig:Utilityc4}
     \end{subfigure}
    \caption{Sample contagion equilibria and \revision{Cobb-Douglas} utility achieved by resident and mutant strategies as a function of the fraction of individuals following the mutant strategy $f$. We consider a resident with reproduction number $\Rrg= \Ropt = 4$ in all panels.  (a,c): Endemic equilibria of the good and bad contagion for the cases of relative infectiousness and mutant reproduction number $c = 0.25$, $\Rmg= \revision{5}$ (panel (a)) and $c = 4$, $\Rmg= 3$ (panel (c)). (b,d) Plots of \revision{Cobb-Douglas} utility $U_r(f)$ and $U_m(f)$ achieved at contagion equilibrium for individuals following resident and mutant strategy, will parameters $c = 0.25$, $\Rmg= \revision{5}$ (panel(b)) or $c = 4$, $\Rmg= 3$ (panel(d)). \revision{For both cases, we use a Cobb-Douglas utility function with weight parameter $\alpha = 0.75$.} In both cases, the utility of the mutant type (green curve) always exceeds the utility of the resident type (blue curve), and the replicator equation will favor fixation to an all-mutant composition. }
     \label{fig:replicatorcless1}
\end{figure}

We now use this replicator equation to study the dynamics of pairwise competition between two sociality strategies. In Figure \ref{fig:replicatorcless1}, we display the endemic equilibria of the two-type contagion dynamics and the Cobb-Douglas utilities at the contagion equilibrium for both the socially-optimal level of sociality and a chosen mutant strategy across the possible compositions of $f$ mutant and $1-f$ resident sociality strategists. We consider an optimal sociality strategy $\Ropt = 4$, and choose a mutant strategy $\Rmg= 7$ when the good contagion spreads more readily (Figure \ref{fig:replicatorcless1}b, $c = 0.25$) and a mutant strategy $\Rmg= 3$ when the bad contagion spreads more readily (Figure \ref{fig:replicatorcless1}d, $c = 4$). In both cases, we see that $U_m(f) > U_r(f)$ for all $f \in [0,1]$, so it follows from Equation \eqref{eqn:replicator} that the mutant proportion will always increase for the pairs of $\Rmg$ and $\Rrg$ we consider under replicator dynamics. Therefore, these choices of mutant strategies are able to successfully invade populations otherwise composed of the socially-optimal strategy, and  will eventually fix to the composition $f=1$. As a result, we see that strategies that maximize collective utility can lose out in evolutionary competition against rival strategies, showing that dynamics of Equation \eqref{eqn:replicator} produce a social dilemma in the evolution of sociality strategies.

\begin{figure}[ht!]
    \centering
    \includegraphics[width = 0.48\linewidth]{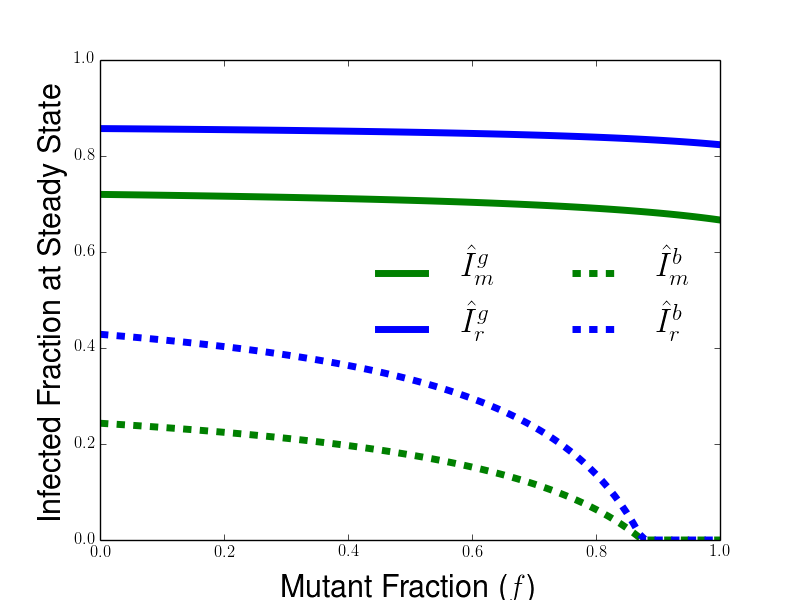}
    \includegraphics[width = 0.48\linewidth]{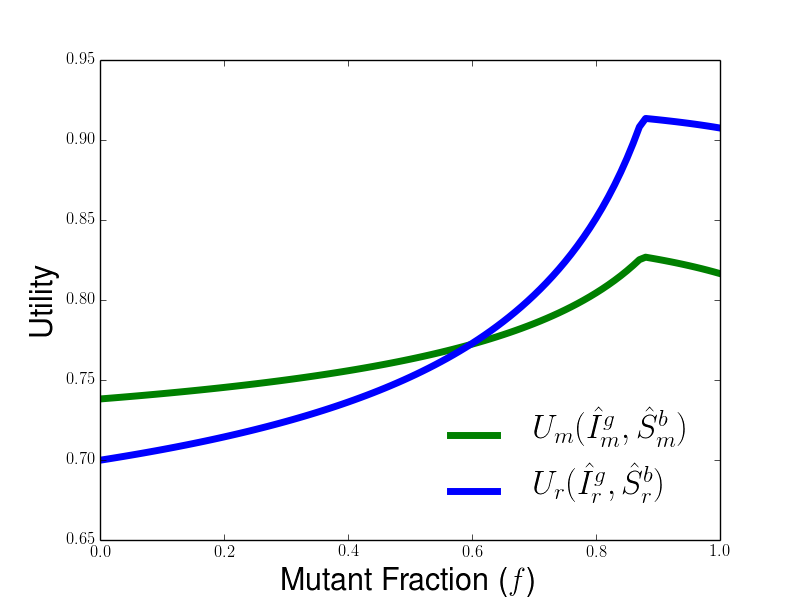}
    \caption{Endemic equilibria of the two contagions for the resident and mutant populations (left) and the Cobb-Douglas utility (right) for a case in which the resident and mutant sociality stategies lie on opposite sides of the social optimum. The utility for the resident strategy (blue curve) and the mutant strategy (green curve) intersect at a single fraction of mutants $f$, which is the equilibrium of the replicator equation at which the two types will coexist. \revision{The Cobb-Douglas utility has weight parameter $\alpha0.25$, the relative infectiousness of the bad contagion is $c = 0.25$, and the resident and mutant reproduction numbers are given by $\Rrg = 7$ and $\Rmg = 3$, respectively.} 
    }
    \label{fig:replicatorcoexistence}
\end{figure}

\revision{In addition to the cases of pairwise dominance between strategies seen in Figures \ref{fig:replicatorcless1}, we now demonstrate that there are pairs of sociality strategies for which stable coexistence can be achieved at an interior equilibrium under the replicator equation. In Figure \ref{fig:replicatorcoexistence}, we provide an example of mutant and resident sociality strategies for which there is a fraction of mutants $f_{eq}$ around 0.6 such that the mutant and resident have equal utilities $U_m(f_{eq}) = U_r(f_{eq})$. Because the mutants obtain higher utility than the residents for $f < f_{eq}$ and the residents outcompete the mutants for $f > f_{eq}$, we see that $f_{eq}$ will be a stable interior steady state for the replicator equation, and a population with initial composition featuring a mix of mutants and residents will result in long-time coexistence of the two types composed of $f_{eq}$ fraction mutants and $1 - f_{eq}$ fraction residents.}

\revision{Having seen different possible behaviors under the replicator dynamics through figures \ref{fig:replicatorcless1} and \ref{fig:replicatorcoexistence}, we now turn to an analytical characterization of the possible scenarios for pairwise competition in the case of Cobb-Douglas utility. In Proposition \ref{prop:replicatorlongtime}, we show that there are three possible long-time behaviors -- dominance of the mutant strategy, dominance of the resident strategy, and long-time coexistence of the resident and mutant strategies at a unique interior equilibrium-- and that we can characterize which of these behaviors occurs by comparing the relative utility of each strategy in the limits in which one strategy is rare ($f = 0$ and $f =1$). Notably, we find that it is not possible for the replicator equation to support bistability of the all-resident and all-mutant equilibria at $f=0$ and $f=1$ in the case of Cobb-Douglas utility. We provide a proof of Proposition \ref{prop:replicatorlongtime} in Appendix A.3.}

\begin{restatable}{proposition}{replicatorresult}
\label{prop:replicatorlongtime}
\revision{Suppose that the resident and mutant types have sociality strategies featuring reproduction numbers $\Rr^{(g)} \geq 1$ and $\Rmg \geq 1$ for the good contagion, with $\Rr^{(g)} \ne \Rmg$ and at least one of these reproduction numbers strictly greater than $1$. Then, for any $c > 0$ and for any resident and mutant types with reproduction numbers $\Rr^{(b)} = c \Rr^{(g)}$ and $\Rmb = c \Rr^{(g)}$ for the bad contagion, the difference of Cobb-Douglas log-utilities $\log\left[U_m(f) \right] - \log\left[U_r(f) \right]$ is a decreasing function of $f$. As a consequence, the long-time behavior can be determined by the relative values of $U_m(f)$ and $U_r(f)$ at the endpoints $f = 0$ and $f = 1$. The three possible cases are the following:}
\begin{itemize}
    \item{$U_m(0) > U_r(0)$ and $U_m(1) > U_r(1)$:} \revision{$f = 1$ is globally stable and the mutant will fix in the population.}
    
    \item{$U_m(0) < U_r(0)$ and $U_m(1) < U_r(1)$:} \revision{$f = 0$ is globally stable and the resident will fix in the population.} 
    
    \item{$U_m(0) > U_r(0)$ and $U_m(1) < U_r(1)$:} \revision{There exists a unique interior equilibrium $\hat{f} \in (0,1)$ that is globally stable, and mutant and resident will coexist in the long-time population.}
\end{itemize}
\end{restatable}

\revision{This result tells us that pairwise analysis of invasibility between strategies is sufficient to determine the long-time dominance or coexistence between pairs of sociality strategies in the case of Cobb-Douglas utility. This allows us to see that if a given sociality strategy that dominates a rival strategy at the endpoints $f = 0$ and $f=1$, this strategy will always achieve dominance under the dynamics of the replicator equation. The fact that the long-time behavior of pairwise competition between sociality strategies reduces to a comparison of utilities at the two endpoints provides additional motivation for using the adaptive dynamics framework and a pairwise invasibility analysis to calculate the long-time evolution of the social interaction rates in the population.}

\subsection{Adaptive Dynamics} \label{sec:adaptivedynamics}

Having considered the evolutionary competition between two sociality strategies, we now turn to try to understand how the level of socialization changes over longer evolutionary timescales. To do this, we adopt the framework of adaptive dynamics, looking to characterize evolutionarily-stable levels of socialization. First, we study the contagion dynamics in the limit of infinitessimal mutant frequency (Section \ref{sec:adaptiveendemic}). \revision{Then} we characterize the socially-optimal and evolutionarily-stable outcomes of the long-time adaptive dynamics under Cobb-Douglas utility (Section \ref{sec:Cobb-Douglas}). After illustration the social dilemmas of sociality strategies for this choice of utility functions (Section \ref{sec:exploringsocial}), we study a more general family of utility functions to understand the broadest set of assumptions under which \revision{there exists a unique socially-optimal level of social interaction and in which} we can expect to see a discrepancy between self-interested and socially-optimal behavior under evolutionary dynamics depending on our coupled good and bad contagion processes (Section \ref{sec:genutility}). \revision{In particular, we derive natural conditions required of utility functions to guarantee that the adaptive dynamics will support greater levels of social interaction than is socially optimal when the good contagion spreads more readily than the bad contagion ($c < 1$) and such that the adaptive dynamics produce less social interaction than optimal when the bad contagion spreads more readily ($c > 1$).}

\subsubsection{Endemic Equilibrium in Mutant Population} \label{sec:adaptiveendemic}

In the adaptive dynamics limit, we consider the case in which mutants are initially rare in the population. This allows us to consider the limit in which the fraction $f$ of individuals with the mutant strategy tends to zero in our model for dimorphic contagion dynamics of \revision{Equations \eqref{eq:rgooddimorphic} and \eqref{eq:mgooddimorphic}}. In this limit, the vanishing presence of the mutant population results in the dynamics for the resident population reducing to the monomorphic dynamics governed by Equation \eqref{eq:SISmonomorphicR}. The resident population will then converge to to the long-time equilibrium from Equation \eqref{eq:equilibriapiecewise}. For the mutant population, we see that the dynamics of the good contagion reduces to the following equation
\begin{equation} \label{eq:mutantadaptivecontagion}
    \revision{\frac{1}{\gamma_g}}  \dv{I_m^{(g)}}{t} =  \Rmg\left( 1 - I_m^{(g)} \right) I_r^{(g)} - I_m^{(g)},
\end{equation}
where $\Rrg$ and $\Rmg$ are the reproduction numbers for the resident and mutant populations under the good contagion. Because, when $\Rrg> 1$, the equilibrium fraction $\hat{I}^{(g)}_r$ of residents with the good contagion is given by $\hat{I}^{(g)}_r = 1 - \frac{1}{\Rrg}$, we find from Equation \eqref{eq:mutantadaptivecontagion} that the equilibrium fraction $\hat{I}^{(g)}_m$ of mutants with the good contagion must satisfy 
$$\Rmg\left( 1 - \hat{I}^{(g)}_m \right) \hat{I}_r^{(g)} - \hat{I}_m^{(g)} = \Rmg\left(1 - \frac{1}{\Rrg}\right) (1 - \hat{I}_m^{(g)}) - \hat{I}_m^{(g)} = 0.$$
Solving this equation, we can find the following expression for $\hat{I}^{(g)}_m$ as a function of the mutant and resident reproduction numbers $\Rmg$ and $\Rrg$:
\begin{equation}
\hat{I}^{(g)}_m(\Rmg,\Rrg) = \frac{\Rmg(\Rrg- 1)}{\Rmg(\Rrg- 1) + \Rrg}.
\end{equation}
This equilibrium is stable whenever it is biologically feasible, which holds for all $\Rmg\geq 0$ if $\Rrg> 1$. 
For the bad contagion, we can use our assumption that the reproduction numbers for the mutant and resident populations are $\Rm^{b} = c \Rmg$ and $\Rr^{b} = c \Rrg$ to similarly find that the level of infectiousness in the endemic equilibrium for the mutant population under the bad contagion is given by 
\begin{equation}
    \hat{I}_m^{(b)} =  \left\{
     \begin{array}{cr}
       0 & : \Rrg\leq \frac{1}{c} \vspace{2mm} \\ 
 \ds\frac{\Rmg\left( c \Rrg- 1\right)}{\Rmg\left( c \Rrg- 1\right) + \Rrg} & : \Rrg> \frac{1}{c}
     \end{array}
   \right. .
\end{equation}
Because we formulate our utility functions in terms of susceptibility of the bad contagion, it is also helpful for us to use $\hat{S}_m^{(b)} = 1 - \hat{I}_m^{(b)}$ to see that equilibrium fraction of the mutant population that is susceptible to the bad contagion is 
\begin{equation}
    \hat{S}_m^{(b)} = \left\{
     \begin{array}{cr}
       1 & : \Rrg\leq \frac{1}{c} \vspace{2mm} \\ 
 \ds \frac{\Rrg}{\Rmg\left( c \Rrg- 1\right) + \Rrg} & : \Rrg> \frac{1}{c}
     \end{array}
   \right. .
\end{equation}
In our adaptive dynamics analysis, we will need to know how these endemic equilibria vary with the mutant sociality level $\Rmg$, so we compute the partial derivatives
\begin{subequations} \label{eq:endemicpartials}
\begin{align}
    \dsdel{\hat{I}_m^{(g)}(\Rmg,\Rrg)}{\Rmg} &= \frac{\Rrg(\Rrg- 1)}{\left(\Rmg\Rrg+ \Rrg- \Rmg\right)^2} \\
    \dsdel{\hat{S}_m^{(b)}(\Rmg,\Rrg)}{\Rmg} &= \left\{
     \begin{array}{cr}
       0 & : \Rrg\leq \frac{1}{c} \vspace{1mm} \\ 
 \ds \frac{\Rrg\left(c \Rrg- 1\right)}{\left( c \Rmg\Rrg+ \Rrg -\Rmg\right)^2 } & : \Rrg> \frac{1}{c}
     \end{array}
   \right. .
\end{align}
\end{subequations}
In the limit of local mutation in which $\Rmg\to \Rrg$, we may further compute that 
\begin{subequations} \label{eq:localendemicpartials}
\begin{align}
\dsdel{\hat{I}_m^{(g)}(\Rmg,\Rrg)}{\Rmg} \bigg|_{\Rmg= \Rrg} &= \frac{\Rrg-1}{\Rr^3} \\
 \dsdel{\hat{S}_m^{(b)}(\Rmg,\Rrg)}{\Rmg} \bigg|_{\Rmg= \Rrg} &=
 \left\{
     \begin{array}{cr}
       0 & : \Rrg\leq \frac{1}{c} \vspace{2mm} \\ 
 \ds \frac{\revision{c \Rrg- 1}}{c^2 \Rr^3 } & : \Rrg> \frac{1}{c}
     \end{array}
   \right.
\end{align}
\end{subequations}

\subsubsection{Cobb-Douglas Utility} \label{sec:Cobb-Douglas}

For the Cobb-Douglas utility, we have the utility of a mutant with reproduction number $\Rmg$ in a resident population with reproduction number $\Rrg$ is given by 
\begin{equation} U\left(\Rmg,\Rrg\right) = \left(\hat{I}_m^{(g)}(\Rmg,\Rrg)\right)^{\alpha} \left( \revision{ \hat{S}_m^{(b)}(\Rmg,\Rrg)} \right)^{1 - \alpha} %
\end{equation}
It is helpful for \revision{our} analysis to alternatively consider the log-utility
\begin{equation} \label{eq:logutil} \log\left(U\left[\Rmg,\Rrg\right]\right)  = \alpha \log\left[ \hat{I}_m^{(g)}(\Rmg,\Rrg) \right]  + \left(1 - \alpha\right) \revision{\log\left[\hat{S}_m^{(b)}(\Rmg,\Rrg) \right]}
\end{equation}
We can now consider a relative advantage of a mutant in the resident population by considering the quantity
\begin{equation} \label{eq:logdiff} s_{\Rrg}(\Rmg) := \log\left(U\left[\Rmg,\Rrg\right]\right) - \log\left(U\left[\Rrg,\Rrg\right]\right) \end{equation}
\revision{Next}, we can compute the local selection gradient \revision{$s'_{\Rrg}(\Rrg)$} by differentiating the relative advantage with respect to $\Rmg$ and evaluating the derivative when $\Rmg= \Rrg$. \revision{Differentiating both sides of Equation \eqref{eq:logdiff}, we can use our expression for log-utility from Equation \eqref{eq:logutil}, the endemic equilibria from a monomorphic population from Equation \eqref{eq:equilibriapiecewise}, and the partial derivatives from Equation \eqref{eq:localendemicpartials} to see that the local selection gradient takes the following form:}%
\begin{equation}
\label{eq:cdsrprime}
\begin{aligned} s'_{\Rrg}(\Rrg) &:= \dsdel{ s_{\Rrg}(\Rmg)}{\Rmg} \bigg|_{\Rmg= \Rrg} = \dsdel{\log\left(U\left[\Rmg,\Rrg\right]\right)}{\Rmg} \bigg|_{\Rmg= \Rrg}  \\
&= \revision{ \left( \frac{1}{\hat{I}_m^{(g)}(\Rrg,\Rrg)} \right) \dsdel{\hat{I}_m^{(g)}(\Rmg,\Rrg)}{\Rmg} \bigg|_{\Rmg= \Rrg}} \\ &+ \revision{ \left(\frac{1}{\hat{S}_m^{(b)}(\Rrg,\Rrg)} \right) \dsdel{\hat{S}_m^{(b)}(\Rmg,\Rrg)}{\Rmg}   \bigg|_{\Rmg= \Rrg}. } \\
&= \revision{  \left\{ \begin{array}{cr}
      \ds \frac{\alpha}{\left(\Rrg\right)^2} & : \Rrg\leq \frac{1}{c} \vspace{1mm} \\ 
\ds\frac{1}{\left(\Rrg\right)^2} \left[\alpha - \left(1 - \alpha\right) \left(\Rrg- \frac{1}{c} \right) \right] & : \Rrg> \frac{1}{c}
     \end{array}
   \right. }
\end{aligned}
\end{equation}

We note that the local selection gradient is always positive for $\Rrg\leq \frac{1}{c}$ and is a decreasing function of $\Rrg$. Therefore, we deduce that there is unique evolutionary singular strategy satisfying $s'_{\Rrg}(\Rrg) = 0$, which is given by
\begin{equation} \label{eq:cdsingular} \left(\Rrg\right)^* = \frac{\alpha}{1 - \alpha} + \frac{1}{c} \end{equation}
Because we only consider levels of sociality for the good contagion with $\Rrg\geq 1$, the singular strategy given in Equation \eqref{eq:cdsingular} is infeasible when $\left(\Rrg\right)^* < 1$ and, consequently, the selection gradient $s'_{\Rrg}(\Rrg)$ is negative at $\Rrg= 1$. Using Equation \eqref{eq:cdsrprime}, we see that this singular strategy is infeasible when
\begin{equation} \label{eq:selectionatonenegative}
    s'_{\Rrg}(\Rrg) \bigg|_{\Rrg= 1} = \alpha - \left( 1 - \alpha\right) \left( 1 - \frac{1}{c} \right) < 0 \Longrightarrow \left( 1 - 2 \alpha \right) c > 1 - \alpha.
\end{equation}
Notably, this can only occur if $\alpha < \frac{1}{2}$, when individuals are more concerned with avoiding the bad contagion than with acquiring the good contagion. When $\alpha < \frac{1}{2}$ we can rearrange the inequality from Equation \eqref{eq:selectionatonenegative} to obtain the following condition for the infeasibility of the singular strategy $\left(\Rrg\right)^*$:
\begin{equation} \label{eq:cdsocialcollapse}
  c > \frac{1-\alpha}{1- 2 \alpha}. 
\end{equation}
When $\alpha < \frac{1}{2}$ and condition of \eqref{eq:cdsocialcollapse} holds for $c$, we can use this inequality to see that the selection gradient satisfies
\begin{equation} \label{eq:selectioncollapse} s'_{\Rrg}(\Rrg) = \frac{1}{\left(\Rrg\right)^2} \left[\alpha - \left(1 - \alpha\right) \left(\Rrg- \frac{1}{c} \right) \right] \leq \frac{1}{\left(\Rrg\right)^2} \left(1 - \alpha \right) \left(1 - \Rrg\right). \end{equation}
Then, for any interior sociality strategy $\Rrg> 1$, we see from Equation \eqref{eq:selectioncollapse} that
\[ s'_{\Rrg}(\Rrg) < 0 \: \: \mathrm{when} \: \: c > \frac{1-\alpha}{1 - 2 \alpha} \:, \: \Rrg> 1. \]
As a result, the selection gradient is always decreasing and $\Rrg= 1$ is the resulting ESS level of sociality when Equation \eqref{eq:cdsocialcollapse} holds. \revision{Therefore we find that our evolutionarily-stable strategy has the following piecewise characterization}
\begin{equation} \label{eq:RESSCD}
    \revision{\RESS = \ds\max\left(1,\frac{1}{c} + \frac{\alpha}{1-\alpha}\right)}.
\end{equation}
We further study the classification of \revision{$\RESS$} as an evolutionarily-stable strategy and convergence stable strategy using the relevant conditions on the local selection gradient in Appendix B.

 In Table \ref{tab:CDtable}, we compare the evolutionarily-stable and socially-optimal levels of sociality $\RESS$ and $\Ropt$ for the different possible values of the relative spreading abilities for the good and bad contagion $c$ and the weight $\alpha$ placed upon the good and bad contagion in the Cobb-Douglas utility. Across each of the cases, we see from Table \ref{tab:CDtable} that $\RESS < \Ropt$ when $c > 1$, $\RESS = \Ropt$ when $c = 1$, and $\RESS > \Ropt$ when $c < 1$. In particular, this means that the evolutionary-stable level of interactions exceeds the social optimum when the good contagion spreads more effectively than the bad contagion, while the evolutionary-stable level of interaction is less than the social optimum when the bad contagion spreads more effectively than the good contagion. %
\renewcommand{\arraystretch}{3.5}
\begin{table}[ht] 
\centering 
\begin{tabular}{c|c|c}  
& $\alpha c > (1-\alpha)(c-1)$ & $\alpha c < (1-\alpha)(c-1)$ \\
\hline 
$c > 1-\alpha $ & \begin{math}\begin{aligned} \RESS{} &= \frac{1}{c} + \frac{\alpha}{1-\alpha} \\ 
\Ropt &= \frac{1}{1-\alpha}\end{aligned} \end{math} & 
\begin{math}
\begin{aligned} 
\RESS &= 1 \\ 
\Ropt &= \frac{1}{1- \alpha{}} > 1 
\end{aligned} 
\end{math} \\

 \hline

 $c < 1 - \alpha$ & \begin{math}\begin{aligned}\RESS &= \frac{1}{c} + \frac{\alpha}{1-\alpha} > 1 \\ \Ropt &= \frac{1}{c}\end{aligned}\end{math} & N/A \\ 
\end{tabular}
\caption{Evolutionarily-stable and socially-optimal sociality strategies $\RESS$ and $\Ropt$ for different cases on the relative values of the relative weight $\alpha$ of the good contagion under Cobb-Douglas utility and the relative infectiousness $c$ of the \revision{bad} contagion.  }
\label{tab:CDtable}
\end{table}
 \renewcommand{\arraystretch}{1}
 
 \revision{Finally, we can use the relationship between the reproduction number $\Rg{\cdot}$ and the social interaction rate $\sigma{\cdot}$ to describe the socially-optimal and evolutionarily-stable sociality strategies in terms of the underlying rate of social interactions for individuals. Noting that $\sigma_{y} = \ds\frac{\gamma_g \Rg{y}}{p_g}$, we can use Equations \eqref{eq:RoptCD} and \eqref{eq:RESSCD} to provide the following characterization of the socially-optimal $\sigma_{g}^{\mathrm{opt}}$ and $\sigma_{g}^{\mathrm{ESS}}$:}
 \begin{subequations}
 \begin{align}
     \revision{\sigma_{g}^{\mathrm{opt}}} &= \revision{\frac{\gamma_g}{p_g} \max\left(\frac{1}{c}, \frac{1}{1-\alpha}   \right)} \\
     \revision{\sigma_{g}^{\mathrm{ESS}}} &=\revision{ \frac{\gamma_g}{p_g} \max\left(1 , \frac{1}{c} + \frac{\alpha}{1- \alpha} \right)}.
 \end{align}
 \end{subequations}
 \revision{Because the socially-optimal and evolutionarily-stable social interaction strategies are proportional to the equivalent quantities expressed in terms of the reproduction number of the good contagion, we see that the qualitative results and social dilemmas can be seen in either formulation of the sociality strategies.}
 
 \subsubsection{Illustrating the Social Dilemma} \label{sec:exploringsocial}
\begin{figure}[tph!]
    \centering
       \begin{subfigure}[b]{\textwidth}
         \centering
         \includegraphics[width=\textwidth, height = 0.45\textheight]{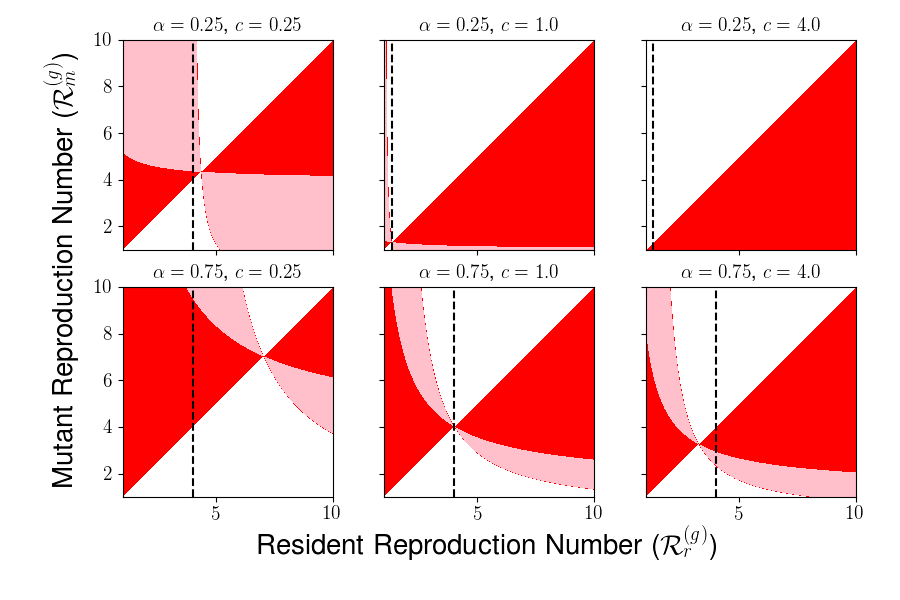}
         \vspace{-10mm}
         \caption{Example Pairwise Invasibility Plots (PIPs): \revision{Colors of regions described in caption for Figure \ref{fig:combined} on next page.}}
         \label{fig:PIPsample} 
     \end{subfigure}
\begin{subfigure}[h]{0.45 \textwidth}
         \centering
         \includegraphics[width=\textwidth]{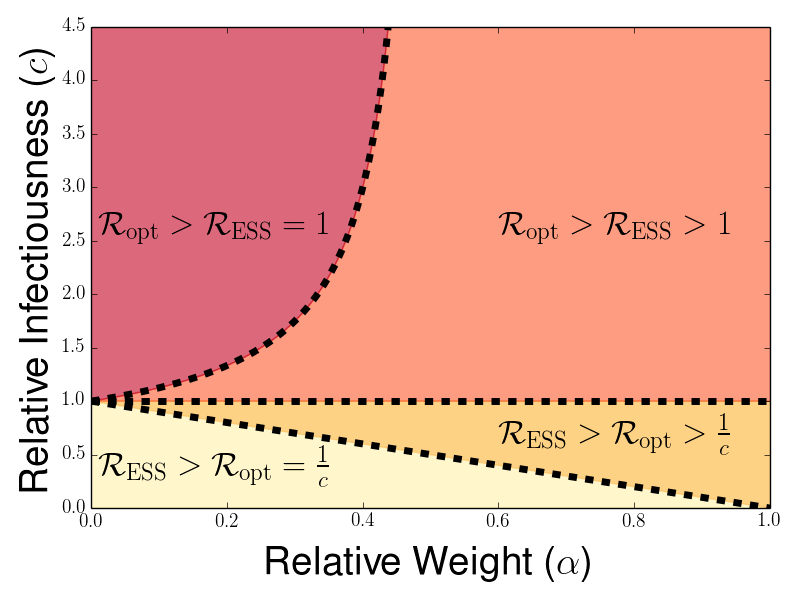}
         \vspace{-5mm}
         \caption{Regimes for $\RESS$ and $\Ropt$ for various $c$ and $\alpha$.}
         \label{fig:ESS_SO_regimes}
     \end{subfigure}  
     \hfill
\begin{subfigure}[h]{0.45 \textwidth}
         \centering
         \includegraphics[width=\textwidth, height = 0.5\textheight]{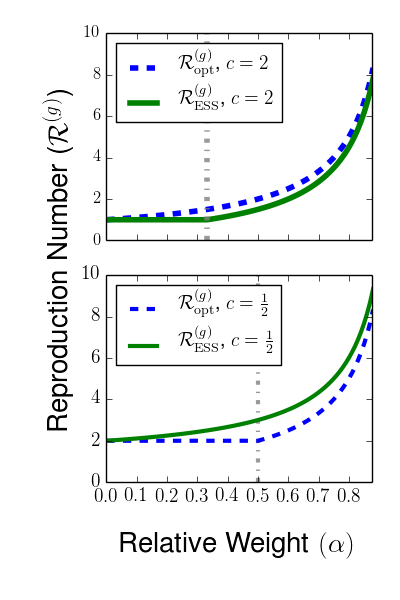}
         \vspace{-10mm}
         \caption{Comparing values of $\RESS$ and $\Ropt$}
         \label{fig:RESSvsRopt}
     \end{subfigure}

    \caption{(Caption next page)}
   
    \label{fig:combined}
\end{figure}

Now, we illustrate how the social dilemma seen in our model with Cobb-Douglas utility depends on $\alpha$ and $c$. One tool for visualizing the relative competitive abilities of different sociality strategies under adaptive dynamics is a pairwise-invasibility plot (PIP), which compares the relative competitive ability of possible mutant and resident strategies in the limit of a rare mutant. In Figure \ref{fig:combined}a, we present example PIPs for various values of $c$ (0.25, 1, and 4) and $\alpha$ (0.25 and 0.75) for the case of Cobb-Douglas utility, displaying the possible combinations of of sociality strategies $\Rrg$ and $\Rmg$ for which $\Rrg$  outcompetes $\Rmg$ when either strategy is rare (shaded in red), for which $\Rmg$ outcompetes $\Rrg$ when either is rare (shaded in white), and for which each of the two strategies can invade the other when rare (shaded in pink). Consistent with the classification using the local selection gradient, we see that $\RESS > \Ropt$ when $c = 0.25$, $\RESS = \Ropt$ when $c=1$, and $\RESS < \Ropt$ when $c = 4$.

In addition, we present in Figure \ref{fig:combined} an illustration of the different possible regimes for $\RESS$ and $\Ropt$ as determined by the parameters $c$ and $\alpha$, highlighting the various cases presented in Table \ref{tab:CDtable}. In Figure \ref{fig:combined}, we show how $\RESS$ and $\Ropt$ vary with the relative weight $\alpha$ for the cases of $c = \frac{1}{2}$ and $c = 2$ for the relative infectiousness of the bad contagion. Through the illustrations in Figure \ref{fig:combined}b,c, we highlight the direction of the social dilemma by showing that evolutionary dynamics promote too much interaction when $c < 1$ and promote too little interaction when $c < 1$. Furthermore, we highlight the extreme forms of the social dilemma that can be achieved for certain parameters, in which either $\RESS = 1$ and the good contagion cannot spread in the population or in which $\Ropt > \frac{1}{c}$ and the bad contagion could be eliminated with socially-optimal behavior but remains present in the evolutionarily-stable strategy. 

\addtocounter{figure}{-1}
\begin{figure} [t!]
  \caption{(Previous page.) (a): Pairwise invasibility plot with mutual invasbility or lack thereof shown. \revision{In each panel, horizontal axis describes the reproduction number $\Rr{}$ of the resident strategy, while the vertical axis describes the reproduction number $\Rm{}$ corresponding to the mutant strategy. The color of a given point describes the outcome of pairwise competition between the resident and mutant strategy. Points displayed in white describe pairs of strategies in which the resident strategy dominates the mutant strategy: a small cohort of the mutant will fail to invade a population primarily consisting of resident strategy, while a small cohort of the resident strategy will successfully invade a population primarily consisting of the mutant strategy. Points in red describe pairs of strategies in which the mutant strategy dominates the resident strategy: the mutant strategy will successfully invade the resident when rare, and a population of the mutant strategy will resist the invasion of a small cohort of the resident strategy. Points displayed in pink describe a case in which neither the resident nor mutant strategy dominates the other: the mutant invades the resident when rare and the resident invades the mutant when rare.} The dashed line describes the socially-optimal strategy $\Ropt$, while the point of intersection of two components of the red region corresponds to the evolutionarily-stable strategy $\RESS$. All areas of mutual invasibility are off the diagonal except when arbitrarily close to the evolutionarily-stable strategy ESS, which implies that dimorphism will not evolve if mutations are small. (b): Illustration of the four possible qualitative behaviors for $\Ropt$ and $\RESS$ across the range of relative levels of infectiousness for the bad contagion $c$ and relative weight placed on the good contagion $\alpha$ under Cobb-Douglas utility. The various regions are defined by the relative size of $\RESS$ and $\Ropt$, as well as whether $\RESS = 1$ or $\RESS > 1$ and whether $\Ropt = \frac{1}{c}$ or $\Ropt > \frac{1}{c}$, with the boundaries between regions as characterized by Table \ref{tab:CDtable}. (c): Reproduction numbers $\RESS$ (solid green line) and $\Ropt$ (dashed blue line) for the good contagion. We plot these reproduction numbers as a function of the relative importance of the good contagion $\alpha$ and for the relative infectiousness values $c = 2$ (top) and $c =\frac{1}{2}$ (bottom).}
\end{figure}

\subsubsection{General Utility Function} \label{sec:genutility}

Here we study the possibility of social dilemmas for a more general family of utility functions 
describing agents who benefit from the good contagion and suffer from the bad contagion. What we would to describe through this family of utility functions is the desired property that, all else being equal, individuals would prefer increased exposure to the good contagion and decreased exposure to the bad contagion. Mathematically, we can formulate this by introducing the utility function \[ U[\Rmg, \Rrg] :=U(\hat{I}^{(g)}_m(\Rmg,\Rrg),\hat{S}^{(b)}_m(c\Rmg,c\Rrg))\] with continuous partial derivatives satisfying
\begin{itemize}
    \item $\frac{\partial U(\cdot,\cdot)}{\partial \hat{I}^{(g)}(\cdot)} > 0$: an individual's utility is improved by increased exposure to the good contagion
    \item $\frac{\partial U(\cdot,\cdot)}{\partial \hat{S}^{(b)}(\cdot)} > 0$: an individual's utility is improved by decreased exposure to the bad contagion
\end{itemize}

First we examine the question of social optimality. If all individuals have the resident level of sociality $\Rrg$, then the levels of contagion converge to the equilibria $\ihat{g}(\Rrg)$ and $\revision{\shat{b}(c \Rrg) = 1 - \ihat{b}(c \Rrg)}$ from Equation \eqref{eq:equilibriapiecewise}, and the utility for individuals can be written as
\begin{equation}
\label{eq:generalutilitymonomorphic}
U[\Rrg,\Rrg] = U\left(\ihat{g}(\Rrg),\revision{\shat{b}(c \Rrg) }\right) \end{equation}
To find potential socially-optimal levels of sociality $\Rrg$, we differentiate Equation \eqref{eq:generalutilitymonomorphic} with respect to $\Rrg$. \revision{To study social optima, we have to consider separately the cases in which both contagions spread (when $\Rrg > \max\{1, \frac{1}{c} \}$), in which either only the good contagion spreads (when $\frac{1}{c} \geq \Rrg > 1$) or only the bad contagion spreads (when $1 \geq \Rrg > \frac{1}{c}$), or in which both contagions do not spread (when $\Rrg \leq \min\{1,\frac{1}{c} \}$). We note that we were able to primarily ignore the latter two cases for the Cobb-Douglas utility function, as $U(\Rrg)$ vanished when $\Rrg \leq 1$ for that family of functions.  }

Using the expressions for the monomorphic contagion equilibria from Equation \eqref{eq:equilibriapiecewise}, we see that
\begin{equation} \label{eq:utilityderivativegeneral}
\begin{aligned}  
\dsdel{U[\Rrg,\Rrg]}{\Rrg} 
&= \frac{\partial U[\Rrg,\Rrg]}{\revision{\partial}\ihat{g}(\Rrg)} \dsdel{\ihat{g}(\Rrg)}{\Rrg} + \frac{\partial U[\Rrg,\Rrg]}{\revision{\partial} \shat{b}(\Rrg)} \dsdel{\shat{b}(\Rrg)}{\Rrg} \\
&= \left\{
     \begin{array}{cr}
     \revision{0} &: \revision{\Rrg \leq 1, \frac{1}{c}} \vspace{2mm} \\
     \revision{-\ds\frac{1}{c \left(\Rrg \right)^2} \frac{\partial U[\Rrg,\Rrg]}{\partial \shat{b}(\Rrg)} }  &: \revision{\frac{1}{c} < \Rrg \leq 1}\\
       \ds\frac{1}{\left(\Rrg\right)^2} \frac{\partial U[\Rrg,\Rrg]}{\revision{\partial} \ihat{g}(\Rrg)} & : \revision{1 < } \Rrg\leq \frac{1}{c} \\
       \ds\frac{1}{\left(\Rrg\right)^2} \left[\frac{\partial U[\Rrg,\Rrg]}{\revision{\partial} \ihat{g}(\Rrg)} - \frac{1}{c} \frac{\partial U[\Rrg,\Rrg]}{\revision{\partial} \shat{b}(\Rrg)}  \right] & : \Rrg> 1, \frac{1}{c}
     \end{array} .
   \right.
\end{aligned} \end{equation}
From our assumption on the partial derivatives of $U[\Rrg,\Rrg]$, we see that $U[\Rrg,\Rrg]$ is increasing \revision{for the case in which $1 <  \Rrg \leq  \tfrac{1}{c}$ (which is possible when $c < 1$), and is decreasing for the case in which $\frac{1}{c} < \Rrg \leq 1$ (only possible when $c > 1$).}. \revision{This means that, for the case in which $c \leq 1$, we can} we look for socially-optimal levels of sociality among $\Rrg\in [\tfrac{1}{c},\infty)$. \revision{For the case in which $c >1$, it is also possible that the maximizer of the social utility is $\Rrg = \frac{1}{c} < 1$, in which case it is socially optimal to have neither contagion spread in the population. We will show in Appendix B that such a social optimum is possible for some linear utility functions and the Constant Elasticity of Substitution (CES) family of utility functions \cite{arrow1961capital,mas1995microeconomic}. Even in these extreme case in which non-interaction can be collectively optimal, we can now study the conditions under which the social utility function will have at least a unique local optimum for sociality strategies $\Rr \in (1,\infty)$ for which the good contagion can spread. }

\revision{Now we look to characterize the existence and uniqueness of maximizers of the utility $U[\Rrg,\Rrg]$ when both contagions are present in the population, which occurs for $\Rrg > \max\{1,\frac{1}{c}\}$.} From the term in square brackets in Equation \eqref{eq:utilityderivativegeneral}, we see that one sufficient condition on $U[\Rrg,\Rrg]$ for uniqueness of a socially-optimal level of sociality $\Rrg$ is that the $\del{U[\Rrg,\Rrg]}{\ihat{g}(\Rrg)}$ is a decreasing function of $\Rrg$ and that $\del{U[\Rrg,\Rrg]}{\shat{b}(\Rrg)}$ is an increasing function of $\Rrg$. 
Under these assumptions, the derivative in Equation \eqref{eq:utilityderivativegeneral} is decreasing in $\Rrg$ and therefore changes sign at most once for $\Rrg\in [\tfrac{1}{c},\infty)$. In that case, either there exists a unique interior social optimum, or the 
social optimal is achieved at one of the endpoints $\Rrg= 1$ (only if $c > 1$), $\Rrg= \frac{1}{c}$,  
$\Rrg= \infty$. 

If we allow for greater regularity on the utility function, we can obtain a sufficient condition for the existence of a unique social optimum featuring a finite rate of social interaction. Taking the following partial derivatives 
\begin{align*}
   \dsdel{}{\Rrg} \left( \dsdel{U[\Rrg,\Rrg]}{\hat{I}^{(g)}(\Rrg)} \right) &=  \doubledelsame{U[\Rrg,\Rrg]}{\left(\hat{I}^{(g)}(\Rrg)\right)} \overbrace{\left(\dsdel{\hat{I}^{(g)}(\Rrg)}{\Rrg}\right)}^{> 0} \\ 
    \dsdel{}{\Rrg} \left( \dsdel{U[\Rrg,\Rrg]}{\hat{S}^{(b)}(\Rrg)} \right) &=  \doubledelsame{U[\Rrg,\Rrg]}{\left(\hat{S}^{(b)}(\Rrg)\right)} \underbrace{\left(\dsdel{\hat{S}^{(b)}(\Rrg)}{\Rrg}\right)}_{< 0},
\end{align*}
we see that taking the assumption that the second partial derivatives above are negative will allow us to deduce from Equation \eqref{eq:utilityderivativegeneral} that $\del{}{\Rrg} U[\Rrg,\Rrg]$ is a strictly decreasing function of $\Rrg$. Therefore we see that a natural sufficient condition for existence of unique, finite social optimum $\Ropt$ is the $U(\cdot,\cdot)$ be twice-differentiable and strictly concave. Such an assumption holds for the \revision{CES utility function \cite{arrow1961capital,mas1995microeconomic}}, as well as for a variety of families of utility functions that contain Cobb-Douglas and CES as special cases \cite{matsuyama2017beyond}. However, a linear utility function $U[\Rrg,\Rrg] = \alpha \hat{I}^{(g)}(\Rrg) + (1-\alpha) \hat{S}^{(b)}(\Rrg)$ is not strictly concave, and we show in Appendix B that, for a such a utility function, infinite social interaction can be collectively optimal for a range of $c$ and $\alpha$.  %

\revision{Furthermore, for the case in which $c \leq 1$ and the bad contagion will not spread if $\Rrg \leq 1$, we can use these regular and concavity assumptions along with Equation \eqref{eq:utilityderivativegeneral} to deduce that this unique local maximum of $U[\Rrg,\Rrg]$ is, in fact, the global maximizer of the social utility. For the case of $c > 1$, the strategy $\Rrg = \frac{1}{c} < 1$ will also be a local maximum because $U[\Rr,\Rr]$ is a decreasing function on $\frac{1}{c} < \Rrg < 1$ (see Equation \eqref{eq:utilityderivativegeneral}). In this case, we need to compare the value of the utility functions at these two points to determine the global maximizer of $U[\Rrg,\Rrg]$. This presence of two local optima also has consequences for evolutionary dynamics, as we will show in Appendix B that it is possible to achieve evolutionary bistability between $\Rrg = \frac{1}{c}$ and a sociality strategy featuring presence of the good contagion under linear and CES utility functions when $c > 1$. }

Now we will consider the question of evolutionarily-stable strategies. To study ESSes, we compute the local selection gradient 
              \begin{equation}
              \label{eq:generalselectiongradientfirst}
              \begin{aligned} s_{\Rrg}'(\Rrg)  
              &= \dsdel{ U[\Rmg,\Rrg]}{\hat{I}^{(g)}(\Rmg,\Rrg)} \dsdel{\hat{I}^{(g)}(\Rmg,\Rrg)}{\Rmg} \bigg|_{\Rmg= \Rrg}  \\  &+ \dsdel{ U[\Rmg,\Rrg]}{\hat{S}^{(b)}(\Rmg,\Rrg)}  \dsdel{\hat{S}^{(b)}(\Rmg,\Rrg)}{\Rmg}  \bigg|_{\Rmg= \Rrg}. %
              \end{aligned}
              \end{equation}
 \revision{We can use Equation \eqref{eq:localendemicpartials} to rewrite out expression for the local selection gradient, which will do by separately considering the cases in which $\Rrg \leq \frac{1}{c}$ and $\Rrg > \frac{1}{c}$. When $\Rrg \leq \frac{1}{c}$, our selection gradient takes the form}
 \begin{subequations}
\label{eq:selectiongradientgeneral}
\begin{equation}
    \label{eq:selectiongradientless}
    \revision{s_{\Rrg}'(\Rrg) = \ds\left(\frac{\Rrg-  1}{\Rr^3} \right) \ds\frac{\partial U[\Rrg,\Rrg]}{\partial \hat{I}^{(g)}(\Rrg,\Rrg)}.} 
\end{equation}
\revision{When $\Rrg > \frac{1}{c}$, we can write our selection gradient as}
\begin{equation}
    \label{eq:selectiongradientgreater}
    \begin{aligned}
        \revision{s_{\Rrg}'(\Rrg) = \ds\frac{1}{\left(\Rrg\right)^2}} & \revision{\left[ \left( \frac{\Rrg- 1}{\Rrg} \right) \frac{\partial U[\Rrg,\Rrg]}{\partial \hat{I}^{(g)}(\Rrg,\Rrg)} \right.}  \\
        &- \revision{ \left. \left( \ds\frac{\Rrg- \frac{1}{c}}{\Rrg} \right) \frac{1}{c} \frac{\partial U[\Rrg,\Rrg]}{ \partial \hat{S}^{(b)}(\Rrg,\Rrg)}\right]}
    \end{aligned}
\end{equation}
 \end{subequations}
%
We notice from our assumption on $\del{U(\cdot)}{I^{(g)}(\cdot)}$ that the selection gradient is always positive for $\Rrg\leq \tfrac{1}{c}$, and we can compute that
\begin{equation} \label{eq:selectiongradient1c} \lim_{\Rrg\to \frac{1}{c}^{+}} s'_{\Rrg}(\Rrg) = c^2 \left( 1 - c \right)  \frac{\partial U[\Rrg,\Rrg]}{\partial \hat{I}^{(g)}(\Rrg,\Rrg)} > 0. \end{equation}
In particular, this means that, for the case in which $\Ropt = \frac{1}{c}$ for $c < 1$, the social optimum cannot be evolutionarily-stable. We will look to find ESS sociality strategies with $\Rrg \in (\frac{1}{c},\infty)$.

To explore social dilemmas when $\Ropt > \frac{1}{c}$, we can look to re-express the local selection gradient in a form providing a comparison to first-order condition for the social optimization problem. We obtain 
\begin{align} s'_{\Rrg}(\Rrg) &= \left( \frac{\Rrg- 1}{\Rr^3 } \right) \left[  \frac{\partial U[\Rrg,\Rrg]}{\partial \hat{I}^{(g)}(\Rrg,\Rrg)} - \frac{1}{c} \frac{\partial U[\Rrg,\Rrg]}{ \partial \hat{S}^{(b)}(\Rrg,\Rrg)}  \right] \\ &+ \frac{1}{c \Rr^3} \left( \frac{1}{c} - 1 \right) \frac{\partial U[\Rrg,\Rrg]}{ \partial \hat{S}^{(b)}(\Rrg,\Rrg)} \nonumber. 
\end{align}
Using Equation \eqref{eq:utilityderivativegeneral}, we can therefore relate the local selection gradient to the derivative of the monomorphic utility through the equation
\begin{equation} \label{eq:selectiongeneralcompare}
\begin{aligned}
 s'_{\Rrg}(\Rrg) &= \left(\revision{\frac{\Rrg - 1}{\Rrg}} \right)\dsdel{U[\Rrg,\Rrg]}{\Rrg} \\ &+ \frac{1}{c \left(\Rrg\right)^3} \left( \frac{1}{c} - 1 \right) \frac{\partial U[\Rrg,\Rrg]}{ \partial \hat{S}^{(b)}(\Rrg,\Rrg)}. 
 \end{aligned}
\end{equation}
From our assumption on the utility function that $\del{U[\cdot]}{S^{(b)}(\cdot)} > 0$ \revision{for $\Rrg > \max\{1,\frac{1}{c}\}$}, we can therefore deduce that, for such $\Rrg$, 
 
 \begin{subequations}
    \begin{alignat}{2}
    \label{eq:gradientinequalities}
        \revision{\dsdel{U[\Rrg,\Rrg]}{\Rrg} \geq 0 \: \:} & \revision{\mathrm{implies} \: \: s'_{\Rrg}(\Rrg)  > 0}  & \textnormal{ for } c < 1 \\ 
         \revision{\dsdel{U[\Rrg,\Rrg]}{\Rrg} \leq 0} \: \: & \revision{\mathrm{implies} \: \: s'_{\Rrg}(\Rrg)   < 0}    & \textnormal{ for } c > 1.
    \end{alignat}
\end{subequations}

In particular, this means that interior local maxima of the monomorphic utility function $U[\Rrg,\Rrg]$ are not evolutionarily-stable strategies unless $c =1$, which is a structurally unstable case. When $c = 1$, the local selection gradient coincides with the derivative of the monomorphic utility function, and we see that social optima coincide with evolutionarily-stable strategies. Furthermore, under the additional assumptions on $U[\Rmg,\Rrg]$ guaranteeing the existence of a unique social optimum $\Ropt$, we can deduce from Equation \eqref{eq:gradientinequalities} that any evolutionarily-stable states $\RESS$ satisfy $\RESS > \Ropt$ for $c < 1$ and $\RESS < \Ropt$ for $c > 1$. In particular, this means that the evolutionary dynamics feature a social dilemma in which the evolutionarily-stable and socially-optimal outcomes disagree when $c \neq 1$, generalizing the qualitative form of the social dilemma that we observed for the case of Cobb-Douglas utility.%

\section{Discussion} \label{sec:discussion}

In this paper, we have considered the question of how individuals choose their level of social interactions in response to the benefits of exposure to a good contagion and the costs of exposure to a bad contagion. Through both the frameworks of shorter-term replicator dynamics and long-time adaptive dynamics, we characterize evolutionarily-stable levels of sociality, and show how these evolutionary outcomes can be misaligned with the sociality strategies that optimize collective utility. Of key importance to the evolutionary dynamics is the relative transmissibility of the good and bad contagion, as the evolutionarily-stable sociality level features too much interaction when the good contagion spreads more readily than the bad contagion, while evolutionary dynamics favor too little interaction when the bad contagion in the alternative scenario. From this analysis, we see that even in a simple model of social interaction and contagion spread, the benefits of good contagion and the fear of bad can result in a social dilemma in the evolution of sociality strategies. 

Most strikingly, there were two extreme examples of the social dilemma in which individually-rational behavior led to different qualitative behavior contagion behavior than is seen in the socially-optimal scenario. When the good contagion spread sufficiently more readily than the bad contagion, socially-optimal sociality strategies can eradicate the bad contagion at equilibrium. However, in such cases, the evolutionarily-stable strategy may still introduce positive levels of the bad contagion due to overpursuit of the benefits of the good contagion. When the bad contagion spreads sufficiently more rapidly than the good contagion, the evolutionarily-stable outcome featured no social interaction whatsoever, even though the socially-optimal sociality strategy always features a positive level of social interaction. Under the Cobb-Douglas utility function, a strategy featuring zero social interaction constitutes a utility minimizer, so the evolutionary dynamics end up achieving the worst possible outcome when the bad contagion is sufficiently contagious.    

Having identified a social dilemma in the evolution of sociality strategies, a natural follow-up question is what mechanisms can be used to help mitigate the suboptimal outcomes achieved under evolutionary dynamics. We can draw inspiration from the literature of the evolution of cooperation to study how additional effects of population structure including assortment, reciprocity, and multilevel selection can help to promote efficient levels of socialization \cite{nowak2006five,nowak2010evolutionary}. In particular, the case of a complete collapse of social interactions as an evolutionarily-stable outcome suggests that there may be scenarios in which the tradeoff between good and bad contagion would require additional mechanisms beyond well-mixed individual-level selection in order to permit the existence of groups featuring social interactions. As a first attempt to explore one such mechanism, we show in Appendix B that an assortative process preferencing interactions with same-strategy individuals can help to mitigate the effects of the social dilemma and more closely align the evolutionarily-stable and socially-optimal levels of sociality. %
By matching together more frequently individuals who interact too much (respectively too little) when the good contagion spreads more (respectively less) readily than the bad contagion, this assortative process helps to internalize the negative externalities generated by suboptimal levels of social interaction.

The continuous-strategy nature of our model of sociality also bears resemblance to recent work on the sustainable management of common-pool resources like fisheries, which has shown that social pressures can help to maintain sustainable levels of extraction effort as a stable social norm \cite{tavoni2012survival,schluter2016robustness,tilman2017maintaining}. Further work on the two-contagion model can look to more deeply connect our social dilemma of sociality to the broader literature on evolutionary game theory and on game-theoretic models of host behavior and infectious disease. In particular, the nonlinear dependence of utility upon the composition of sociality strategies in the population and the endemic equilibria of the dimorphic contagion dynamics bears similarity to game-theoretic models of vaccination \revision{both in well-mixed and spatial populations} \cite{bauch2003group,fu2011imitation,chen2019imperfect} as well a models of nonlinear public goods games from evolutionary game theory \cite{santos2011risk,archetti2018analyze}. These game-theoretic scenarios arising from the collective behavior of an interacting population may be a more realistic representation of social dilemmas one encounters in everyday life than the stylized models of pairwise interactions such as the Prisoners' Dilemma.  

There are also many natural directions for future research regarding how social dilemmas of sociality may arise under more complicated models of disease dynamics or social network structure. While this paper restricts attention to a pair of simple contagions with bilinear incidence functions, many models of social transmission  explore the spread of information via complex contagion \cite{dodds2004universal,osborne2018complex}. In addition, our assumption that the two contagions spread independently in the population could be relaxed to consider a variety of possible interacting dynamics between a pair of contagions, with examples ranging from the coupled spread of a disease and awareness of the disease outbreak \cite{xia2019new} to the cultural transmission of a risky or careful behavior that impacts the likelihood of exposure to infectious disease \cite{tanaka2002coevolution}. Because both complex contagion and the superinfection of  simple contagions often produce behaviors including bistability or Hopf bifuractions in disease dynamics \cite{dodds2004universal,gao2016coinfection}, it may be possible to observe more complicated evolutionary behaviors like evolutionary cycling in sociality strategies \cite{dercole2002ecological} if the good contagion features a sigmoidal incidence function or if infection with the good contagion impacts the ability to acquire or avoid the bad contagion. In addition, while sociality strategies are modeled here through the rate of well-mixed interactions that individuals have, it is also reasonable to consider how the benefits and costs of social interaction can impact how individuals choose neighbors in a network-structured population. Tools such as pair approximations \cite{gross2008robust,marceau2010adaptive} or graphons \cite{vizuete2020graphon,erol2020contagion,aurell2021finite} can be used to model the spread of couple contagions on graph-structured populations, and evolutionary questions could also explore the evolution of modular network structure in the presence of infectious disease \cite{sah2017unraveling,saad2020dynamics,silk2021role}. 

In a recent paper, Ashby and Farine also study the evolutionary dynamics of sociality strategies depending upon the costs and benefits induced by the joint spread of a pair SIS contagions in a well-mixed population \cite{ashby2020social}. Focusing on the case of an infectious disease and an informational contagion, the authors assume that exposure to the informational contagion reduces the rate of death due to the infectious disease. Considering the effects of birth and both natural and disease-dependent death, Ashby and Farine apply an adaptive dynamics approach study pairwise invasibility of sociality strategies at demographic-epidemiological equilibria and to study the coevolution of host sociality strategy and virulence of the infectious disease. One advantage of their approach is that it allows explicit study of the ecoevolutionary dynamics for a specific tradeoff between the good and bad contagion. By contrast, our use of a simpler pair of SIS contagions and a flexible utility function measuring the costs and benefits of social interaction makes possible comparisons between socially-optimal and evolutionarily-stable sociality strategies. A shared feature of the two models is the key role played by the relative transmissibility of the two contagions (encoded in our model by the parameter $c$), which helps to shape the sociality strategies supported by the long-time evolutionary dynamics \cite{ashby2020social}.

While many models for the evolution of social animal groups involve network or groups structures \cite{sah2018disease,udiani2020disease,silk2021role,fu2015risk,gosak2021endogenous}, our model for the evolutionary dynamics of social strategies provides a simple, well-mixed baseline model for understanding the tension between the benefits and costs of informational and disease transmission via social interactions. This social dilemma of sociality motivates further study into mechanisms that can help promote socially-optimal rates of social interaction \cite{nowak2006five} and the establishment of social groups featuring long-time interaction. This misalignment between individual and collective interests is also reminiscent of the social dilemmas of social distancing explored in modern disease outbreaks \cite{morin2015social,glaubitz2020oscillatory,martcheva2021effects,cho2020mean, peng2021multilayer}, in which individuals may choose to interact more than is collectively optimal in pursuit of economic or personal benefits of social interactions. 
Because infectious disease and social learning has been attributed as factors driving social evolution in settings ranging from the development of modular population structure \cite{sah2018disease} and division-of-labor \cite{udiani2020disease} in social insects to the cultural evolution of collectivist social norms in human populations \cite{fincher2008pathogen}, we hope that this model highlights the challenges faced by individuals and populations in light of the inherent benefits and costs of social interaction.

\renewcommand{\abstractname}{Author contributions}
\begin{abstract} 
DBC, DHM, and PR conceived the study and developed the model. DBC and DHM analyzed the model and produced figures. DBC and DHM wrote the first draft of the paper, which all authors edited. All authors contributed to the interpretation and contextualization of the results.
\end{abstract}

\renewcommand{\abstractname}{Acknowledgments}
\begin{abstract} 
 This research arose from a collaboration through the Princeton-Humboldt Cooperation and Collective Cognition Network (CoCCoN). We are grateful to all CoCCoN participants for helpful discussions during early presentations of this work. We thank Sarah Kocher, Joshua B. Plotkin, Juan Bonachela, Chadi M. Saad-Roy, and Olivia J. Chu for additional helpful discussions. DBC gratefully acknowledges support from NSF Grant DMS-1514606, ARO Grant W911NF-18-1-0325, and the Simons Foundation Math + X grant. 
\end{abstract}

\renewcommand{\abstractname}{Acknowledgments}
\begin{abstract} 
All code used to generate figures, numerical results (figures 2 and 5), and symbolic calculations (Appendix B.4) is archived on Github (\href{https://github.com/dbcooney/Social-Dilemmas-of-Sociality-due-to-Beneficial-and-Costly-Contagion}{https://github.com/dbcooney/Social-Dilemmas-of-Sociality-due-to-Beneficial-and-Costly-Contagion}) and licensed for reuse, with appropriate attribution/citation, under a BSD 3-Clause Revised License. 
\end{abstract}

\bibliography{adaptivesociality}

\begin{thebibliography}{10}

\bibitem{rubenstein2017evolution}
D.~R. Rubenstein and P.~Abbot, ``The evolution of social evolution,'' {\em
  Comparative Social Evolution}, pp.~1--18, 2017.

\bibitem{alexander1974evolution}
R.~D. Alexander, ``The evolution of social behavior,'' {\em Annual Review of
  Ecology and Systematics}, vol.~5, no.~1, pp.~325--383, 1974.

\bibitem{kermack1927contribution}
W.~O. Kermack and A.~G. McKendrick, ``A contribution to the mathematical theory
  of epidemics,'' {\em Proceedings of the Royal Society of London. Series {A},
  Containing papers of a mathematical and physical character}, vol.~115,
  no.~772, pp.~700--721, 1927.

\bibitem{bass1969new}
F.~M. Bass, ``A new product growth for model consumer durables,'' {\em
  Management Science}, vol.~15, no.~5, pp.~215--227, 1969.

\bibitem{rogers2010diffusion}
E.~M. Rogers, {\em Diffusion of Innovations}.
\newblock Simon and Schuster, 2010.

\bibitem{romano2020stemming}
V.~Romano, A.~J. Macintosh, and C.~Sueur, ``Stemming the flow: Information,
  infection, and social evolution,'' {\em Trends in Ecology \& Evolution},
  vol.~35, no.~10, pp.~849--853, 2020.

\bibitem{kashima2013fission}
K.~Kashima, H.~Ohtsuki, and A.~Satake, ``Fission-fusion bat behavior as a
  strategy for balancing the conflicting needs of maximizing information
  accuracy and minimizing infection risk,'' {\em Journal of Theoretical
  Biology}, vol.~318, pp.~101--109, 2013.

\bibitem{brannstrom2013hitchhiker}
{\AA}.~Br{\"a}nnstr{\"o}m, J.~Johansson, and N.~Von~Festenberg, ``The
  hitchhiker’s guide to adaptive dynamics,'' {\em Games}, vol.~4, no.~3,
  pp.~304--328, 2013.

\bibitem{diekmann2002beginners}
O.~Diekmann, ``A beginners guide to adaptive dynamics,'' {\em Summer School on
  Mathematical Biology}, pp.~63--100, 2002.

\bibitem{geritz1998evolutionarily}
S.~A. Geritz, {\'E}.~Kisdi, G.~Mesz{\'e}na, and J.~A. Metz, ``Evolutionarily
  singular strategies and the adaptive growth and branching of the evolutionary
  tree,'' {\em Evolutionary Ecology}, vol.~12, no.~1, pp.~35--57, 1998.

\bibitem{fenichel2011adaptive}
E.~P. Fenichel, C.~Castillo-Chavez, M.~G. Ceddia, G.~Chowell, P.~A.~G. Parra,
  G.~J. Hickling, G.~Holloway, R.~Horan, B.~Morin, C.~Perrings, M.~Springborn,
  L.~Velazquez, and C.~Villalobos, ``Adaptive human behavior in epidemiological
  models,'' {\em Proceedings of the National Academy of Sciences}, vol.~108,
  no.~15, pp.~6306--6311, 2011.

\bibitem{morin2013sir}
B.~R. Morin, E.~P. Fenichel, and C.~Castillo-Chavez, ``{SIR} dynamics with
  economically driven contact rates,'' {\em Natural Resource Modeling},
  vol.~26, no.~4, pp.~505--525, 2013.

\bibitem{morin2014disease}
B.~R. Morin, C.~Perrings, S.~Levin, and A.~Kinzig, ``Disease risk mitigation:
  The equivalence of two selective mixing strategies on aggregate contact
  patterns and resulting epidemic spread,'' {\em Journal of theoretical
  biology}, vol.~363, pp.~262--270, 2014.

\bibitem{berdahl2019dynamics}
A.~Berdahl, C.~Brelsford, C.~De~Bacco, M.~Dumas, V.~Ferdinand, J.~A. Grochow,
  L.~H{\'e}bert-Dufresne, Y.~Kallus, C.~P. Kempes, and A.~Kolchinsky,
  ``Dynamics of beneficial epidemics,'' {\em Scientific Reports}, vol.~9,
  no.~1, pp.~1--9, 2019.

\bibitem{reluga2009sis}
T.~C. Reluga, ``An {SIS} epidemiology game with two subpopulations,'' {\em
  Journal of Biological Dynamics}, vol.~3, no.~5, pp.~515--531, 2009.

\bibitem{Zhan2018CouplingDO}
X.-X. Zhan, C.~Liu, G.~Zhou, Z.-K. Zhang, G.-Q. Sun, J.~J.~H. Zhu, and Z.~Jin,
  ``Coupling dynamics of epidemic spreading and information diffusion on
  complex networks,'' {\em Applied Mathematics and Computation}, vol.~332,
  pp.~437 -- 448, 2018.

\bibitem{perra2011towards}
N.~Perra, D.~Balcan, B.~Gon{\c{c}}alves, and A.~Vespignani, ``Towards a
  characterization of behavior-disease models,'' {\em PloS One}, vol.~6, no.~8,
  2011.

\bibitem{peng2021multilayer}
K.~Peng, Z.~Lu, V.~Lin, M.~R. Lindstrom, C.~Parkinson, C.~Wang, A.~L. Bertozzi,
  and M.~A. Porter, ``A multilayer network model of the coevolution of the
  spread of a disease and competing opinions,'' {\em Mathematical Models and
  Methods in Applied Sciences}, pp.~1--40, 2021.

\bibitem{tanaka2002coevolution}
M.~M. Tanaka, J.~Kumm, and M.~W. Feldman, ``Coevolution of pathogens and
  cultural practices: a new look at behavioral heterogeneity in epidemics,''
  {\em Theoretical Population Biology}, vol.~62, no.~2, pp.~111--119, 2002.

\bibitem{bauch2005imitation}
C.~T. Bauch, ``Imitation dynamics predict vaccinating behaviour,'' {\em
  Proceedings of the Royal Society B: Biological Sciences}, vol.~272, no.~1573,
  pp.~1669--1675, 2005.

\bibitem{papst2022modeling}
I.~Papst, K.~P. O’Keeffe, and S.~H. Strogatz, ``Modeling the interplay
  between seasonal flu outcomes and individual vaccination decisions,'' {\em
  Bulletin of Mathematical Biology}, vol.~84, no.~3, pp.~1--17, 2022.

\bibitem{cascante2022disease}
J.~Cascante-Vega, S.~Torres-Florez, J.~Cordovez, and M.~Santos-Vega, ``How
  disease risk awareness modulates transmission: coupling infectious disease
  models with behavioural dynamics,'' {\em Royal Society open science}, vol.~9,
  no.~1, p.~210803, 2022.

\bibitem{greenhalgh2015awareness}
D.~Greenhalgh, S.~Rana, S.~Samanta, T.~Sardar, S.~Bhattacharya, and
  J.~Chattopadhyay, ``Awareness programs control infectious disease--multiple
  delay induced mathematical model,'' {\em Applied Mathematics and
  Computation}, vol.~251, pp.~539--563, 2015.

\bibitem{nunn2015sociality}
C.~L. Nunn, M.~E. Craft, T.~R. Gillespie, M.~Schaller, and P.~M. Kappeler,
  ``The sociality--health--fitness nexus: synthesis, conclusions and future
  directions,'' {\em Philosophical Transactions of the Royal Society B:
  Biological Sciences}, vol.~370, no.~1669, p.~20140115, 2015.

\bibitem{udiani2020disease}
O.~Udiani and N.~H. Fefferman, ``How disease constrains the evolution of social
  systems,'' {\em Proceedings of the Royal Society B}, vol.~287, no.~1932,
  p.~20201284, 2020.

\bibitem{mcleod2014sexually}
D.~V. McLeod and T.~Day, ``Sexually transmitted infection and the evolution of
  serial monogamy,'' {\em Proceedings of the Royal Society B: Biological
  Sciences}, vol.~281, no.~1796, p.~20141726, 2014.

\bibitem{evans2020infected}
J.~C. Evans, M.~J. Silk, N.~J. Boogert, and D.~J. Hodgson, ``Infected or
  informed? {Social} structure and the simultaneous transmission of information
  and infectious disease,'' {\em Oikos}, vol.~129, no.~9, pp.~1271--1288, 2020.

\bibitem{bonds2005higher}
M.~H. Bonds, D.~D. Keenan, A.~J. Leidner, and P.~Rohani, ``Higher disease
  prevalence can induce greater sociality: a game theoretic coevolutionary
  model,'' {\em Evolution}, vol.~59, no.~9, pp.~1859--1866, 2005.

\bibitem{ashby2020social}
B.~Ashby and D.~R. Farine, ``Social information use shapes the coevolution of
  sociality and virulence,'' {\em bioRxiv}, 2020.

\bibitem{epstein2008coupled}
J.~M. Epstein, J.~Parker, D.~Cummings, and R.~A. Hammond, ``Coupled contagion
  dynamics of fear and disease: mathematical and computational explorations,''
  {\em PLoS One}, vol.~3, no.~12, p.~e3955, 2008.

\bibitem{epstein2021triple}
J.~M. Epstein, E.~Hatna, and J.~Crodelle, ``Triple contagion: a two-fears
  epidemic model,'' {\em Journal of the Royal Society Interface}, vol.~18,
  no.~181, p.~20210186, 2021.

\bibitem{sandholm2010population}
W.~H. Sandholm, {\em Population Games and Evolutionary Dynamics}.
\newblock MIT Press, 2010.

\bibitem{hofbauer1998evolutionary}
J.~Hofbauer and K.~Sigmund, {\em Evolutionary Games and Population Dynamics}.
\newblock Cambridge University Press, 1998.

\bibitem{arrow1961capital}
K.~J. Arrow, H.~B. Chenery, B.~S. Minhas, and R.~M. Solow, ``Capital-labor
  substitution and economic efficiency,'' {\em The Review of Economics and
  Statistics}, vol.~43, no.~3, pp.~225--250, 1961.

\bibitem{mas1995microeconomic}
A.~Mas-Colell, M.~D. Whinston, and J.~R. Green, {\em Microeconomic Theory},
  vol.~1.
\newblock Oxford University Press New York, 1995.

\bibitem{matsuyama2017beyond}
K.~Matsuyama and P.~Ushchev, ``Beyond {CES}: Three alternative cases of
  flexible homothetic demand systems,'' {\em Global Poverty Research Lab
  Working Paper}, no.~17-109, 2017.

\bibitem{nowak2006five}
M.~A. Nowak, ``Five rules for the evolution of cooperation,'' {\em Science},
  vol.~314, no.~5805, pp.~1560--1563, 2006.

\bibitem{nowak2010evolutionary}
M.~A. Nowak, C.~E. Tarnita, and T.~Antal, ``Evolutionary dynamics in structured
  populations,'' {\em Philosophical Transactions of the Royal Society B:
  Biological Sciences}, vol.~365, no.~1537, pp.~19--30, 2010.

\bibitem{tavoni2012survival}
A.~Tavoni, M.~Schl{\"u}ter, and S.~Levin, ``The survival of the conformist:
  social pressure and renewable resource management,'' {\em Journal of
  Theoretical Biology}, vol.~299, pp.~152--161, 2012.

\bibitem{schluter2016robustness}
M.~Schl{\"u}ter, A.~Tavoni, and S.~Levin, ``Robustness of norm-driven
  cooperation in the commons,'' {\em Proceedings of the Royal Society B:
  Biological Sciences}, vol.~283, no.~1822, p.~20152431, 2016.

\bibitem{tilman2017maintaining}
A.~R. Tilman, J.~R. Watson, and S.~Levin, ``Maintaining cooperation in
  social-ecological systems,'' {\em Theoretical Ecology}, vol.~10, no.~2,
  pp.~155--165, 2017.

\bibitem{bauch2003group}
C.~T. Bauch, A.~P. Galvani, and D.~J. Earn, ``Group interest versus
  self-interest in smallpox vaccination policy,'' {\em Proceedings of the
  National Academy of Sciences}, vol.~100, no.~18, pp.~10564--10567, 2003.

\bibitem{fu2011imitation}
F.~Fu, D.~I. Rosenbloom, L.~Wang, and M.~A. Nowak, ``Imitation dynamics of
  vaccination behaviour on social networks,'' {\em Proceedings of the Royal
  Society B: Biological Sciences}, vol.~278, no.~1702, pp.~42--49, 2011.

\bibitem{chen2019imperfect}
X.~Chen and F.~Fu, ``Imperfect vaccine and hysteresis,'' {\em Proceedings of
  the Royal Society B}, vol.~286, no.~1894, p.~20182406, 2019.

\bibitem{santos2011risk}
F.~C. Santos and J.~M. Pacheco, ``Risk of collective failure provides an escape
  from the tragedy of the commons,'' {\em Proceedings of the National Academy
  of Sciences}, vol.~108, no.~26, pp.~10421--10425, 2011.

\bibitem{archetti2018analyze}
M.~Archetti, ``How to analyze models of nonlinear public goods,'' {\em Games},
  vol.~9, no.~2, p.~17, 2018.

\bibitem{dodds2004universal}
P.~S. Dodds and D.~J. Watts, ``Universal behavior in a generalized model of
  contagion,'' {\em Physical Review Letters}, vol.~92, no.~21, p.~218701, 2004.

\bibitem{osborne2018complex}
M.~Osborne, X.~Wang, and J.~Tien, ``Complex contagion leads to complex dynamics
  in models coupling behaviour and disease,'' {\em Journal of Biological
  Dynamics}, vol.~12, no.~1, pp.~1035--1058, 2018.

\bibitem{xia2019new}
C.~Xia, Z.~Wang, C.~Zheng, Q.~Guo, Y.~Shi, M.~Dehmer, and Z.~Chen, ``A new
  coupled disease-awareness spreading model with mass media on multiplex
  networks,'' {\em Information Sciences}, vol.~471, pp.~185--200, 2019.

\bibitem{gao2016coinfection}
D.~Gao, T.~C. Porco, and S.~Ruan, ``Coinfection dynamics of two diseases in a
  single host population,'' {\em Journal of Mathematical Analysis and
  Applications}, vol.~442, no.~1, pp.~171--188, 2016.

\bibitem{dercole2002ecological}
F.~Dercole, R.~Ferri{\`e}re, and S.~Rinaldi, ``Ecological bistability and
  evolutionary reversals under asymmetrical competition,'' {\em Evolution},
  vol.~56, no.~6, pp.~1081--1090, 2002.

\bibitem{gross2008robust}
T.~Gross and I.~G. Kevrekidis, ``Robust oscillations in {SIS} epidemics on
  adaptive networks: Coarse graining by automated moment closure,'' {\em EPL
  (Europhysics Letters)}, vol.~82, no.~3, p.~38004, 2008.

\bibitem{marceau2010adaptive}
V.~Marceau, P.-A. No{\"e}l, L.~H{\'e}bert-Dufresne, A.~Allard, and L.~J.
  Dub{\'e}, ``Adaptive networks: Coevolution of disease and topology,'' {\em
  Physical Review E}, vol.~82, no.~3, p.~036116, 2010.

\bibitem{vizuete2020graphon}
R.~Vizuete, P.~Frasca, and F.~Garin, ``Graphon-based sensitivity analysis of
  {SIS} epidemics,'' {\em IEEE Control Systems Letters}, vol.~4, no.~3,
  pp.~542--547, 2020.

\bibitem{erol2020contagion}
S.~Erol, F.~Parise, and A.~Teytelboym, ``Contagion in graphons,'' {\em
  Available at SSRN}, 2020.

\bibitem{aurell2021finite}
A.~Aurell, R.~Carmona, G.~Dayanikli, and M.~Lauriere, ``Finite state graphon
  games with applications to epidemics,'' {\em arXiv preprint
  arXiv:2106.07859}, 2021.

\bibitem{sah2017unraveling}
P.~Sah, S.~T. Leu, P.~C. Cross, P.~J. Hudson, and S.~Bansal, ``Unraveling the
  disease consequences and mechanisms of modular structure in animal social
  networks,'' {\em Proceedings of the National Academy of Sciences}, vol.~114,
  no.~16, pp.~4165--4170, 2017.

\bibitem{saad2020dynamics}
C.~M. Saad-Roy, N.~S. Wingreen, S.~A. Levin, and B.~T. Grenfell, ``Dynamics in
  a simple evolutionary-epidemiological model for the evolution of an initial
  asymptomatic infection stage,'' {\em Proceedings of the National Academy of
  Sciences}, 2020.

\bibitem{silk2021role}
M.~J. Silk and N.~H. Fefferman, ``The role of social structure and dynamics in
  the maintenance of endemic disease,'' {\em Behavioral Ecology and
  Sociobiology}, vol.~75, no.~8, pp.~1--16, 2021.

\bibitem{sah2018disease}
P.~Sah, J.~Mann, and S.~Bansal, ``Disease implications of animal social network
  structure: a synthesis across social systems,'' {\em Journal of Animal
  Ecology}, vol.~87, no.~3, pp.~546--558, 2018.

\bibitem{fu2015risk}
F.~Fu, S.~D. Kocher, and M.~A. Nowak, ``The risk-return trade-off between
  solitary and eusocial reproduction,'' {\em Ecology Letters}, vol.~18, no.~1,
  pp.~74--84, 2015.

\bibitem{gosak2021endogenous}
M.~Gosak, M.~U. Kraemer, H.~H. Nax, M.~Perc, and B.~S. Pradelski, ``Endogenous
  social distancing and its underappreciated impact on the epidemic curve,''
  {\em Scientific reports}, vol.~11, no.~1, pp.~1--10, 2021.

\bibitem{morin2015social}
B.~R. Morin, C.~Perrings, A.~Kinzig, and S.~Levin, ``The social benefits of
  private infectious disease-risk mitigation,'' {\em Theoretical ecology},
  vol.~8, no.~4, pp.~467--479, 2015.

\bibitem{glaubitz2020oscillatory}
A.~Glaubitz and F.~Fu, ``Oscillatory dynamics in the dilemma of social
  distancing,'' {\em Proceedings of the Royal Society A}, vol.~476, no.~2243,
  p.~20200686, 2020.

\bibitem{martcheva2021effects}
M.~Martcheva, N.~Tuncer, and C.~N. Ngonghala, ``Effects of social-distancing on
  infectious disease dynamics: an evolutionary game theory and economic
  perspective,'' {\em Journal of Biological Dynamics}, vol.~15, no.~1,
  pp.~342--366, 2021.

\bibitem{cho2020mean}
S.~Cho, ``Mean-field game analysis of {SIR} model with social distancing,''
  {\em arXiv preprint arXiv:2005.06758}, 2020.

\bibitem{fincher2008pathogen}
C.~L. Fincher, R.~Thornhill, D.~R. Murray, and M.~Schaller, ``Pathogen
  prevalence predicts human cross-cultural variability in
  individualism/collectivism,'' {\em Proceedings of the Royal Society B:
  Biological Sciences}, vol.~275, no.~1640, pp.~1279--1285, 2008.

\bibitem{diekmann1990definition}
O.~Diekmann, J.~A.~P. Heesterbeek, and J.~A. Metz, ``On the definition and the
  computation of the basic reproduction ratio {$\mathcal{R}_0$} in models for
  infectious diseases in heterogeneous populations,'' {\em Journal of
  Mathematical Biology}, vol.~28, no.~4, pp.~365--382, 1990.

\bibitem{vdd2002reproduction}
P.~van~den Driessche and J.~Watmough, ``Reproduction numbers and sub-threshold
  endemic equilibria for compartmental models of disease transmission,'' {\em
  Mathematical Biosciences}, vol.~180, no.~1-2, pp.~29--48, 2002.

\bibitem{yorke1978dynamics}
J.~Yorke, H.~Hethcote, and A.~Nold, ``Dynamics and control of the transmission
  of gonorrhea,'' {\em Sexually Transmitted Diseases}, vol.~5, no.~2, p.~51,
  1978.

\bibitem{may1988transmission}
R.~M. May and R.~M. Anderson, ``The transmission dynamics of human
  immunodeficiency virus ({HIV}),'' {\em Philosophical Transactions of the
  Royal Society B: Biological Sciences}, vol.~321, no.~1207, pp.~565--607,
  1988.

\bibitem{scala2001small}
A.~Scala, L.~N. Amaral, and M.~Barth{\'e}l{\'e}my, ``Small-world networks and
  the conformation space of a short lattice polymer chain,'' {\em EPL
  (Europhysics Letters)}, vol.~55, no.~4, p.~594, 2001.

\bibitem{lloyd2001viruses}
A.~L. Lloyd and R.~M. May, ``How viruses spread among computers and people,''
  {\em Science}, vol.~292, no.~5520, pp.~1316--1317, 2001.

\bibitem{pastor2002immunization}
R.~Pastor-Satorras and A.~Vespignani, ``Immunization of complex networks,''
  {\em Physical Review E}, vol.~65, no.~3, p.~036104, 2002.

\bibitem{pastor2002epidemic}
R.~Pastor-Satorras and A.~Vespignani, ``Epidemic dynamics in finite size
  scale-free networks,'' {\em Physical Review E}, vol.~65, no.~3, p.~035108,
  2002.

\bibitem{koutsoupias1999worst}
E.~Koutsoupias and C.~Papadimitriou, ``Worst-case equilibria,'' in {\em Annual
  Symposium on Theoretical Aspects of Computer Science}, pp.~404--413,
  Springer, 1999.

\bibitem{papadimitriou2001algorithms}
C.~Papadimitriou, ``Algorithms, games, and the internet,'' in {\em Proceedings
  of the Thirty-Third Annual {ACM} Symposium on Theory of Computing},
  pp.~749--753, 2001.

\bibitem{christodoulou2005price}
G.~Christodoulou and E.~Koutsoupias, ``The price of anarchy of finite
  congestion games,'' in {\em Proceedings of the Thirty-Seventh Annual ACM
  Symposium on Theory of Computing}, pp.~67--73, 2005.

\bibitem{carmona2019price}
R.~Carmona, C.~V. Graves, and Z.~Tan, ``Price of anarchy for mean field
  games,'' {\em ESAIM: Proceedings and Surveys}, vol.~65, pp.~349--383, 2019.

\bibitem{hethcote1987epidemiological}
H.~W. Hethcote and J.~W. Van~Ark, ``Epidemiological models for heterogeneous
  populations: proportionate mixing, parameter estimation, and immunization
  programs,'' {\em Mathematical Biosciences}, vol.~84, no.~1, pp.~85--118,
  1987.

\bibitem{grafen1979hawk}
A.~Grafen, ``The hawk-dove game played between relatives,'' {\em Animal
  Behaviour}, vol.~27, pp.~905--907, 1979.

\bibitem{bergstrom2003algebra}
T.~C. Bergstrom, ``The algebra of assortative encounters and the evolution of
  cooperation,'' {\em International Game Theory Review}, vol.~5, no.~03,
  pp.~211--228, 2003.

\bibitem{van2017hamilton}
M.~van Veelen, B.~Allen, M.~Hoffman, B.~Simon, and C.~Veller, ``Hamilton's
  rule,'' {\em Journal of Theoretical Biology}, vol.~414, pp.~176--230, 2017.

\bibitem{cornforth2012synergy}
D.~M. Cornforth, D.~J. Sumpter, S.~P. Brown, and {\AA}.~Br{\"a}nnstr{\"o}m,
  ``Synergy and group size in microbial cooperation,'' {\em The American
  Naturalist}, vol.~180, no.~3, pp.~296--305, 2012.

\bibitem{coder2018effects}
K.~Coder~Gylling and {\AA}.~Br{\"a}nnstr{\"o}m, ``Effects of relatedness on the
  evolution of cooperation in nonlinear public goods games,'' {\em Games},
  vol.~9, no.~4, p.~87, 2018.

\bibitem{iyer2020evolution}
S.~Iyer and T.~Killingback, ``Evolution of cooperation in social dilemmas with
  assortative interactions,'' {\em Games}, vol.~11, no.~4, p.~41, 2020.

\bibitem{vasconcelos2019consensus}
V.~V. Vasconcelos, S.~A. Levin, and F.~L. Pinheiro, ``Consensus and
  polarization in competing complex contagion processes,'' {\em Journal of the
  Royal Society Interface}, vol.~16, no.~155, p.~20190196, 2019.

\end{thebibliography}
\bibliographystyle{ieeetr}

\appendix
\appendixpage
\addappheadtotoc

\section{Properties of Contagion Dynamics and Evolutionary Dynamics in Dimorphic Populations}\label{sec:dimorphicproperties}

In this section, we discuss additional properties of the contagion and evolutionary dynamics in the case of two competing sociality strategies. In Section \ref{appsec:dimorphic-contact-rates}, we derive the contact rates experienced in the two-type contagion dynamics in a population under various possible compositions of the population and the possible sociality strategies of the resident and mutant type. In Section \ref{sec:R0calculation}, we then characterize the basic reproduction number of the two-strategy contagion dynamics, exploring how the pair of sociality strategies can determine whether the long-time contagion dynamics will converge to a contagion-free state or to a unique endemic equilibrium. Finally, in Section \ref{appsec:replicatorlongtime}, we study the behavior of the replicator equation for the evolutionary dynamics for a given pair of sociality strategies. \revision{We provide a proof of Proposition \ref{prop:replicatorlongtime}, showing} that the qualitative behavior of the long-time composition of resident and mutant sociality strategies can be determined by analyzing the ability of each strategy to invade the other when initially rare in the population. 

\subsection{Dimorphic Contact Rates}\label{appsec:dimorphic-contact-rates}

To explore the contagion dynamics in a population with two sociality strategies, we start with a density-dependent description of the population. We assume that the total population has size $N$, of whom $N_m = f N$ follow a mutant strategy making $\sigma_m$ social contacts per unit time and $N_r = (1-f) N$ follow a resident strategy with an analogous social contact rate of $\sigma_m$. Focusing first on a single contagion, we describe for, each contagion $x \in \{ g,b\}$, the disease states in the resident population by the densities $\Irdens^{(x)}$ and $\Srdens^{(x)} = N_r - \Irdens^{(x)}$ and in the mutant population by $\Imdens^{(x)}$ and $\Smdens^{(x)} = N_m - \Imdens^{(x)}$. Under these sociality strategies, infectious resident and mutants individuals collectively have $\sigma_r \Irdens^{(x)}$ and $\sigma_m \Imdens^{(x)}$ social contacts per unit time, while the rate of social contacts for the whole population is $\sigma_m f N + \sigma_r (1-f) N$. Therefore the probability that an individuals meets an infectious individual in a given social interaction can be written as 
\[ \frac{\sigma_m \Imdens^{(x)} + \sigma_r \Irdens^{(x)}}{N\left(\sigma_m f N + \sigma_r (1-f)\right)}. \]
Further assuming that social interactions produce infection with probability $p_x$ and that infectious individuals recover with rate $\gamma_x$, the density of infectious mutant and resident individuals evolve according to the system of ODEs
\begin{subequations} \label{eq:dimorphicdensitycontagion}
\begin{align}
    \dv{\Imdens^{(x)}}{t} & = \sigma_m p_x \Big[\frac{\sigma_m  \Imdens^{(x)} + \sigma_r \Irdens^{(x)}}{N\left(\sigma_m f + \sigma_r (1-f)\right)}\Big] \Smdens^{(x)} - \gamma_x \Imdens^{(x)} \\
    \dv{\Irdens^{(x)}}{t} & = \sigma_r p_x \Big[\frac{\sigma_m \Imdens^{(x)} + \sigma_r  \Irdens^{(x)}}{N\left(\sigma_m f + \sigma_r (1-f)\right)}\Big] \Srdens^{(x)}  -  \gamma_x \Irdens^{(x)}
\end{align}
\end{subequations}
Using the reproduction numbers $\Rmx = \tfrac{\sigma_m p_x}{\gamma}$ and $\Rrg = \tfrac{\sigma_r p}{\gamma_x}$ and denoting the fractions $I^{(x)}_x := \tfrac{\mathbb{I}^{(x)}_y}{N_y}$ and $S^{(x)}_y := \tfrac{\mathbb{S}^{(x)}_y}{N_y}$ of infectious and susceptible individuals in the populations $y \in \{r,m\}$, we can further obtain the following frequency-dependent analogue of Equation \eqref{eq:dimorphicdensitycontagion}
\begin{subequations} \label{eq:dimorphicfrequencycontagion}
\begin{align}
   \label{eq:mdimorphicfrequency} \frac{1}{\gamma_x} \dv{I^{(x)}_m}{t} & = \Rmg \Big[\frac{\Rmg f I^{(x)}_m  + \Rrg (1-f) I_r^{(x)}}{\Rmx f + \Rrx (1-f)}\Big] S_m^{(x)} - I_m^{(x)} \\
   \label{eq:rdimorphicfrequency} \frac{1}{\gamma_x} \dv{I_r^{(x)}}{t} & = \Rrx \Big[\frac{\Rmx f I_m^{(x)} + \Rrx  (1-r)I_r^{(x)}}{\Rmx f + \Rrx (1-f)}\Big] S_r^{(x)}  - I_r^{(x)}.
\end{align}
\end{subequations}
Finally, rescaling time and using the fact that $S^{(x)}_y = 1 - I^{(x)}_y$ allows us to obtain the frequency-depending dimorphic contagion dynamics of \revision{Equations \eqref{eq:rgooddimorphic} and \eqref{eq:mgooddimorphic}} for the good contagion and an analogous system for the bad contagion.

\subsection{Basic Reproduction Number for Dimorphic Contagion Dynamics} \label{sec:R0calculation}
To gain insight into the contagion process with two levels of sociality and to assess the stability of the disease-free equilibrium, we use the next-generation matrix method of Diekmann and colleagues \cite{diekmann1990definition} and van den Driessche and Watmough \cite{vdd2002reproduction} to find the overall basic reproduction number of the contagion process. \revision{For this analysis, we multiply both sides of Equations \eqref{eq:mdimorphicfrequency} and \eqref{eq:rdimorphicfrequency} by $\gamma_x$, which allows to rewrite the coupled contagion dynamics for the resident and mutant populations as}

\begin{equation}
\begin{aligned}
    \revision{\dv{I_m}{t}} &= \mathcal{F}_m(I_m, I_r) - \mathcal{V}_m(I_m, I_r)
      \\
    \vspace{2mm} \revision{\dv{I_r}{t}} &= \mc{F}_r(I_m, I_r) - \mc{V}_r(I_m, I_r),
\end{aligned}
\label{eqn:next-gen-mat-diff-eq}
\end{equation}

\revision{where}
\begin{equation*}
\begin{aligned}
\mc{F}_m(I_m^{(x)}, I_r^{(x)}) &= \revision{\gamma_x} \left\{ \Rmx \Big[\frac{\Rrx (1-f) I_r^{(x)} + \Rmx f I_m^{(x)}}{\Rrx (1-f) + \Rmx f}\Big] (1 - I_m^{(x)}) \right\} \\ \\
\mc{F}_r(I_m^{(x)}, I_r^{(x)}) &= \revision{\gamma_x} \left\{ \Rrx \Big[\frac{\Rrx (1-f) I_r^{(x)} + \Rmx f I_m^{(x)}}{\Rrx (1-f) + \Rmx f}\Big] (1 - I_r^{(x)}) \right\} \\ \\
\mc{V}_m(I_m^{(x)}, I_r^{(x)}) &= \revision{\gamma_x} I_m^{(x)} \\ \\
\mc{V}_r(I_m^{(x)}, I_r^{(x)}) &= \revision{\gamma_x} I_r^{(x)}
\end{aligned}
\end{equation*}

and $\mc{F}_m$ and $\mc{F}_r$ represent rates of new infections, \revision{while} $V_m$ and $V_r$ represent rates of recoveries.

The next generation matrix is the matrix $FV^{-1}$ where the \revision{matrices $F$ and $V$ are given by}
\begin{equation} \label{eq:Fmatrix}
F = \left. \begin{pmatrix}
\dsdel{\mc{F}_m}{I_m^{(x)}} & \dsdel{\mc{F}_m}{I_r^{(x)}} \\ \\
\dsdel{\mc{F}_r}{I_m^{(x)}} & \dsdel{\mc{F}_r}{I_r^{(x)}}
\end{pmatrix}\right|_{(I_m^{(x)}, I_r^{(x)}) = (0, 0)} = \begin{pmatrix}
\dfrac{ \revision{\gamma_x} f\left(\Rmx\right)^2}{\Rhat^{(x)}} & \dfrac{\revision{\gamma_x}(1 - f)\Rmx \Rrx}{\Rhat^{(x)}} \\ \\
\dfrac{\revision{\gamma_x}f\Rmx \Rrx}{\Rhat^{(x)}} & \dfrac{\revision{\gamma_x}(1 - f)\left(\Rrx\right)^2}{\Rhat^{(x)}}
\end{pmatrix}
\end{equation}

\begin{equation}
V = \begin{pmatrix}
\dsdel{V_m}{I_m^{(x)}} & \dsdel{V_m}{I_r^{(x)}} \\ \\
\dsdel{V_r}{I_m^{(x)}} & \dsdel{V_r}{I_r^{(x)}}
\end{pmatrix}_{{(I_m^{(x)}, I_r^{(x)}) = (0, 0)}} = \begin{pmatrix}
\revision{\gamma_x} & 0 \\ \\
0 & \revision{\gamma_x} 
\end{pmatrix},
\end{equation}
\revision{with $\Rhat^{(x)} := \Rmx (1-f) + \Rmx f$. We may then compute the next generation matrix, which takes the following form:}
\begin{equation} \label{eq:next-generation}
    F V^{-1} = \begin{pmatrix}
\dfrac{ \revision{\gamma_x} f\left(\Rmx\right)^2}{\Rhat^{(x)}} & \dfrac{\revision{\gamma_x}(1 - f)\Rmx \Rrx}{\Rhat^{(x)}} \\ \\
\dfrac{\revision{\gamma_x}f\Rmx \Rrx}{\Rhat^{(x)}} & \dfrac{\revision{\gamma_x}(1 - f)\left(\Rrx\right)^2}{\Rhat^{(x)}}
\end{pmatrix} 
\begin{pmatrix} \ds\frac{1}{\gamma_x} & 0 \\ 0 & \ds\frac{1}{\gamma_x} \end{pmatrix} = \begin{pmatrix}
\dfrac{ f\left(\Rmx\right)^2}{\Rhat^{(x)}} & \dfrac{(1 - f)\Rmx \Rrx}{\Rhat^{(x)}} \\ \\
\dfrac{f\Rmx \Rrx}{\Rhat^{(x)}} & \dfrac{(1 - f)\left(\Rrx\right)^2}{\Rhat^{(x)}}
\end{pmatrix}
\end{equation}

 \revision{We note that, in this case, our next generation matrix given by Equation \ref{eq:next-generation} is a rank one matrix, as the two columns are scalar multiples of each other. The ratio between the first column and the second column is $\frac{f \Rmx{}}{(1-f) \Rrx}$, which is a consequence of our assumptions that the difference between the resident and mutant strategies is their relative rate of social contacts $\sigma_m$ and $\sigma_r$ and that each social interaction follows an unbiased sampling of interaction partners from the pool of available contacts (with probabilities of interaction between sociality strategies determined by the relative contact rates and relative abundances of the resident and mutant strategies).}

The quantity $\Rnet^{(x)}$ given by the spectral radius of the next generation matrix is the overall basic reproduction number of the contagion process. When $\Rnet^{(x)} > 1$, the contagion-free equilibrium is unstable, and the results of 
\revision{Hethcote and Yorke} \cite{yorke1978dynamics} then imply that there exists a unique stable endemic equilibrium. When $\Rnet^{(x)} < 1$, the contagion-free equilibrium is stable, and \revision{as shown by Hethcote and Yorke} \cite{yorke1978dynamics}, this implies that no endemic equilibrium exists.

The eigenvalues of $F$ are $0$ and $\frac{f\left(\Rmx\right)^2 + (1-f)\left(\Rrx\right)^2}{f \Rmx + (1-f) \Rrx}$. Since the second eigenvalue is always non-negative for $0 \le f \le 1$ and $\Rmx, \Rrx \ge 0$ (with at least one greater than 0), it is the spectral radius, so

\begin{equation} 
    \Rnet = \frac{f\left(\Rmx\right)^2 + (1-f)\left(\Rrx\right)^2}{f \Rmx + (1-f)\Rrx}.
    \label{eqn:rnet}
\end{equation}

\Rnet is the ratio of the second moment to the first moment for the subpopulation reproduction numbers: $$\Rnet^{(x)}= \frac{E(\R{}^2)}{E(\R{})}$$
This formula is analogous to the basic reproduction number $\R{\mathrm{RG}}$ studied for SIS models with interactions taking place on random graphs with heterogeneous distributions over individual degree $k$ \cite{may1988transmission,scala2001small,lloyd2001viruses,pastor2002immunization,pastor2002epidemic}, given by 
\begin{equation} \label{eq:networkSISR_0}
\R{\mathrm{RG}} = \frac{\beta}{\gamma} \frac{\langle k \rangle}{\langle k^2 \rangle} .
\end{equation}
From this correspondence, we can see that our results for the two-group SIS model with different rates of interaction for the resident and mutant groups can be reinterpreted as a model for interactions occurring on a stochastic block model with two groups of individuals having different linking probabilities.

\begin{figure}[tph!]
    \centering
    \includegraphics[width=\textwidth]{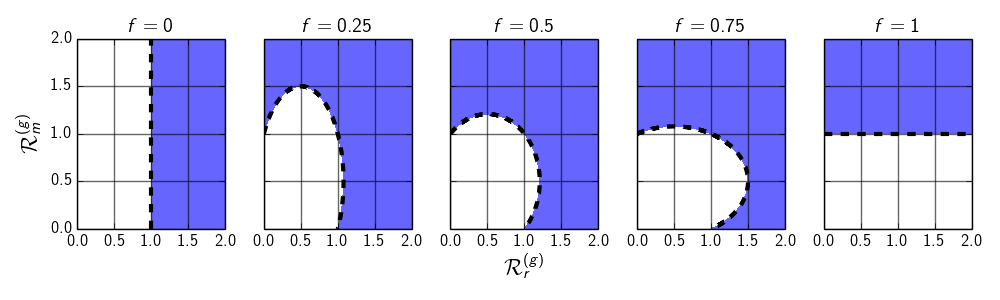}
    \caption{Regions for which the net reproduction number $\Rnet^{(x)} > 1$ (blue) for various resident and mutant reproduction numbers $\Rrx$ and $\Rmx$, for various fractions of mutant population $f$. }
    \label{fig:stability-region}
\end{figure}

When $0 < f <1$, the condition $\Rnet^{(x)} < 1$ for stability of the contagion-free equilibrium can be rewritten as %
\begin{equation}
4f \left(\Rmx - \frac{1}{2}\right)^2 + 4(1-f)\left(\Rrx - \frac{1}{2}\right)^2 < 1 
    \label{eqn:rnet-ellipse}.
\end{equation}
This stability boundary characterizes 
an ellipse in the $\Rrx, \Rmx$ plane centered at $(\Rrx, \Rmx) = (\frac{1}{2}, \frac{1}{2})$ with an $\Rmx$-axis length of $\frac{1}{\sqrt{f}}$ and an $\Rrx$-axis length of $\frac{1}{\sqrt{1-f}}$, which we illustrate in Figure \ref{fig:stability-region}. 

\begin{remark} We can use Equation \eqref{eqn:rnet} to deduce the following properties of $\Rnet$. 
\begin{itemize}
    \item In the limiting cases of $f = 0$ and $f=1$, the net reproduction number reduces to $\Rnet^{(x)} = \Rrx$ and $\Rnet^{(x)} = \Rrx$, respectively. 
    \item If $\Rmx, \Rrx > 1$, then $\Rnet^{(x)} > 1$ and contagion-free equilibrium is unstable.
    \item If $0 \le \Rmx, \Rrx < 1$, then $\Rnet^{(x)} \leq 1$ and the contagion-free equilibrium is stable.
    \item If $\Rrx = 1$, then the signs of $\Rnet^{(x)} - 1$ and $\Rmx - 1$ agree. In other words, if the fully-resident population has a marginally stable contagion-free equilibrium, than the stability of the contagion-free equilibrium under the dimorphic dynamics is determined by the sign of $\Rmx - 1$. 
    \item If $\Rrx < 1$, then $\Rnet^{(x)} > 1$ provided that
    \[\Rmx > \frac{1}{2} + \ds\sqrt{\frac{1}{4} + \left(\frac{1-f}{f}\right) \Rrx \left(1 - \Rrx\right)}. \]
    This threshold is maximized when $\Rrx = \frac{1}{2}$.
\end{itemize}
\end{remark}

\subsection{Analysis of Replicator Equation for Cobb-Douglas Utility}\label{appsec:replicatorlongtime}

Here we consider the long-time behavior of the replicator equation for pairwise competition for sociality strategies from the Section \ref{sec:dimorphicevolutionary} Section in the case of Cobb-Douglas utility. \revision{The main output of this section is the proof of Proposition \ref{prop:replicatorlongtime}, which characterizes the possible long-time outcomes for the replicator equation in the case of Cobb-Douglas utility.}

\revision{To understand the stability of equilibria of the replicator equation with Cobb-Douglas utility}, it is convenient to study a following modified form of Equation \eqref{eqn:replicator} based upon log-transformed utilities. In this case, given resident and mutant sociality strategies featuring good contagion reproductive numbers $\Rmg$ and $\Rmg$, the fraction $f$ of individuals with the mutant strategy evolves according to

\begin{equation} \label{eq:replicatorlog} 
    \dsddt{f} = f(1-f) \left[ \log\left( U_m(f)\right) - \log\left(U_r(f) \right) \right],
\end{equation}
where 
\begin{equation} \label{eq:logutil-replicator}
    \begin{aligned}
        \log\left(U_m(f) \right) &= \alpha \log\left[ U\left( \hat{I}^{(g)}_m(\Rmg,\Rrg,f), \hat{S}^b_m(\Rmg,\Rrg,f)\right) \right] \\
         \log\left(U_r(f) \right) &= \alpha \log\left[ U\left( \hat{I}^{(g)}_r(\Rmg,\Rrg,f), \hat{S}^b_r(\Rmg,\Rrg,f) \right)\right] 
    \end{aligned}
\end{equation}
and the quantities $\hat{I}^x_y(\cdot)$ and $\hat{S}^x_y(\cdot)$ are equilibrium values of the dimorphic contagion dynamics as describe by \revision{Equations \eqref{eq:rgooddimorphic} and \eqref{eq:mgooddimorphic}}. 

Solutions to Equation \eqref{eq:replicatorlog} will have the same long-time behavior as solutions to Equation \eqref{eqn:replicator}. Therefore we can characterize the long-time behavior of Equation \eqref{eqn:replicator} using a useful monotonicity property for the relative log-utility as a function of the fraction of mutants $f$. \revision{Next, we recall the statement of Proposition \ref{prop:replicatorlongtime}, in which we show that that the only possible long-time behaviors for the replicator equation under Cobb-Douglas utility are dominance of the mutant strategy, dominance of the resident strategy, and coexistence of the two strategies at a unique interior equilibrium. Furthermore, the behavior for a given pair of strategies can be determined by evaluating the utilities of the resident and mutant strategies at the endpoints $f = 0$ and $f = 1$ in which the population has an all-resident or all-mutant composition, so dominance under pairwise invasibility analysis correspond to dominance under competition at relative frequencies of the mutant and resident at all $f \in [0,1]$.}
\vspace{2.5mm}

\renewcommand{\theproposition}{\arabic{proposition}}
\setcounter{proposition}{0}

\begin{proposition} \label{prop:replicatorlongtimeappendix}
Suppose that the resident and mutant types have sociality strategies featuring reproduction numbers $\Rrg \geq 1$ and $\Rmg \geq 1$ for the good contagion, with $\Rrg\ne \Rmg$ and at least one of these reproduction numbers strictly greater than $1$. Then, for any $c > 0$ and for any resident and mutant types with reproduction numbers $\Rrb = c \Rrg$ and $\Rmb = c \Rmg$ for the bad contagion, the difference of Cobb-Douglas log-utilities $\log\left[U_m(f) \right] - \log\left[U_r(f) \right]$ is a decreasing function of $f$. As a consequence, the long-time behavior can be determined by the relative values of $U_m(f)$ and $U_r(f)$ at the endpoints $f = 0$ and $f = 1$. The three possible cases are the following:
\begin{itemize}
    \item{$U_m(0) > U_r(0)$ and $U_m(1) > U_r(1)$:} $f = 1$ is globally stable and the mutant will fix in the population. 
    
   \item{$U_m(0) < U_r(0)$ and $U_m(1) < U_r(1)$:} $f = 0$ is globally stable and the resident will fix in the population. 
    
  \item{$U_m(0) > U_r(0)$ and $U_m(1) < U_r(1)$:} There exists a unique interior equilibrium $\hat{f} \in (0,1)$ that is globally stable, and mutant and resident will coexist in the long-time population. 
\end{itemize}
\end{proposition}
\vspace{2.5mm}
We note that the fact that the difference in log-utilities is decreasing rules out the possibility that $U_m(0) > U_r(0)$ and $U_m(1) < U_r(0)$, and therefore it is impossible for the replicator dynamics to achieve bistability of a full-resident and full-mutant population under Cobb-Douglas utility. 

\begin{proof}

It is useful to write the dimorphic SIS dynamics of \revision{Equations \eqref{eq:rgooddimorphic} and \eqref{eq:mgooddimorphic}} in terms of the transmission function $\Lambda_x(I_m^{(x)}, I_r^{(x)})$:

\begin{equation} \label{eq:LambdaI}
    \Lambda_x(I_m^{(x)}, I_r^{(x)}) = 
    \frac{\Rmx f I_m^{(x)} + \Rrx (1-f) I_r^{(x)}}{\Rmx f + \Rrx (1-f)} \geq 0.
\end{equation}

We can use this transmission function to find the implicit expressions for nonzero contagion equilibria 

\begin{equation} 
\begin{aligned}
    \ibar{m}^{(x)} &= \frac{\Rmx\Lambda_x(\ibar{m}^{(x)},\ibar{r}^{(x)})}{1 + \Rmx\Lambda_x(\ibar{m}^{(x)},\ibar{r}^{(x)})} &&= 1 - \frac{1}{1 + \Rmx\Lambda_x(\ibar{m}^{(x)},\ibar{r}^{(x)})} \\
    \ibar{r}^{(x)} &= \frac{\Rrx\Lambda_x(\ibar{m}^{(x)},\ibar{r}^{(x)})}{1 + \Rrx\Lambda_x(\ibar{m}^{(x)},\ibar{r}^{(x)})} &&= 1 - \frac{1}{1 + \Rrx\Lambda_x(\ibar{m}^{(x)},\ibar{r}^{(x)})}
\end{aligned}.
\label{eqn:ibars-dimorphic}
\end{equation}

From our assumptions on the reproduction numbers $\Rmg$ and $\Rrg$, we know that the good contagion will be present at equilibrium for both resident and mutant: $\ibar{m}^{(g)} >0$ and $\ibar{r}^{(g)} >0$. For the bad contagion, either the equilibrium fraction of susceptible individuals is $1$ for both groups (when $\Rnet^{(g)} < \frac{1}{c}$)) or it can expressed in terms of Equation \eqref{eqn:ibars-dimorphic} via $(\sbar{r}^{(b)}, \sbar{m}^{(b)}) = (1 - \ibar{r}^{(b)}, 1 - \ibar{m}^{(b)})$ (when $\Rnet^{(g)} \geq \frac{1}{c}$. Using these properties of the equilibria for both contagions, we see that the \revision{difference between Cobb-Douglas log-utilities} for the mutant and resident populations has the following piecewise characterization
\begin{equation} \label{eq:logdiffutil}
  \log(U_m(f)) - \log(U_r(f)) =   \left\{
    \begin{array}{cr}
         \alpha \log\left( \ds\frac{\ibar{m}^{(g)}}{\ibar{r}^{(g)}} \right) &: \Rnet^{(g)} < \ds\frac{1}{c}  \\
          \alpha \log\left( \ds\frac{\ibar{m}^{(g)}}{\ibar{r}^{(g)}} \right) + \left(1 - \alpha \right) \log\left( \ds\frac{\sbar{m}^{(b)}}{\sbar{r}^{(b)}} \right) &: \Rnet^{(g)} \geq \ds\frac{1}{c}
    \end{array}
    \right.
\end{equation}

Now we look to study how the difference in log-utilities changes with mutant fraction $f$.  Using Equation \eqref{eq:LambdaI}, we compute that, for each contagion $x \in \{g,b\}$, 
\begin{equation}
    \pdv{\Lambda_x}{f} = \frac{\Rmx \Rrx (\ibar{m}^{(x)} - \ibar{r}^{(x)})}{\left(f \Rmx + (1-f) \Rrx\right)^2}
    \label{eqn:ddf-lambda}
\end{equation}

Using Equation \eqref{eqn:ibars-dimorphic}, the ratio of endemic equilibrium levels for the good contagion is given by 
\begin{equation} \label{eq:zetaratio} \zeta_g := \log \left(\frac{\ibar{m}^{(g)}}{\ibar{r}^{(g)}}\right) = \log\left[\left(\frac{\Rmg}{\Rrg}\right) \frac{1 + \Rrg\Lambda_g}{1 + \Rmg\Lambda_g}\right]. \end{equation} 
We can then differentiate to see that 
\begin{equation} \label{eq:zetaderiv} \dsdel{\zeta_g}{f} = \frac{\Rrg}{1 + \Rrg \Lambda_g} \pdv{\Lambda}{f} - \frac{\Rmg}{1 + \Rmg\Lambda_g} \pdv{\Lambda}{f} = \pdv{\Lambda_g}{f}\left(\frac{\Rrg}{1 + \Rrg \Lambda_g} - \frac{\Rmg}{1 + \Rmg \Lambda_g}\right) = \pdv{\Lambda_g}{f} \frac{1}{\Lambda_g} (\ibar{r}^{(g)} - \ibar{m}^{(g)}) < 0, \end{equation}
where we deduce the direction of the inequality by using Equation \eqref{eqn:ddf-lambda} to note that $\pdv{\Lambda_x}{f}$ and $\ibar{r}^{(x)}-\ibar{m}^{(x)}$ always have opposite signs (for both the good and bad contagion) and note that the inequality is strict because of the assumption that $\Rmg \ne \Rrg$.

Similarly, we can use Equation \eqref{eqn:ibars-dimorphic} to compute the log-ratio for the susceptible fractions at the nontrivial endemic equilibrium for the bad contagion as 
\begin{equation} \label{eq:etaration} \nu_b := \log\left(\frac{\sbar{m}^{(b)}}{\sbar{r}^{(b)}}\right)  = \log\left(\frac{1 - \ibar{m}^{(b)}}{1 - \ibar{r}^{(b)}}\right) = \log\left(\frac{1 + c \Rrg \Lambda_b}{1 + c \Rmg \Lambda_b} \right). \end{equation}%
Then we can differentiate to see that
\begin{equation} \label{eq:etaderiv} \pdv{\nu_b}{f} %
= \pdv{\Lambda_b}{f}\left(\frac{c\Rrg}{1 + c\Rrg \Lambda_b} - \frac{c\Rmg}{1 + c\Rmg \Lambda_b}\right) = \pdv{\Lambda_b}{f} \frac{1}{\Lambda_b} (\ibar{r}^{(b)} - \ibar{m}^{(b)}) 
< 0. \end{equation}
Differentiating Equation \eqref{eq:logdiffutil} with respect to $f$, we see from Equations \eqref{eq:zetaderiv} and \eqref{eq:etaderiv} that 
\begin{equation} \label{eq:logutilderiv}
\dsdel{}{f} \left[\log(U_m(f)) - \log(U_f(f)) \right] =   \left\{
    \begin{array}{cr}
        \alpha \dsdel{\zeta_g}{f}  &: \Rnet^g < \ds\frac{1}{c}  \\
          \alpha \dsdel{\zeta_g}{f} + \left(1 - \alpha \right) \dsdel{\nu_b}{f} &: \Rnet^g \geq \ds\frac{1}{c}
    \end{array}
    \right\} < 0.
\end{equation}
Therefore we see that the difference in log-utilities is a differentiable, monotonically decreasing function of the mutant fraction $f$, implying that the there will be a unique stable equilibrium of the replicator dynamics of Equation \eqref{eqn:replicator} and \eqref{eq:replicatorlog} and that the trifold alternative described above holds for the longtime behavior of these replicator equations.
\end{proof}

\section{Additional Analysis of the Adaptive Dynamics Limit}
\label{sec:additionaladaptive}

In this section, we further \revision{explore} our adaptive dynamics analysis of the social dilemma of sociality. \revision{In Section \ref{sec:CDoptimum}, we provide the derivation of the formula for the socially optimum level of social interaction for the case of Cobb-Douglas utility.} In Section \ref{sec:PoA}, we the Price of Anarchy to quantify the gap between social utility for populations following the socially-optimal and evolutionarily-stable sociality strategies. In Section \ref{appsec:convergencestability}, we further characterize the evolutionary and convergence stability of sociality strategies under adaptive dynamics, showing that the sociality strategy corresponding to $\RESS$ is the unique evolutionarily-stable and convergence stable strategy for the Cobb-Douglas utility. In Section \ref{sec:linearutility}), we analyze the socially-optimal and evolutionarily-stable sociality strategies in the case of a linear utility function, showing that it possible to achieve infinite sociality levels and bistable evolutionary dynamics in this case. \revision{In Section \ref{sec:CESutility}, we perform a similar analysis for the Constant Elasticity of Substitution (CES) family of utility functions, show, for the case in which the bad contagion spreads more readily than the good contagion ($c > 1$), that there is broad range of utility parameters for which the evolutionary dynamics achieve bistability between an ESS featuring spread of both contagions and an ESS at $\frac{1}{c}$ in which both contagions are eliminated.} Finally, in Section \ref{sec:assortment}, we consider the role of assortative interactions in which individuals preferentially interact with individuals with the same sociality strategy. This assortative mechanism helps to internalize negative externalities generated by subpoptimal rates of social interaction, and we show that assortment helps to mitigate the social dilemma and produces evolutionarily-stable levels of social interaction that are closer to the social optimum.

\subsection{Derivation of Socially Optimal Level of Sociality for Cobb-Douglas Utility} \label{sec:CDoptimum}

\revision{For completeness, we provide in this section the derivation of the socially optimum interaction rate $\Ropt$ for the case of Cobb-Douglas utility. The main effort involved in this derivation is checking the conditions under which the utility for the population is maximized by a level of social interaction at which the bad contagion is unable to spread $\frac{1}{c}$ or at a level of social interaction at which both the good and bad contagion spread in the population.}

\revision{Since $U(\Rg{}) = 0$ for $\Rg{} \le 1$, it suffices to optimize $U$ on the interval $(1, \infty)$. We can equivalently maximize $\logur$; differentiating that log-utility yields
\begin{equation} \label{eq:derivCDmonopiecewise}
\dsdel{}{\Rg{}} \logur = \left\{
     \begin{array}{cr}
      \dfrac{\alpha }{\Rg{} \left(\Rg{} - 1\right)} & : 1 \leq \Rg{} < \dfrac{1}{c} \\[2em]
     \dfrac{1}{\Rg{}} \left[ \dfrac{\alpha}{\Rg{} - 1} - \left(1 - \alpha\right) \right] & :   \Rg{} >  \dfrac{1}{c}
     \end{array}
   \right.   .
\end{equation}
The log-utility is not differentiable at $\Rg{} = \frac{1}{c}$, but it has left and right derivatives at $\frac{1}{c}$ given by the expressions from the cases $\Rg{} < \frac{1}{c}$ and $\Rg{} > \frac{1}{c}$, respectively.}

\revision{From Equation \eqref{eq:derivCDmonopiecewise}, we see that the log-utility is increasing for $\Rg{} < \frac{1}{c}$, so its maximizer must reside in $[\frac{1}{c}, \infty)$. We also see from \eqref{eq:derivCDmonopiecewise} that the log-utility has a local maximum  at $(\Rg{})^* = \frac{1}{1-\alpha}$ provided that $(\Rg{})^* > \frac{1}{c}$, a condition that is satisfied when $c > 1 - \alpha$. Furthermore, in this case, log-utility is increasing for $\Rg{} \in [\frac{1}{c},(\Rg{})^*)$ and decreasing for $\Rg{} > (\Rg{})^*$, so $(\Rg{})^*$ maximizes the log-utility when $c > 1 - \alpha$.} 

\revision{For the alternative case in which  $ c \leq 1 - \alpha$, we can see from Equation \eqref{eq:derivCDmonopiecewise} that, for $c < 1$ and $\Rg{} > \frac{1}{c}$,
\begin{dmath}
    {\dsdel{\logur}{\Rg{}} \leq c \left[ \frac{\alpha}{c^{-1} - 1} - (1-\alpha) \right] = \left(\frac{c}{1-c}\right) \left[ \alpha c - (1-c)(1 - \alpha)\right]} \\ {= \left( \frac{c}{1-c} \right)\left[ \alpha + c - 1\right] < 0},
\end{dmath}
and therefore we can combine this with the fact that log-utility is increasing for $\Rg{} < \frac{1}{c}$ to deduce that the log-utility is maximized at $\Rg{} = \frac{1}{c}$ when $c \leq 1 - \alpha$. To summarize, under the Cobb-Douglas utility, the socially-optimal level of sociality $\Ropt$ is given by
\begin{equation} \label{eq:RoptCDappendix}
    \Ropt =  \max\left( \frac{1}{c}, \; \frac{1}{1 - \alpha} \right).%
\end{equation}
}

\subsection{Quantifying the Social Dilemma via the Price of Anarchy}
\label{sec:PoA}

\begin{figure}[tph!]
    \centering
    \includegraphics[width = 0.48\textwidth,height = 0.48\textwidth]{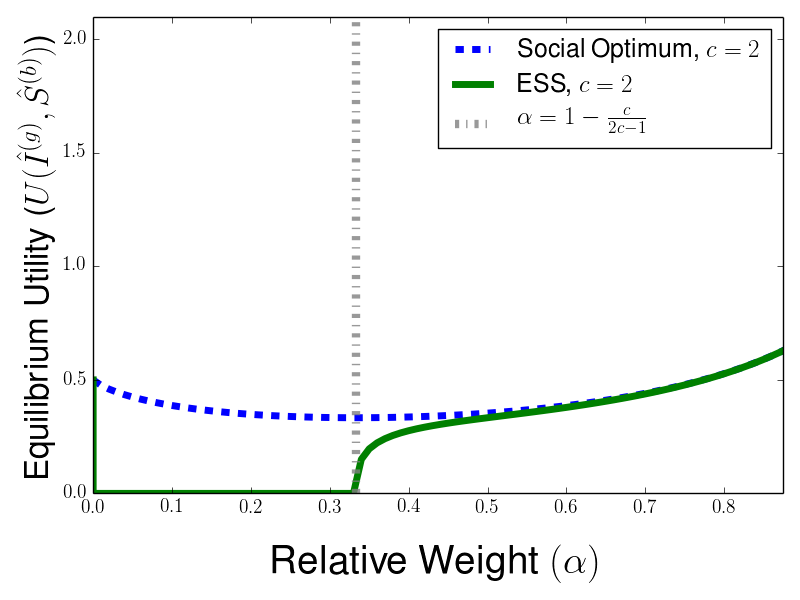}
    \includegraphics[width = 0.48\textwidth,height = 0.48\textwidth]{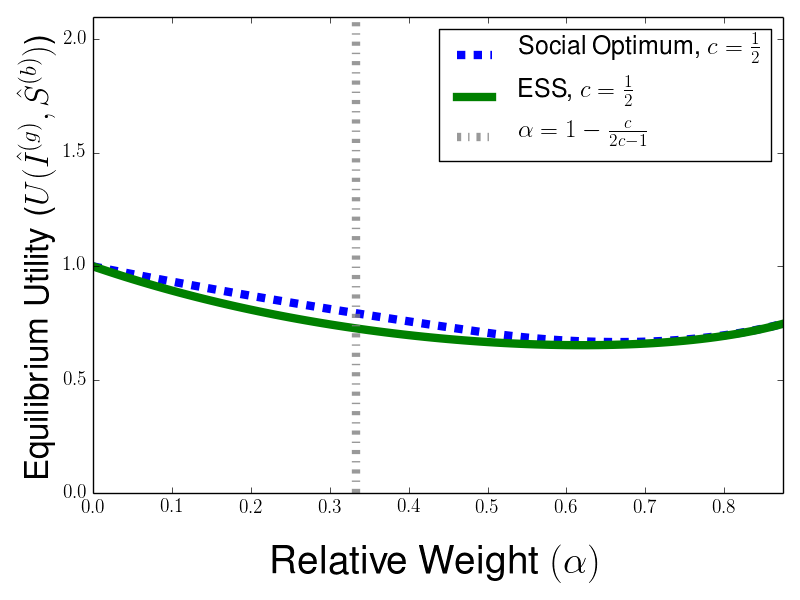}
    \caption{Utility for individuals at evolutionarily-stable and socially-optimal levels of sociality. We plot these utilities as a function of the relative importance of the good contagion $\alpha$ and for the relative infectiousness values $c = 2$ (left) and $c =\frac{1}{2}$ (right).}
    \label{fig:dilemma_utility}
\end{figure}

In addition to our analysis considered in Section \ref{sec:adaptivedynamics}, we can compare the evolutionarily-stable and socially-optimal outcomes based upon the utility levels achieved by monomorphic populations featuring each strategy. In Figure \ref{fig:dilemma_utility}, we plot the Cobb-Douglas utilities obtained by populations following strategies $\RESS$ and $\Ropt$ given in Table \ref{tab:CDtable} for the same values of $\alpha$ and $c$ as in Figure \ref{fig:combined}c. Notably, the utility of the ESS strategy when $c = 2$ is increasing in $\alpha$, while the utility of the social optimum for $c=2$ and both strategies for $c = \frac{1}{2}$ achieve a minimal utility for intermediate values of $\alpha$. This means that, for the evolutionary dynamics with $c = 2$, caring more about exposure to good contagion produces a better outcome long-run outcome even though the bad contagion spreads more readily than the good contagion.

\begin{figure}[ht!]
    \centering
    \includegraphics[width = 0.8\textwidth]{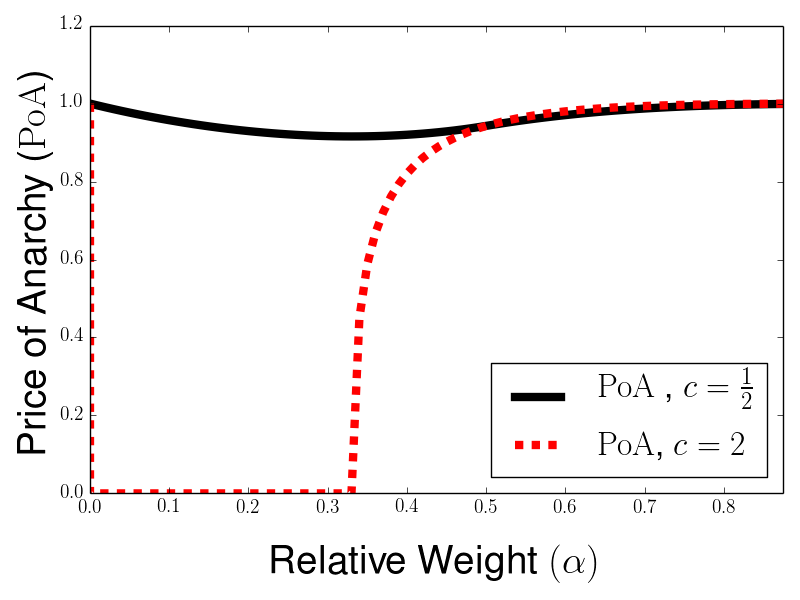}
    \caption{The Price-of-Anarchy (PoA) characterizing the efficiency of the evolutionarily-stable sociality $\RESS$ relative to the socially-optimal sociality $\Ropt$. The PoA is plotted as a function of $\alpha$, and we consider the cases $c = \frac{1}{2}$ (black solid line) and $c = 2$ (red dashed line).}
    \label{fig:dilemma_PoA}
\end{figure}

Another quantity often used in game theory to compare the efficiency of Nash equilibria and social optima is the Price of Anarchy (PoA), \revision{introduced by Papadimitriou \cite{koutsoupias1999worst,papadimitriou2001algorithms}}, which measures the ratio of the utilities between such outcomes \cite{koutsoupias1999worst,papadimitriou2001algorithms,christodoulou2005price,carmona2019price}. In our context, we can compare the evolutionarily-stable and socially-outcomes by defining the PoA as 
\begin{equation}
    \mathrm{PoA} := \frac{U[\RESS,\RESS]}{U[\Ropt,\Ropt]}.
\end{equation}
Because the Cobb-Douglas utility is non-negative and $\Ropt$ maximizes $U[\Rmg,\Rrg]$, the PoA takes on values between 0 and 1. In Figure \ref{fig:dilemma_PoA}, we plot the PoA as a function of $\alpha$ for the cases $c = \frac{1}{2}$ and $c = 2$. For $c = 2$, the PoA is $0$ for $\alpha < \frac{1}{3}$, when the evolutionarily-stable population is achieving the worst possible payoff. The PoA turns out to be a non-decreasing function of $\alpha$ when $c = 2$, while the inefficiency of the evolutionarily-stable outcome is maximized at an intermediate value of $\alpha$ when $c = \frac{1}{2}$.

\subsection{Evolutionary and Convergence Stability for Cobb-Douglas Utility} \label{appsec:convergencestability}

Now we will further examine the evolutionary and convergence stability of the sociality strategy 
$\RESS = \max\left(1, \tfrac{1}{c} + \tfrac{\alpha}{1-\alpha} \right)$ in the case of Cobb-Douglas utility using the classification criteria for evolutionarily singular strategies \cite{geritz1998evolutionarily,diekmann2002beginners,brannstrom2013hitchhiker}. For a strategy $(\Rrg)^*$ to be evolutionarily-stable, the strategy must be a local maximizer of the relative utility function $s_{(\Rrg)^*}(\Rmg)$ \cite{brannstrom2013hitchhiker}. This always holds in the boundary case when $(\Rrg)^* = 1$ because $s_{(\Rrg)^*}(\Rmg)$ is decreasing for $\Rmg$ near $1$. For the interior case \revision{in which $(\Rrg)^* = \frac{1}{c} + \frac{\alpha}{1-\alpha}$}, we will apply the second derivative test. We see that
\begin{equation}
\begin{aligned} 
& \frac{\partial^2 s_{\Rrg}(\Rmg)}{\partial \left(\Rmg\right)^2} \bigg|_{\Rmg = \Rrg} \\ &=  \left[ \dsdel{}{\Rmg} \left( \dsdel{\log\left[U \left(\Rmg,\Rrg \right) \right]}{\Rmg}\right) \right] \bigg|_{\Rmg = \Rrg} \\ 
&=  \left[ \dsdel{}{\Rmg} \left( \frac{\alpha \Rrg }{\Rmg \left(\Rmg \Rrg + \Rrg - \Rmg \right)}  - \frac{\left(1 - \alpha \right) \left( c \Rrg - 1 \right)}{c \Rmg \Rrg + \Rrg - \Rmg} \right) \right] \bigg|_{\Rmg = \Rrg}  \\ 
&= \left[ -\frac{\alpha \Rrg \left( 2 \Rmg \Rrg + \Rrg - 2 \Rmg \right)}{\left(\Rrg (\Rmg)^2 + \Rmg \Rrg - (\Rmg)^2 \right)^2} + \frac{\left(1-\alpha\right) \left( c \Rrg - 1 \right)^2}{\left(c \Rrg \Rmg + \Rrg - \Rmg \right)^2} \right] \bigg|_{\Rmg = \Rrg}  \\ &= \frac{1}{(\Rmg)^4} \left[ - \alpha \left(2 \Rmg - 1 \right) + \left(\frac{1-\alpha}{c^2}\right) \left(c \Rmg - 1\right)^2 \right]
\label{eq:cdm2deriv}
\end{aligned}
\end{equation}
Then, evaluating the derivative at the interior singular strategy $\left(\Rmg\right)^* = \frac{\alpha}{1-\alpha} + \frac{1}{c}$, we get that 
\begin{equation} 
\label{eq:cdm2deriv-eval-at-eq}
\frac{\partial^2 s_{\Rrg}(\Rmg)}{\partial \left(\Rmg\right)^2} \bigg|_{\Rmg = \Rrg = \frac{\alpha}{1-\alpha} + \frac{1}{c}} = -\frac{\alpha^2}{1 - \alpha} - \alpha \left( \frac{2}{c} - 1 \right) \end{equation}

The interior singular strategy $(\Rrg)^*$ is an ESS when the righthand side of Equation \eqref{eq:cdm2deriv} is negative, which occurs when 
\begin{equation} \label{eq:cdESScondition} c < 2 \left(\frac{1-\alpha}{1 - 2 \alpha} \right) \end{equation}
Because we know from Equation \eqref{eq:cdsocialcollapse} that $(\Rrg)^*$ is infeasible for $c > \frac{1-\alpha}{1 - 2 \alpha}$, it follows that the interior singular strategy $(\Rrg)^*$ is an ESS whenever it corresponds to a feasible replication number for the good contagion. 

Having shown that $(\Rrg)^*$ is evolutionarily-stable under local mutation, we can now address the convergence stability of the equilibrium (i.e. whether such a stable equilibrium could actually be achieved starting under evolution from a nearby sociality strategy). For an endpoint singular strategy, convergence stability follows from the sign of the local selection gradient near the boundary. To demonstrate convergence stability for interior singular strategies, we require that that the local selection gradient is an increasing function at the evolutionarily singular strategy \cite{geritz1998evolutionarily,diekmann2002beginners,brannstrom2013hitchhiker}. For the singular strategy $(\Rrg)^* = \tfrac{1}{c} + \frac{\alpha}{1-\alpha}$ to be convergence stable, we need to verify that
\begin{equation} \label{eq:convergencestability} s''_{\Rrg}(\Rrg) = \frac{\partial^2 \log\left[U(\Rmg,\Rrg)\right]}{\partial \left(\Rmg\right)^2} \bigg|_{\Rmg = \Rrg = (\Rrg)^*}+ \frac{\partial^2 \log\left[U(\Rmg,\Rrg)\right]}{\partial \Rmg \partial \Rrg} \bigg|_{\Rmg = \Rrg = (\Rrg)^*} < 0  \end{equation}
For the second term, we calculate that

\begin{dmath} 
\frac{\partial^2 \log\left[U(\Rmg,\Rrg)\right]}{\partial \Rmg \partial \Rrg} \bigg|_{\Rmg = \Rrg} \\
{= \dsdel{}{\Rrg} \left(\frac{\alpha}{\left( \Rrg - 1 \right)\left( \Rmg \Rrg + \Rrg - \Rmg \right)} - \frac{\left(1 - \alpha\right) \Rmg}{\Rrg \left(c \Rmg \Rrg + \Rrg - \Rmg \right)} \right) \bigg|_{\Rmg= \Rrg} }
= \left[ \frac{-\alpha}{\left(\Rmg \Rrg + \Rrg - \Rmg \right)^2} - \frac{1-\alpha}{\left( c \Rmg \Rrg + \Rrg - \Rmg\right)^2} \right] \bigg|_{\Rmg = \Rrg} \nonumber \\ 
= -\frac{1}{\left(\Rrg\right)^4} \left[  \alpha + \left(\frac{1-\alpha}{c^2} \right) \right]
\label{eq:cdrm2deriv} 
\end{dmath}

Combining Equations \eqref{eq:cdm2deriv} and \eqref{eq:cdrm2deriv}, we can see that 
\begin{align} s''_{\Rrg}(\Rrg) &= \frac{1}{\left(\Rrg\right)^4} \left[-\alpha \left( 2 \Rrg - 1 \right) + \left(\frac{1-\alpha}{c} \right) \left( c \Rrg - 1 \right)^2 - \left(\alpha + \frac{1-\alpha}{c^2} \right) \right] \bigg|_{\Rrg = (\Rrg)^*} \nonumber \\ &= \frac{1}{\left(\Rrg\right)^3} \left[ c \left(1-\alpha\right) (\Rrg)^*  - 2\left( 1 -\alpha + \alpha c) \right) \right] \label{eq:cdcssplugin}  \end{align}
Therefore we see that $s''_{\Rrg}(\Rrg) < 0$ when 
\begin{equation} \label{eq:cdconvergencestableRcondition} (\Rrg)^* < \frac{2(1-\alpha) + 2 \alpha c}{c(1-\alpha)} = 2 \left(\frac{\alpha}{1-\alpha} + \frac{1}{c} \right)\end{equation}
Our ESS $(\Rrg)^* = \frac{\alpha}{1-\alpha} + \frac{1}{c}$ satisfies this condition, so we can conclude that $(\Rrg)^*$ is convergence stable whenever it is \revision{biologically} feasible. %

\subsection{Linear Utility Function} \label{sec:linearutility}

We can also consider a utility function which places a convex combination of weight on the fractions of time spent informed for the good contagion and susceptible to the bad contagion. This utility function takes the form
\begin{equation} 
\label{eq:linearutil} U[\Rmg,\Rrg] := \alpha \hat{I}_m^{(g)}(\Rmg,\Rrg) + \left(1-\alpha\right) \left(\revision{\hat{S}_m^{(b)}(\Rmg,\Rrg)} \right). %
\end{equation}
For a monomorphic population, \revision{we can use the expressions from Equation \eqref{eq:equilibriapiecewise} for the endemic equilibria to see that the linear utility function takes the following form:}
\begin{equation} \label{eq:Umonomorphiclinear}
    U[\Rrg,\Rrg] =  \left\{
     \begin{array}{cr}
     \revision{1-\alpha} &: \revision{\Rrg > 1, \dfrac{1}{c}} \\
       \alpha \left( 1 - \dfrac{1}{\Rrg}\right) + \left(1-\alpha\right) & :  \revision{1 <} \Rrg \leq \dfrac{1}{c}\\
       \revision{\left( 1 - \alpha \right) \left( \dfrac{1}{c \Rrg} \right)} &: \revision{\frac{1}{c} < \Rrg \leq 1} \\
       \alpha \left( 1 - \dfrac{1}{\Rrg}\right) + \left(1-\alpha\right) \left(\dfrac{1}{c \Rrg} \right) & :  \Rrg > \revision{1,\dfrac{1}{c}}
     \end{array}
   \right. . 
\end{equation}
Because $U[\Rrg,\Rrg]$ is a piecewise affine function of $\frac{1}{\Rrg}$, we see that social optimum will be achieved at one of the endpoints $\Rrg \in \{1,\frac{1}{c}, \infty\}$. This utility function is always increasing when $1 < \Rrg < \tfrac{1}{c}$, \revision{always decreasing when $\frac{1}{c} < \Rrg < 1$}, and is increasing for $\Rrg > \tfrac{1}{c}$ when $\alpha c > 1 - \alpha$. When the latter condition holds, we see that the socially-optimal sociality strategy is $\Ropt = \infty$. If instead, $\alpha c < 1 - \alpha$, then $U[\Rrg,\Rrg]$ is decreasing for $\Rrg > \tfrac{1}{c}$, \revision{$\Ropt = \frac{1}{c}$. For the case in which $\Ropt = \frac{1}{c}$, this social optimum will feature long-time survival of the good contagion and elimination of the bad contagion when $c < 1$, and will feature long-time extinction of both contagions when $c \geq 1$.}  %

Turning to the question of evolutionary stability, we consider the following expression for the relative advantage of a mutant over a resident 
\begin{equation} \label{eq:relativeutilitylinear}
    s_{\Rrg}(\Rmg) := U\left[\Rmg,\Rrg \right] - U\left[\Rrg,\Rrg \right].
\end{equation}
This allows us to compute the local selection gradient as
\begin{equation} \label{eq:linaerlocalselectionfirst}
    s'_{\Rrg}(\Rrg) = \alpha \: \dsdel{\hat{I}_m^{(g)}(\Rmg,\Rrg)}{\Rmg} \bigg|_{\Rmg = \Rrg} \revision{+ \left(1-\alpha\right) \dsdel{\hat{S}_m^{(b)}(\Rmg,\Rrg)}{\Rmg} \bigg|_{\Rmg = \Rrg} }.
\end{equation}
Using Equation \eqref{eq:localendemicpartials}, we can further write the selection gradient as 
\begin{equation}
    s'_{\Rrg}(\Rrg) = \left\{
     \begin{array}{cr}
     0 \vspace{2mm} &: \Rrg \leq 1 , \ds\frac{1}{c} \\
      \revision{\ds\frac{1}{(\Rrg)^3} \left( \Rrg - 1\right)} & : \revision{1 < \Rrg} \leq  \dfrac{1}{c}\\
      \revision{-\ds\frac{1}{c^2 (\Rrg)^3} \left( c \Rrg - 1 \right)} &: \revision{\ds\frac{1}{c} < \Rrg \leq 1} \\
       \ds\frac{1}{c^2 (\Rrg)^3} \left[ \left( \alpha c + \alpha - 1 \right) c \Rrg + 1 - \alpha - \alpha c^2 \right] & :  \Rrg > 1, \dfrac{1}{c}
     \end{array}
   \right. .
\end{equation}
\revision{Notably, this means that the selection gradient is always increasing if $\Rrg \in (\frac{1}{c},1)$ and always descreasing if $\Rrg \in (1,\frac{1}{c})$, but that the sign of the selection gradient for $\Rrg > \max\{1,\frac{1}{c}\}$ will depend on the sign of the slope for the term in square brackets that is linear in $\Rrg$. For the case in which both contagions spread (when $\Rrg  > \max\{1,\frac{1}{c}\}$ we see that there is} a possible interior evolutionarily singular strategy given by \begin{equation}
    \Rint^{(g)} := 1 + \frac{\left(1 - \alpha\right)\left(c-1\right)}{1 - \alpha - \alpha c},
\end{equation}
which satisfies the feasibility condition $\Rint^{(g)} \geq 1$ when the signs of $c-1$ and $1 - \alpha - \alpha c$ agree. %
\revision{As a result, the the evolutionary stability of the strategies $\Rrg = \infty$, and $\Rint^{(g)}$ %
both} depend on the signs of $c - 1$ and $\alpha c - (1-\alpha)$.  

\begin{figure}[!ht]
    \centering
    \includegraphics[width = 0.7\textwidth]{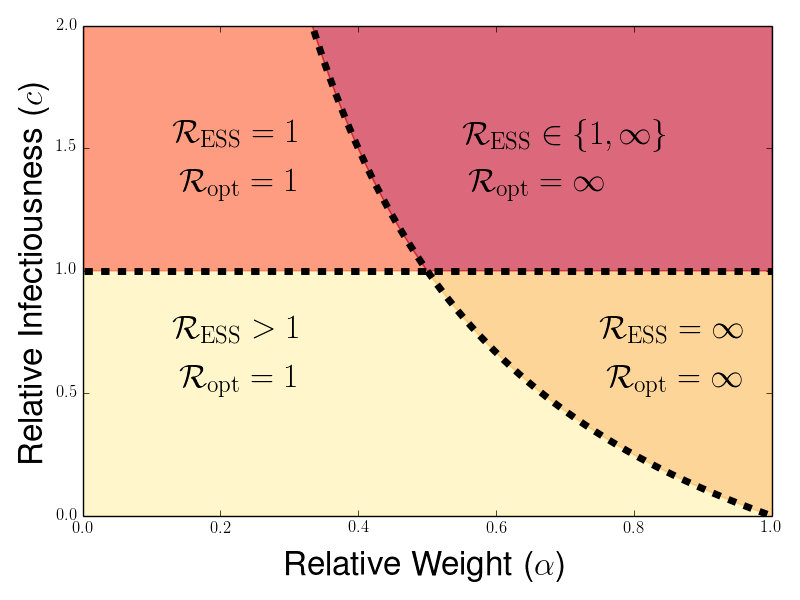}
    \caption{Illustration of the four possible qualitative behaviors for $\Ropt$ and $\RESS$ across the range of relative levels of infectiousness for the bad contagion $c$ and relative weight placed on the good contagion $\alpha$ under linear utility. First, parameter space is divided into two regions, one in which $\alpha c < 1 - \alpha$ and $\Ropt = 1$ (light yellow and orange), and the other in which $\alpha c >  1 - \alpha$ and $\Ropt = \infty$ (dark yellow and red). We further subdivide the region in which $\alpha c < 1 - \alpha$ into the cases in which $c < 1$ and $\RESS > 1$ (light yellow) and in which $c > 1$ and $\RESS \geq 1$ (orange). The region in which $\alpha c > 1 - \alpha$ is further subdivided into regions in which $c < 1$ and $\RESS = \infty$ (dark yellow) and a region in which $c > 1$ and the evolutionary dynamics feature bistability between $\RESS = 1$ and $\RESS = \infty$ (red). 
    } 
    \label{fig:ESSvsSOregimeslinear}
\end{figure}

In Table \ref{tab:linearutility}, we compare the evolutionarily-stable and socially-optimal sociality strategies $\RESS$ and $\Ropt$ for the possible ranges of values of the relative spreading ability $c$ of the two contagions and the weight $\alpha$ placed on infection with the good contagion under the linear utility function. For all of the cases in the table, we see that either the evolutionarily-stable outcome coincides with the social optimum, or there is a social dilemma featuring a discrepancy between the levels of sociality whose direction coincides with the social dilemma in the Cobb-Douglas case (e.g. $\RESS \leq \Ropt$ for $c > 1$ and $\RESS \leq \Ropt$ when $c < 1$). Notably, when $c > 1$ and $c > \frac{1-\alpha}{\alpha}$, we find that the evolutionary dynamics support bistability of \revision{$\frac{1}{c}$} and $\infty$ as evolutionarily-stable states, whose basins of attraction are separated by an evolutionarily unstable state $\Rint^{(g)}$. In Figure \ref{fig:ESSvsSOregimeslinear}, we further illustrate the various parameter regimes in which these socially-optimal and evolutionarily-stable are achieved.

 \renewcommand{\arraystretch}{3}
\begin{table}[ht] 
\centering 
\begin{tabular}{c|c|c} 
& $c > \dfrac{1-\alpha}{\alpha}$ & $c < \dfrac{1-\alpha}{\alpha}$ \\
\hline
$c > 1$ &  $\makecell{\RESS \in \left\{\revision{\frac{1}{c}},\infty\right\} \\ \Ropt = \infty}$ & $\makecell{ \RESS = \revision{\frac{1}{c}}  \\ \Ropt = \revision{\frac{1}{c}}}$\\
 \hline
 $c < 1$ & $\makecell{\RESS = \infty \\ \Ropt = \infty}$ & $\makecell{\RESS = 1 + \dfrac{(1-\alpha)(1-c)}{c(1 - \alpha - \alpha c)} > \revision{\frac{1}{c}} \\ \Ropt = \revision{\frac{1}{c}}}$ \\ 
\end{tabular}
\caption{Evolutionarily-stable and socially-optimal sociality strategies $\RESS$ and $\Ropt$ for different cases on the relative values of the relative weight $\alpha$ of the good contagion under linear utility and the relative infectiousness $c$ of the good contagion.}
\label{tab:linearutility}
\end{table}
 \renewcommand{\arraystretch}{1}

\subsection{Constant Elasticity of Substitution (CES) Utility}
\label{sec:CESutility}

\revision{In this section, we consider the evolutionary dynamics for the CES utility function, introduced by Arrow \cite{arrow1961capital,mas1995microeconomic}, which takes the following form}
\begin{equation}
    \label{eq:CESutility}
    \revision{U\left[\Rmg,\Rrg \right] := \left[ \alpha \left( I_m^{(g)}(\Rmg,\Rrg) \right)^{\rho} + (1 - \alpha) \left(S_m^{(b)}(\Rmg,\Rrg) \right)^{\rho} \right]^{\ds\frac{1}{\rho}}, }
\end{equation}
\revision{where we consider values of the substitution parameter $\rho \in (0,1)$. We note that the CES utility function takes the form of the linear utility function from Equation \eqref{eq:linearutil} when we set $\rho = 1$, and that we recover the Cobb-Douglas utility function in the form of Equation \eqref{eq:CDexample} in the limit as $\rho \to 0$. The parameter $\alpha$ governs the relative weight placed on acquiring the good contagion versus avoiding the bad contagion, and we see that this parameter retains this interpretation when we obtain the linear and Cobb-Douglas utility functions in the limits as $\rho \to 1$ and $\rho \to 0$, respectively.}

\revision{For a monomorphic population following the sociality strategy $\Rrg$, we can use the expressions from Equation \eqref{eq:equilibriapiecewise} for the endemic equilibria for the good and bad contagion to write the following piecewise characterization of the CES utility:}
\begin{equation} \label{eq:CESRrg}
    U\left[\Rrg,\Rrg\right] = 
    \left\{ \begin{array}{cr}
          (1-\alpha)^{\frac{1}{\rho}} &: \Rrg < 1, \frac{1}{c} \\
          (1-\alpha)^{\frac{1}{\rho}} \left[ \ds\frac{1}{c \Rrg}  \right] &: \frac{1}{c} < \Rrg \leq 1 \\
          \alpha^{\frac{1}{\rho}} \left[1 - \ds\frac{1}{\Rrg} \right] &: 1 < \Rrg \leq \frac{1}{c} \\
          \left[ \alpha \left(1 - \ds\frac{1}{\Rrg} \right)^{\rho} + (1-\alpha) \left[\ds\frac{1}{c \Rrg} \right]^{\rho} \right]^{\frac{1}{\rho}} &: \Rrg > 1,\frac{1}{c}
    \end{array}
    \right. .
\end{equation}
\revision{Because the CES utility function is twice-differentiable and strictly concave for $\rho \in (0,1)$, we know from our analysis in the \ref{sec:genutility} section that the utility function $U[\Rrg,\Rrg]$ has a unique local maximizer for $\Rrg \geq \frac{1}{c}$. For $c < 1$, this local maximizer is actually the global maximizer for biologically feasible $\Rrg$ satisfying $\Rrg \geq 0$.} 

\revision{However, we notice also notice that $U\left[\Rrg,\Rrg\right]$ is decreasing on $(\frac{1}{c},1)$ for the case in which $c > 1$. In this case, $\Rrg = \frac{1}{c}$ is also a local maximizer of $U\left[\Rrg,\Rrg\right]$, which has utility $U[\frac{1}{c},\frac{1}{c}] = (1-\alpha)^{\frac{1}{\rho}}$. By considering the limit $\lim_{\Rrg \to \infty} U \left[\Rrg,\Rrg\right] = \alpha^{\frac{1}{\rho}}$, we can deduce that $\alpha > \frac{1}{2}$ (and correspondingly $\alpha > 1 - \alpha$) is a sufficient condition for the maximizer $U\left[ \Rrg,\Rrg \right]$ be achieved by a sociality strategy $\Ropt > 1$ in which the good contagion is present at equilibrium. In particular, this means that, whenever more emphasis is placed on the good contagion than the bad contagion, the socially-optimal sociality strategy will feature spread of both the good and bad contagion when $c < 1$.  }

\begin{figure}[!ht]
    \centering
    \includegraphics[width = 0.7\textwidth]{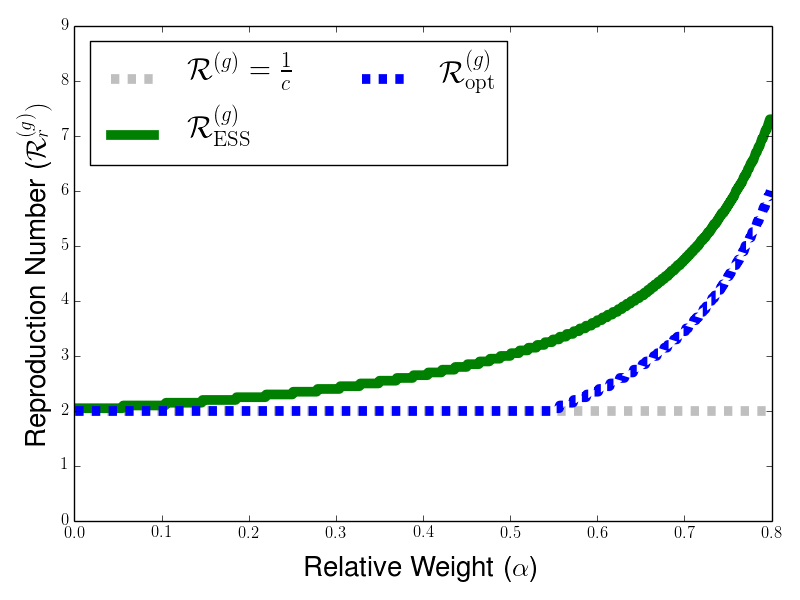}
    \caption{\revision{Illustration of social dilemma for adaptive dynamics under CES utility function when $c = \frac{1}{2}$. The reproduction numbers $\Rrg$ are presented for both evolutionarily-stable sociality strategy $\RESS$ (solid green line) and the socially-optimal sociality strategy $\Ropt$ (dashed blue line), and are plotted as functions of the weight parameter $\alpha$ describing the relative importance placed upon the good contagion. The dashed gray line represents $\Rrg = \frac{1}{c}$, the reproduction number corresponding to the highest level of social interaction at which the bad contagion does not spread. The substitution parameter is fixed at $\rho = 0.5$, and the weight parameter $\alpha$ ranges between $0$ and $0.8$. } } 
    \label{fig:CESequilibrialess}
\end{figure}

\revision{Next, we turn to the adaptive dynamics analysis for studying the evolution of sociality strategies under CES utility. When $c = 1$, we can use Equation \eqref{eq:selectiongeneralcompare} to see that the derivative of the monomorphic utility and selection gradient coincide, and therefore $\RESS = \Ropt$ in this case. When $c < 1$, we know from the discussion in Section \ref{sec:genutility} that there will be a unique socially-optimal strategy $\Ropt \geq \frac{1}{c} > 1$, and that any evolutionarily stable strategy $\RESS$ will satisfy $\RESS > \Ropt$. In Figure \ref{fig:CESequilibrialess}, we illustrate the socially-optimal and evolutionarily-stable sociality strategies $\Ropt$ and $\RESS$ for the CES utility function when $c = \frac{1}{2}$. We see that $\RESS > \Ropt$ when any weight is placed on acquiring the good contagion (i.e. $\alpha > 0$), and the form of the social dilemma resembles the qualitative behavior seen for Cobb-Douglas utility in Figure \ref{fig:combined}(c,bottom) in the case of $c < 1$.}

\revision{Finally, we turn to analyze the case of $c > 1$, in which there are two local maxima for the social utility $U[\Rrg,\Rrg]$. Using Equation \eqref{eq:selectiongradientgreater} and the fact that and $\dsdel{U[\Rrg,\Rrg]}{\shat{(b)}(\Rrg,\Rrg)} > 0$, $\dsdel{\shat{(b)}(\Rrg,\Rrg)}{\Rrg} < 0$, and $\dsdel{\ihat{(g)}(\Rrg,\Rrg)}{\Rrg} = 0$ for $c >1$ and  $\Rrg < 1$, we see that, in this regime, 
\[ s'_{\Rrg}(\Rrg) = \dsdel{U[\Rrg,\Rrg]}{\ihat{(g)}(\Rrg,\Rrg)} \dsdel{\ihat{(g)}(\Rrg,\Rrg)}{\Rrg}  + \dsdel{U[\Rrg,\Rrg]}{\shat{(b)}(\Rrg,\Rrg)}\dsdel{\shat{(b)}(\Rrg,\Rrg)}{\Rrg} < 0.  \]
Because the local selection gradient  $s'_{\Rrg}(\Rrg)$ is negative for when $\frac{1}{c} < \Rrg < 1$, we can deduce that the sociality strategy $\Rrg = \frac{1}{c}$ will be evolutionarily stable. }

\revision{To further understand the basin of attraction of this state under adaptive dynamics, we now look to calculate the selection gradient for $\Rrg > 1$. Our goal is to use Equation \eqref{eq:selectiongeneralcompare} to calculate the local selection gradient for the CES utility when $c > 1$ and $\Rrg > 1$, and we are particularly interested in the behavior of the adaptive dynamics in the case as $\Rrg \to 1^+$. We recall from Equation \eqref{eq:selectiongeneralcompare} that the local selection gradient for $\Rrg > \max\{1,\frac{1}{c} \}$ takes the following form}
\begin{align*} \revision{s'_{\Rrg}(\Rrg)} &= \revision{\left(\frac{\Rrg - 1}{\Rrg} \right)\dsdel{U[\Rrg,\Rrg]}{\Rrg}} \\ &+ \revision{\frac{1}{c \left(\Rrg\right)^3} \left( \frac{1}{c} - 1 \right) \frac{\partial U[\Rrg,\Rrg]}{ \partial \hat{S}^{(b)}(\Rrg,\Rrg)}}. \end{align*}
\revision{We will now compute the main quantities that appear in this expression for the local selection gradient. We first differentiate the expression for CES utility from Equation \eqref{eq:CESRrg} in the case of $\Rrg > \max\{1,\frac{1}{c} \}$, which allows us to see that}
\begin{equation} \label{eq:CESutilityderiv}
\begin{aligned}
   \revision{ \dsdel{U\left[\Rrg,\Rrg\right]}{\Rrg} } &= \revision{\left[ \alpha \left( 1 - \frac{1}{\Rrg} \right)^{\rho} + \left(1 - \alpha\right) \left( \frac{1}{c \Rrg} \right)^{\rho} \right]^{\frac{1}{\rho} - 1} }\\ \\ &\times \revision{\left( \frac{1}{\Rrg} \right)^2 \left[ \alpha \left( \frac{\Rrg-1}{\Rrg } \right)^{\rho - 1} -   \left( \frac{1 - \alpha}{c} \right) \left( \frac{1}{c \Rrg} \right)^{\rho - 1}  \right] .
   }
   \end{aligned}
\end{equation}
\revision{In the limit as $\Rrg \to 1^+$, we can use our expression from Equation \eqref{eq:CESutilityderiv} and the fact that we are considering CES utility parameters with substitution parameters $\rho > 0$ to deduce that}
\begin{equation} \label{eq:CESderiv1}
\begin{aligned}
& \revision{\lim_{\Rrg \to 1^+} \left[ \left( \frac{\Rrg - 1}{\Rrg} \right) \dsdel{U\left[\Rrg,\Rrg\right]}{\Rrg} \right]} \\ &= \revision{\left(\frac{1-\alpha}{c} \right)^{\frac{1}{\rho} - 1} \lim_{\Rrg \to 1^+} \left[ \alpha \left( \frac{\Rrg - 1}{\Rr} \right)^{\rho} - \left( \frac{1 - \alpha}{c} \right)  \left(\frac{1}{c \Rrg} \right)^{\rho - 1} \left( \frac{\Rrg - 1}{\Rr} \right)\right] = 0}
\end{aligned}
\end{equation}
\revision{Next, we consider the partial derivative of utility with respect to the equilibrium level of the bad contagion. We use Equations \eqref{eq:equilibriapiecewise} and \eqref{eq:CESutility} to see that}
\begin{equation} \label{eq:CESSbpartial}
    \begin{aligned}
    \revision{\dsdel{U\left[\Rrg,\Rrg\right]}{\shat{b}(\Rrg,\Rrg)}} &= 
    \revision{ \left[ \alpha \left( \ihat{(g)}(\Rrg,\Rrg) \right)^{\rho} + (1-\alpha) \left( \shat{(b)}(\Rrg,\Rrg) \right)^{\rho}  \right]^{\frac{1}{\rho} - 1} } 
    \\ &\times \revision{\left( 1 - \alpha \right)  \left( \shat{(b)}(\Rrg,\Rrg) \right)^{\rho - 1}} \\
    &= \revision{ (1-\alpha) \left[ \alpha \left( 1 - \frac{1}{\Rrg} \right)^{\rho} + (1-\alpha) \left( \frac{1}{c\Rrg} \right)^{\rho}  \right]^{\frac{1}{\rho} - 1} \left(\frac{1}{c \Rrg} \right)^{\rho - 1} } ,
    \end{aligned}
\end{equation}
\revision{and we see that, in the limit as $\Rrg \to 1^+$}
\begin{equation} \label{eq:CESSbpartial1}
    \revision{\dsdel{U\left[\Rrg,\Rrg\right]}{\shat{b}(\Rrg,\Rrg)} \bigg|_{\Rrg \to 1^+} = (1 - \alpha)^{\frac{1}{\rho}} > 0. }
\end{equation}
\revision{We can now use Equations \eqref{eq:selectiongeneralcompare}, \eqref{eq:CESderiv1}, and \eqref{eq:CESSbpartial1} to see that, for $c > 1$, the local selection gradient satisfies}
\begin{equation}
   \revision{ s'_{\Rrg}(\Rrg) \bigg|_{\Rrg \to 1^+} =  \frac{1}{c} \left( \frac{1}{c} - 1 \right) \left(1 - \alpha\right)^{\frac{1}{\rho}} < 0. }
\end{equation}
\revision{From the continuity of all of the quantities appearing in the local selection gradient of Equation \eqref{eq:selectiongeneralcompare}, we can now deduce that, for $c < 1$, there will always some sociality strategies $\Rrg > 1$ that will be in the basin of attraction of the boundary state $\frac{1}{c} < 1$ under the adaptive dynamics.  This means that either $\frac{1}{c}$ may be a global attractor under the evolutionary dynamics, or that it will coexist in multistability with an evolutionary attractor featuring positive levels of both the good and bad contagion at equilibrium. Now, we will use a numerical approach explore these two possibilities of evolutionary collapse of social interaction and bistability of two evolutionarily-stable sociality strategies.}

\begin{figure}[tph!]
    \centering
    \includegraphics[width = 0.48\textwidth,height = 0.36\textwidth]{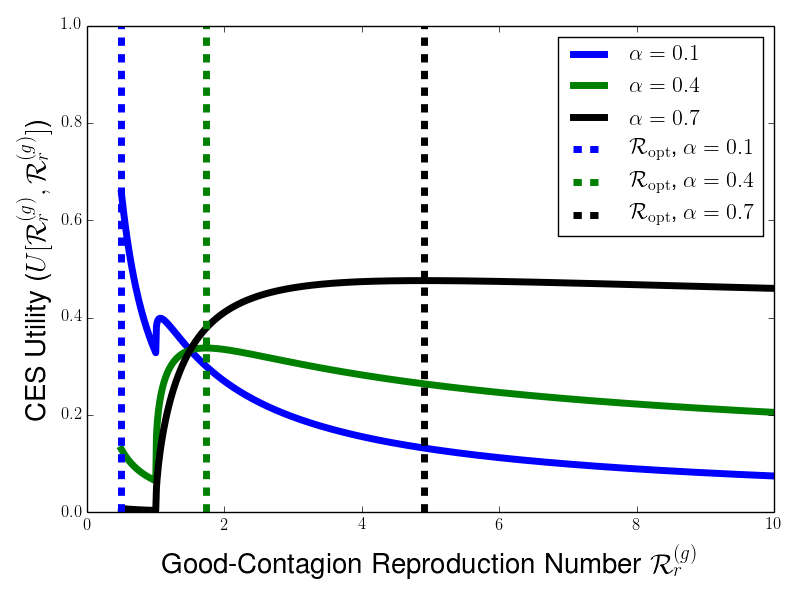}
    \includegraphics[width = 0.48\textwidth,height = 0.36\textwidth]{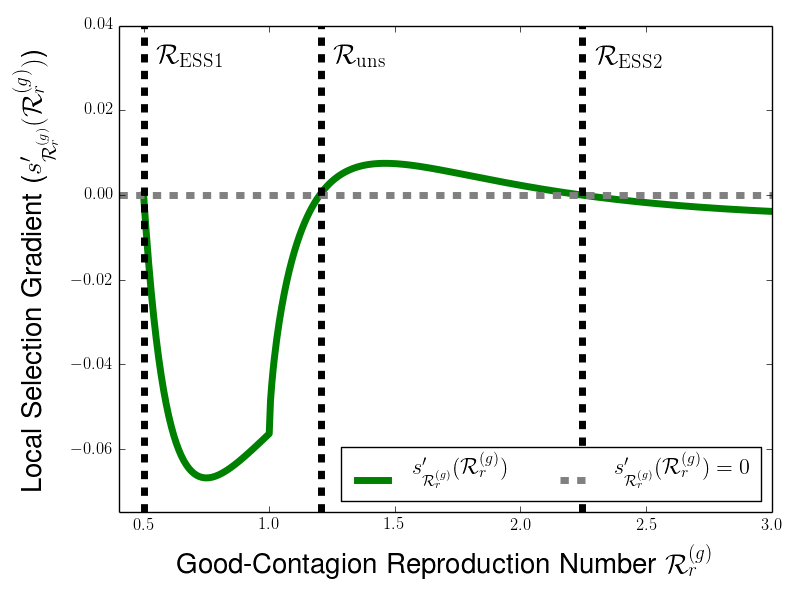}
    \caption{\revision{Illustration of the social utility (left) and selection gradient (right) for the CES utility function when the bad contagion spreads more readily than the good contagion ($c = 2$). Left: Social utility plotted as a function of the good-contagion reproduction number $\Rrg$ for weight parameters $\alpha = 0.1$ (solid blue line), $\alpha = 0.4$ (solid green line), and $\alpha = 0.7$ (solid black line). Dashed lines describe the socially-optimal values of $\Rrg$, with colors corresponding to the color of the utility function with the same $\alpha$ parameter. Right: Plot of the local selection gradient $s'_{\Rrg}(\Rrg)$ (solid green line), plotted as a function of the reproduction number $\Rrg$. Dashed, horizontal gray line corresponds to a selection gradient $s'_{\Rrg}(\Rrg) = 0$, and the vertical dashed black lines describe, from left to right, the ESS sociality strategy $\mathcal{R}_{\mathrm{ESS1}} = \frac{1}{c} < 1$ featuring no long-time contagion, the evolutionarily-unstable strategy $\mathcal{R}_{\mathrm{uns}} > 1$, and the ESS sociality strategy $\mathcal{R}_{\mathrm{ESS2}} > \mathcal{R}_{\mathrm{uns}}$ under which both contagions survive at equilibrium. Because selection gradient is negative for some sociality strategies $\Rrg > 1$, there is a regime of strategies in which the good contagion is eliminated under the adaptive dynamics. The substitution value for is $\rho = 0.5$ for both panels, and the weight parameter is $\alpha = 0.525$ in the right panel.}}
    \label{fig:CES_bistability_example}
\end{figure}

\revision{In Figure \ref{fig:CES_bistability_example}, we illustrate the CES utility function and an example of the selection gradient for the adaptive dynamics under CES utility. In Figure \ref{fig:CES_bistability_example}(left), we display the utility function for three different values of the weight parameter, presenting one example in which utility is maximized by the sociality strategy $\Rrg = \frac{1}{c}$ (with $\alpha = 0.1$) in which neither contagion spreads, as well as two examples in which the socially-optimal sociality strategy is achieved by a strategy $\Ropt > 1$ under which both contagions survive at equilibrium. In Figure \ref{fig:CES_bistability_example}(right), we illustrate an example selection gradient for the CES utility function. We find that the selection gradient vanishes at three points. At two of these points (labeled $\mathcal{R}_{\mathrm{ESS1}} = \frac{1}{c}$ and $\mathcal{R}_{\mathrm{ESS2}}$), the selection gradient decreases past zero as $\Rrg$ increases beyond these points, so these sociality strategies are evolutionarily stable. The third zero (labeled $\mathcal{R}_{\mathrm{uns}} > 1$) is located at a point at which the local gradient is increasing, and therefore it is an evolutionarily-unstable singular strategy, separating the basins of attraction for the two ESS sociality strategies. Because $\mathcal{R}_{\mathrm{uns}} > 1$, this example shows the possibility that an initial population with a sociality strategy $\Rrg > 1$ can be located in the evolutionary basin of attraction of the strategy $\mathcal{R}_{\mathrm{ESS1}}$ featuring elimination of both contagions, even though a more highly-social population can sustain both contagions under the long-time limit of the adaptive dynamics. }

\begin{figure}[tph!]
    \centering
    \includegraphics[width = 0.48\textwidth,height = 0.36\textwidth]{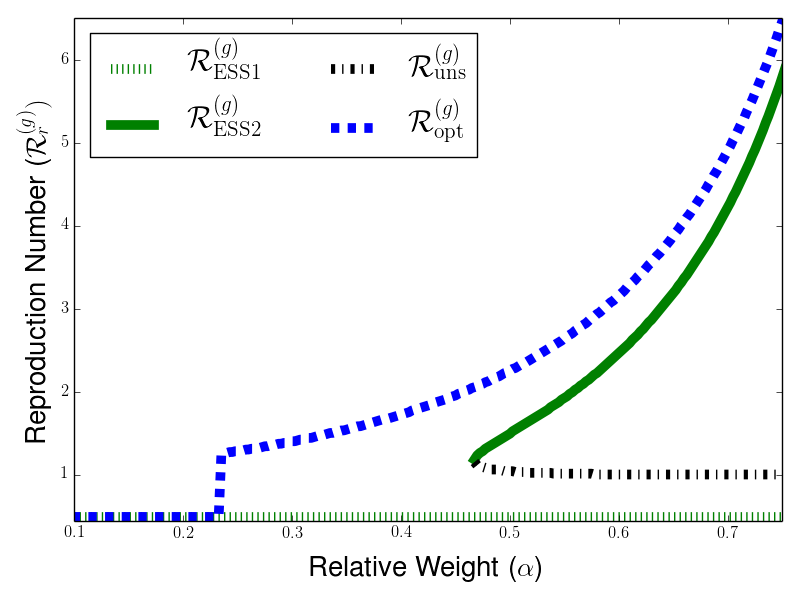}
    \includegraphics[width = 0.48\textwidth,height = 0.36\textwidth]{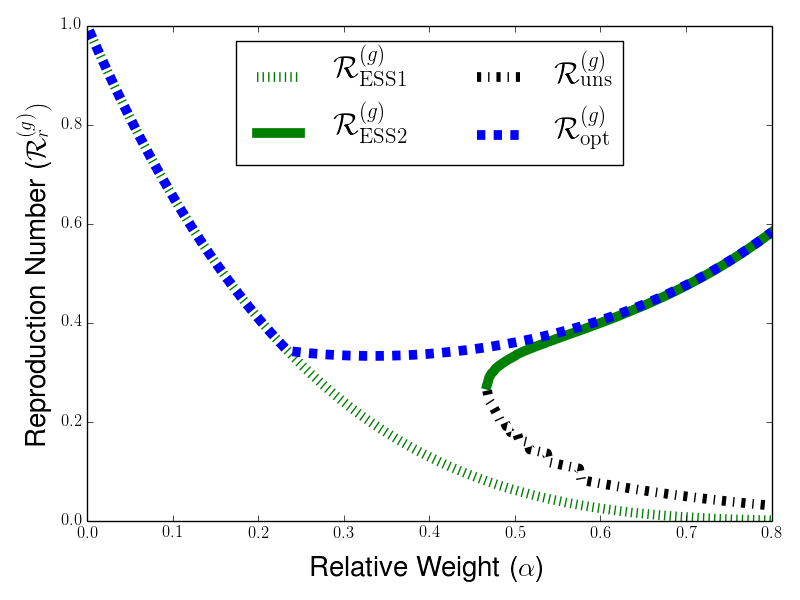}
    \caption{\revision{Illustration of the social dilemma for adaptive dynamics when the bad contagion spreads more readily than the good contagion ($c = 2$). Reproduction numbers $\Rrg$ for sociality strategies (left panel) and associated social utilities $U[\Rrg,\Rrg]$ are plotted as functions of the relative weight parameter $\alpha$ for the CES utility. Strategies presented are the socially-optimal sociality strategy $\Ropt$ (dashed blue line), the evolutionarily-stable strategy $\mathcal{R}_{\mathrm{ESS}1} = \frac{1}{c}$ at which no contagion spreads (dotted green line), the evolutionarily-stable strategy $\mathcal{R}_{\mathrm{ESS}2}$ at which both contagions spread (solid green line), and the evolutionarily-unstable singular strategy $\mathcal{R}_{\mathrm{uns}}$ separating the basins of attraction of the two ESS sociality strategy (dash-dotted black line). CES utility has weight parameters ranging from $\alpha = 0$ to $\alpha = 0.8$, and the substitution parameter is $\rho = 0.25$ for both panels. }}
    \label{fig:CES_equilibriacgreater}
\end{figure}

\revision{In Figure \ref{fig:CES_equilibriacgreater}, we further illustrate the social dilemma by plotting the social optima and evolutionary singular strategies (Figure \ref{fig:CES_equilibriacgreater}, left) and the CES utilities achieved for these sociality strategies (Figure \ref{fig:CES_equilibriacgreater}, right) across a range of weight parameters $\alpha$. We find that the sociality strategy $\mathcal{R}_{\mathrm{ESS1}} = \frac{1}{c}$ featuring no equilibrium contagion is the unique evolutionarily-stable strategy for sufficiently small values of $\alpha$, while an additional pair of singular strategies $\mathcal{R}_{\mathrm{ESS2}}$ and $\mathcal{R}_{\mathrm{uns}}$ appear when $\alpha \approx 0.46$, yielding bistability between $\mathcal{R}_{\mathrm{ESS1}}$ and $\mathcal{R}_{\mathrm{ESS2}}$. By comparing the utilities achieved these sociality strategies, we find that utility achieved by the non-trivial ESS strategy $\mathcal{R}_{\mathrm{ESS2}}$ is less than that achieved by the social optimum $\Ropt$, but that the utility achieved by the equilibrium $\mathcal{R}_{\mathrm{ESS1}}$ featuring no long-term contagion results in even lower utility than the suboptimal ESS outcome achieve by $\mathcal{R}_{\mathrm{ESS2}}$. On top of the social dilemma realized by convergence to the evolutionarily-stable outcome of $\mathcal{R}_{\mathrm{ESS2}}$ achieved by the adaptive dynamics for sufficiently large initial sociality stategy $\Rrg$, we see that the bistability of $\mathcal{R}_{\mathrm{ESS2}}$ and the non-interacting equilibrium $\mathcal{R}_{\mathrm{ESS1}} = \frac{1}{c}$ that the CES utility function can achieve an even worse social dilemma and a long-time collapse of social interaction in the population. }

\subsection{Assortative Interactions and Mitigating the Social Dilemma} \label{sec:assortment}

Now we consider a modified version of our processes in which individuals have a disproportionate probability of interaction with individuals who use the same sociality strategy. This mechanism, often called preferential mixing, has been used to study disease dynamics in multigroup settings \cite{hethcote1987epidemiological}. Similar kinds of assortative interactions have been shown to help promote cooperation in two-strategy games \cite{grafen1979hawk,bergstrom2003algebra,van2017hamilton}, and also help to promote more efficient evolutionarily-stable strategies for continuous-strategy social dilemmas \cite{cornforth2012synergy,coder2018effects,van2017hamilton,iyer2020evolution}. 

For our model of assortative interactions, we again characterize the interaction rate of individuals through their rate of social interaction $\sigma$ and their corresponding reproduction number $\Rg{} := \frac{\sigma \revision{p_g}}{\gamma_g}$ under the good contagion with well-mixed interactions. For each interaction, we assume that an individual is automatically paired with someone following the same sociality strategy with probability $\rho$, and is paired with a randomly chosen member of the group with probability $1 - \rho$. The parameter $\rho$ measures the degree of homophily of social interactions, with $\rho = 0$ corresponding well-mixed interactions and $\rho = 1$ corresponding to the monomorphic interaction case. 

In a population composed of a fraction $f$ following a mutant socaility strategy $\Rmg$ and a fraction $1-f$ following a resident strategy $\Rrg$, we can use the approach from Section \ref{appsec:dimorphic-contact-rates} to derive the dimorphic dynamics for both the good and bad contagion with assortative interactions. For the good contagion, we obtain the following systems of differential equations  
\begin{equation}\label{eq:assortmentgooddimorphic}
\begin{aligned}
    \revision{\frac{1}{\gamma_g}}  \dv{I^{(g)}_r}{t} &= \Rrg \Big[\rho I_r^{(g)} + \left(1 - \rho\right) \frac{\Rrg (1-f) I_r^{(g)} + \Rmg f I^{(g)}_m}{\Rrg (1-f) + \Rmg f}\Big] \left(1 - I^{(g)}_r\right) - I^{(g)}_{r}
    \vspace{4mm} \\ 
     \revision{\frac{1}{\gamma_g}} \revision{\dv{I^{(g)}_m}{t}} &= \Rmg \Big[\rho I_m^{(g)} + \left(1 - \rho \right) \frac{\Rrg (1-f) I^{(g)}_r + \Rmg f I^{(g)}_m}{\Rrg (1-f) + \Rmg f}\Big] \left(1 - I^{(g)}_m \right) - I^{(g)}_m
\end{aligned}.
\end{equation}
We can also derive an analogous system for the bad contagion with reproduction numbers $\Rmb = c \Rmg$ and $\Rrb = c \Rrg$. In the adaptive dynamics limit in which $f \to 0$, the residents see no effect from interactions with mutants and the contagion dynamics for the resident type reduces to the monomorphic dynamics of Equation \eqref{eq:baselineSISmonomorphic}. Therefore the equilibrium levels for the resident strategy are given by $\hat{I}_r^{(g)} = 1 - \frac{1}{\Rrg}$ and $\hat{S}_r^{(b)} = \frac{1}{c\Rrg}$. In the same limit, the contagion dynamics for the mutant population are governed by 
\begin{subequations} \label{eq:mutantadaptivefirst}
\begin{align}
 \revision{\frac{1}{\gamma_g}} \dsddt{I_m^{(g)}} &= \Rmg \left( 1 - I_m^{(g)} \right) \left( \rho I_m^{(g)} + (1-\rho) I_r^{(g)} \right) - I_m^{(g)} \\     
 \revision{\frac{1}{\gamma_b}} \dsddt{I_m^{(b)}} &= c \Rmg \left( 1 - I_m^{(b)} \right) \left(\rho I_m^{(b)} + (1-\rho) I_r^{(b)} \right) - I_m^{(b)}.
\end{align}
\end{subequations}
For simplicity, we will restrict our attention for the adaptive dynamics analysis to the case of $\Rmg > \tfrac{1}{c}$, \revision{so that the} resident dynamics will approach an endemic equilibrium for both the good and bad contagion. Using this assumption and the fact that the resident population evolves independently to its stable equilibrium, we can substitute $\hat{I}_r^{(g)} = 1 - \frac{1}{\Rrg}$ into Equation \eqref{eq:mutantadaptivefirst} to see that the long-time behavior of the mutant population under the good contagion will be determined by
\begin{equation}
\begin{aligned}
    \revision{\frac{1}{\gamma_g}}  \dsddt{I_m^{(g)}} &= - \rho \Rmg (I_m^{(g)})^2 + \left( \rho \Rmg - (1-\rho) \Rmg \hat{I}_r^{(g)} - 1\right) I_m^{(g)} + (1-\rho) \Rmg \hat{I}_r^{(g)} \\ 
    &= \revision{- \rho \Rmg (I_m^{(g)})^2 - \left(1 + \Rmg - 2 \rho \Rmg - (1 - \rho) \frac{\Rmg}{\Rrg} \right) I_m^{(g)} + \Rmg (1-\rho)  \left( 1 - \frac{1}{\Rrg} \right) }.
    \end{aligned}
\end{equation}
Then we see that the only biologically feasible equilibrium for the mutant population under the good contagion is
\begin{equation}
\begin{aligned}
    \hat{I}_m^{(g)} &=  \revision{- \frac{1 + \Rmg - 2 \rho \Rmg - (1 - \rho) \frac{\Rmg}{\Rrg}}{2 \rho \Rmg}} \\
    &+ \revision{\frac{\sqrt{\left(1 + \Rmg - 2 \rho \Rmg - (1 - \rho) \frac{\Rmg}{\Rrg} \right)^2  - 4 \rho(1-\rho) \left(\Rmg\right)^2 \left(1 - \frac{1}{\Rrg}\right)}}{2 \rho \Rmg} }.
    \end{aligned}
\end{equation}
By a similar approach, the mutant population will achieve an equilibrium for the bad contagion of
\begin{equation}
\begin{aligned}
    \hat{I}_m^{(g)} &=  \revision{- \frac{1 + c\Rmg - 2 \rho c \Rmg - (1 - \rho) \frac{\Rmg}{\Rrg}}{2 \rho c\Rmg}} \\
    &+ \revision{\frac{\sqrt{\left(1 + c\Rmg - 2 \rho c \Rmg - (1 - \rho) \frac{\Rmg}{\Rrg} \right)^2  - 4 \rho(1-\rho) c^2 \left(\Rmg\right)^2 \left(1 - \frac{1}{c\Rrg}\right)}}{2 \rho c \Rmg} }.
    \end{aligned}
\end{equation}

Using these equilibrium levels of infectiousness for the mutant population, we can calculate the local selection gradient for the Cobb-Douglas utility. The tedious calculation included in the supplemental Mathematica notebook (available at \href{https://github.com/dbcooney/Social-Dilemmas-of-Sociality-due-to-Beneficial-and-Costly-Contagion}{https://github.com/dbcooney/Social-Dilemmas-of-Sociality-due-to-Beneficial-and-Costly-Contagion}) eventually yields the \revision{local} selection gradient 
\begin{equation} \label{eq:assortmentgradient}
    s_{\Rmg}'(\Rmg) = \frac{-(1-\alpha) c (\Rmg)^2 + \left( 1 - \alpha + c \left[ \alpha + \rho \left( 1 - \alpha \right) \right] \right) \Rmg - \rho}{R \left( R - \rho\right) \left( c R - \rho \right)}
\end{equation}
Notably, we see that the selection gradient coincides with the non-assortative case when $\rho = 0$ and with the derivative of the monomorphic utility function when $\rho = 1$. The numerator of the selection gradient from Equation \eqref{eq:assortmentgradient} is a decreasing function of $\Rmg$, and therefore, for each assortment probability $\rho$, there will be a unique evolutionarily-stable sociality strategy $\RESS(\rho)$. Setting the lefthand side of Equation \eqref{eq:assortmentgradient} equal to $0$, we see that this unique family of ESSes is given by %
\begin{equation} \label{eq:RESSr}
    \RESS(\rho) = \frac{1 + c \left( 1 -\alpha\right)  +   \rho c (1-\alpha) +\sqrt{(\alpha -\alpha  c+\alpha  c \rho-c \rho-1)^2-4 \rho c \left(1 - \alpha \right)}}{2 c \left( 1 - \alpha \right)}.
\end{equation}

\begin{figure}[tph!]
    \centering
    \includegraphics[width = 0.48\textwidth]{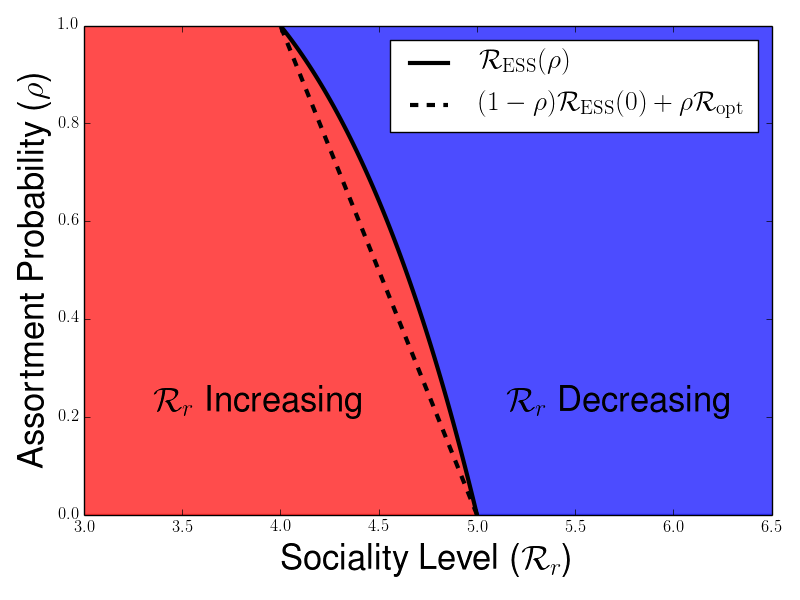}
    \includegraphics[width = 0.48\textwidth]{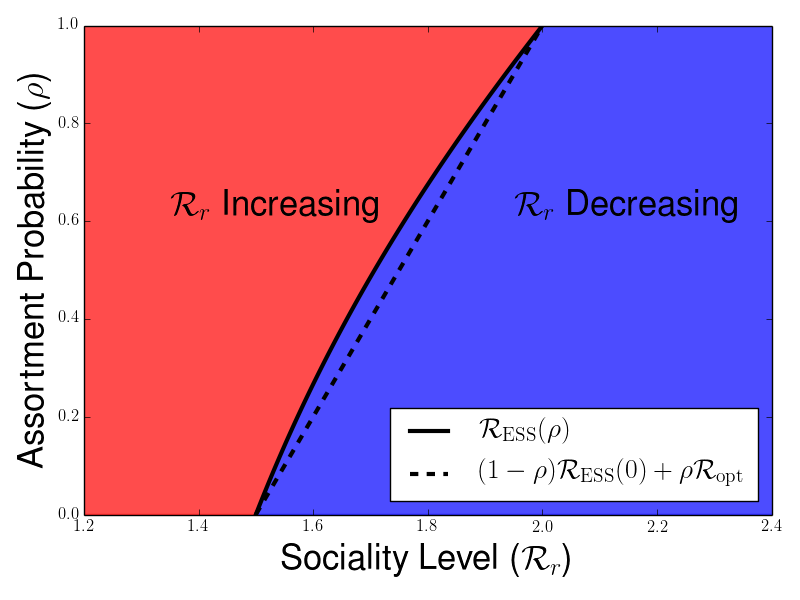}
    \caption{Plot of regions in which the local selection gradient is increasing (red) and decreasing (blue) for varying values of sociality strategy $\Rmg$ and assortment probability $\rho$, with $c = \frac{1}{2}$ (left) and $c = 2$ (right). Solid line black line is the evolutionarily-stable level of sociality $\RESS(\rho)$ for assortment probability $r$, and the dashed black line corresponds to $(1-\rho) \RESS + \rho \Ropt$, linearly interpolating between the ESS under well-mixed interactions and the socially-optimal level of sociality.}
    \label{fig:assortmentESSplots}
\end{figure}

We can also visualize how the evolutionarily-stable level of sociality varies with $\rho$ by plotting the sign of selection gradient for varying values of $\rho$ and $\Rmg$ \cite{saad2020dynamics}. In Figure \ref{fig:assortmentESSplots}, we plot the regions of positive selection gradient in red and of negative selection gradient in blue for examples corresponding to both $c = \frac{1}{2}$ (left) and $c = 2$ (right). The border of the red and blue regions is given by the solid black line the two regions given by the expression from Equation \eqref{eq:RESSr} for the evolutionarily-stable strategy $\RESS(\rho)$ for given $\rho$. As a point of comparison, we plot the black dashed line which interpolates linearly in $r$ between the evolutionarily-stable sociality $\RESS$ when assortment $\rho = 0$ and the socially-optimal level of sociality $\Ropt$ when $\rho= 1$. We see that, in both cases, $\RESS(\rho)$ moves towards the social optimum more slowly than this linear interpolation, which tells us that, while the addition of assortment does help to mitigate the social dilemma, it takes a relatively large level of assortment to make an appreciable impact on improving the social efficiency of the evolutionarily-stable outcome.  

\section{Microfoundation for Population Utility Function} \label{sec:utilitymicrofoundation}

So far, we have consider the utility functions for our evolutionary dynamics as given functions of the equilibrium levels of the contagion in the population. Here we offer a derivation of how linear and Cobb-Douglas utility functions can arise as the expected or time-averaged utility for an individual living in a population at steady state under the contagion dynamics. This serves as a means for understanding how different tradeoffs between the benefits of the good contagion and the costs of the bad contagion at the population level are connected to the underlying payoffs achieved by individuals following various sociality strategies in the presence of the coupled contagion dynamics.   

We calculate the expected utility from payoffs associated with an individual's state under the good and bad contagion, treating the fractions susceptible and infectious at steady state as the probabilities that a given individual is in that state. We describe the status of individuals for two contagion by the pair of types $T  = (T^g, T^b) \in \{S^{(g)}, I^{(g)} \} \times \{S^{(b)}, I^{(g)} \} $. For the good contagion, we assume that a contribution of $a_g > 0$ is made to individual utility when an individual is infected, while a contribution of $0$ is made when the individual is susceptible. For the bad contagion, we assume that a contribution of $0$ is made to individual utility when the individual is infected, while a contribution of $a_b > 0$ is made when the individual is susceptible. If we consider that an individual's utility is the sum of its contributions from the good and bad contagion, then we can see that the utility is given for each of four possible disease states as

\begin{equation} \label{eq:additiveutilitychart}
\begin{blockarray}{ccc}
& S^{(g)} & I^{(g)} \\
\begin{block}{c[cc]}
S^{(b)} & a_b & a_b + a_g \\
I^{(b)} & 0 & a_g \\
\end{block}
\end{blockarray}
\end{equation}

Because the two contagions spread independently through the population, the average utility of a population with monomorphic interaction rates are given by
\begin{equation}
\begin{aligned}u(I^{(g)},S^{(b)}) &= S^{(b)} \left( 1 - I^{(g)} \right) a_b + S^{(b)} I^{(g)} \left( a_b + a_g \right) \\ &+ \left(1 - S^{(b)} \right) \left( 1 - I^{(g)}\right) \left( 0 \right) +\left( 1 - S^{(b)} \right) I^{(g)} a_g \\ &= a_b S^{(b)} + a_g I^{(g)},
\end{aligned}
\end{equation}
and choosing parameters $a_g = \alpha  \in [0,1]$ and $a_b = 1 - \alpha \in [0,1]$, we recover the linear utility function for a monomorphic population
\begin{equation}
U(I^{(g)},S^{(b)}) = \alpha I^{(g)} + \left( 1 - \alpha \right) S^{(b)}
\end{equation}

If instead, we consider individual utility given by the product of the contributions from the good contagion and the bad contagion, the utility for the four pairs of disease states is given by 

\begin{equation} \label{eq:multiplicativeutilitychart}
\begin{blockarray}{ccc}
& S^{(g)} & I^{(g)} \\
\begin{block}{c[cc]}
S^{(b)} & 0 & a_b a_g \\
I^{(b)} & 0 & 0 \\
\end{block}
\end{blockarray}
\end{equation}
Because the two contagion spread independently, the average utility in the population is given by
\[ u(I^{(g)},S^{(b)}) = a_b a_g I^{(g)} S^{(b)}, \]
and the choice $a_b = a_g = 1$ provides us with the utility function takes the form 
\[U(I^{(g)},S^{(b)}) = I^{(g)} S^{(b)} \]
corresponding to a Cobb-Douglas utility with positive returns to scale and symmetric dependence on its inputs. 

In a similar manner, we can derive Cobb-Douglas utilities with asymmetric weight placed on $I^{(g)}$ and $S^{(b)}$ by considering the possibility that the utility of an individual itself depends upon social interactions taking place at the equilibrium contagion frequencies. For example, we can obtain a utility of the form $U(I^{(g)},S^{(b)}) = (I^{(g)})^2 S^{(b)}$ under a process in which individuals are paired with a random partner, and an individual only receives a positive payoff if they are susceptible to the bad contagion and both they and their partner are infected by the good contagion. Analogously, we can derive a utility of the form $U(I^{(g)},S^{(b)}) = I^{(g)} \left(S^{(b)}\right)^2$ in a scenario in which an individual only obtains positive payoff if they are infected with the good contagion and if they and a randomly chosen partner are both susceptible to the bad contagion. 

By considering more complex matching processes and rules for assigning positive payoff, similar Cobb-Douglas utilities can be derived, highlighting the different ways in which the enjoyment of access to beneficial contagions and absence of harmful contagions can depend on the state of the whole population. While our underlying contagion processes are simple SIS contagions, the presence of utility functions of this form means that the evolutionary dynamics can depend on the population state in a manner reminiscent of complex contagion process and of models of evolutionary games with imitation protocols resembling complex contagion \cite{vasconcelos2019consensus}.

\end{document}